**Title:**

# Superconducting nanowire for single-photon detection: progress, challenges and opportunities

*Itamar Holzman and Yachin Ivry**

I. Holzman Author 1, Prof. Y. Ivry Author 2

1. Department of Materials Science & Engineering, Technion – Israel Institute of Technology, Haifa, 32000, Israel.

2. Solid State Institute, Technion – Israel Institute of Technology, Haifa, 32000, Israel.

E-mail: ivry@technion.ac.il

# Table of contents

## Abstract

Single-photon detectors and nanoscale superconducting devices are two major candidates for realizing quantum technologies. Superconducting-nanowire single-photon detectors (SNSPDs) comprise these two solid-state and optic aspects enabling high-rate (1.3 GBit $s^{-1}$) quantum key distribution over long distances (>400 km), long-range (>1200 km) quantum communication as well as space communication (239,000 miles). The attractiveness of SNSPDs stems from competitive performance in the four single-photon relevant characteristics at wavelengths ranges from UV to the mid IR: high detection efficiency, low false-signal rate, low uncertainty in photon time arrival and fast reset time. However, to-date, these characteristics cannot be optimized simultaneously. In this review, we present the mechanisms that govern these four characteristics and demonstrate how they are affected by material properties and device design as well as by the operating conditions, allowing aware optimization of SNSPDs. Based on the evolution in the existing literature and state-of-the-art, we propose how to choose or design the material and device for optimizing SNSPD performance, while we also highlight possible future opportunities in the SNSPD technology.

## 1. Introduction: Quantum materials for quantum sensing

There has been recently a global rapid-pace growing interest in quantum technologies. In particular, the importance of systems for quantum communication, quantum encryption and quantum key distribution has been pronounced.[1–7] Single-photons are quantum creatures, which are attractive candidates for serving as the medium of these technologies.[8–12] Single photon detectors are thus a pivot technology for realizing the potential of quantum photonic systems. From the technological perspective, single-photon detectors are characterized by two main properties, each of which is divided to two sub-categories, so that there are four main characteristics. The first category includes the efficiency at which the device detects photons—the probability of detecting a single photon (device efficiency, DE) and the rate at which false signals are recorded (hereafter, dark-count rate, DCR). The second category concerns the speed or the time performance—the uncertainty in time detection of an arriving photon (timing jitter) as well as the time duration after one photon has been detected, while the detector is not ready yet sense the following photon (reset time).

The four characteristics of a single-photon detector dictate a set of requirements from the device physics. Examples for such requirements are the spectral absorption of the material at the relevant wavelength (e.g. 1550 nm for fiber or silicon-compatible communication), fast translation of the arriving photon to a detectable electric signal, fast recovery time after each detection event, low thermal and electric fluctuations, reliable and easy processing as well as low-processing and low-operation costs. Hence, by understanding the mechanisms that govern this device physics, we can deduce the requirements on the material properties and design of the device for optimizing the single-photon detection technology.

Developed by Gol’tsman et al. in 2001,[13] superconducting-nanowire single-photon detectors

(SNSPDs) are an excellent example for an already-working technology that comprises two important aspects of the quantum world—quantum light (single photons) and quantum matter (nano-superconducting structures). SNSPDs outperform competing single-photon detection technologies (for complementary review, see e.g. Hadfield[14] and Eisaman et al.[15]). The demonstrated advantageous efficiency and timing performance of SNSPDs comprise 93% system detection efficiency,[16] $10^{-4}$ counts per second (cps) dark-count rate,[17] 4.6-ps timing jitter[18] and 119-ps reset time[19] at the communication-relevant wavelength (1550-nm), while these characteristics are improved for photons of a shorter wavelength.[18,20–22] To-date, there is no single device that can show all these characteristics simultaneously, encumbering the realization of the great potential of SNSPDs (for a summary of current state of SNSPD performance, see representative data in **Table 1)**. Thus, the goal of this review is to discuss the progress of SNSPDs, explain the origin of their advantageous characteristics and point out on the opportunities they encompass.

The attractive potential of SNSPDs covers several specific applications within and beyond the quantum world, some of which have already been implemented, at least partially. Such applications that have been progressed significantly in the past few years include high-rate (1.3 Gbit $s^{-1}$ [23]) and long-distance (404 km[24]) quantum key distribution,[25] long-distance quantum communication (>1200 km[26]), satellite laser ranging[27] as well as high-definition video transmission from space (20 Mbps for 239,000 miles).[28] Moreover, SNSPDs are implemented in quantum cryptography and quantum tomography[29] as well as long-distance (>1 km[30]) imaging, while they have been used for characterizing fundamental quantum-mechanic properties, e.g. in loophole-free Bell tests[31] and non-classical single photon and phonons correlation.[32] Lastly, SNSPDs have also been utilized also for various non-quantum technologies, such as failure analysis of integrated circuits,[33] absolute temperature rapid measurements,[34] molecular spectroscopy.[34] See **Figure 1** for examples. A detailed survey of applications of SNSPDs can be find, e.g. in reviews by Natarajan et al.[35] and by Yamashita et al..[36] We should note that the great technological requirements for SNSPDs beyond the academic world has raised the need to make such devices available commercially so that today, SNSPDs are supplied by companies, such as SCONTEL, Single Quantum, PhotonSpot, Quantum Opus and Photech.

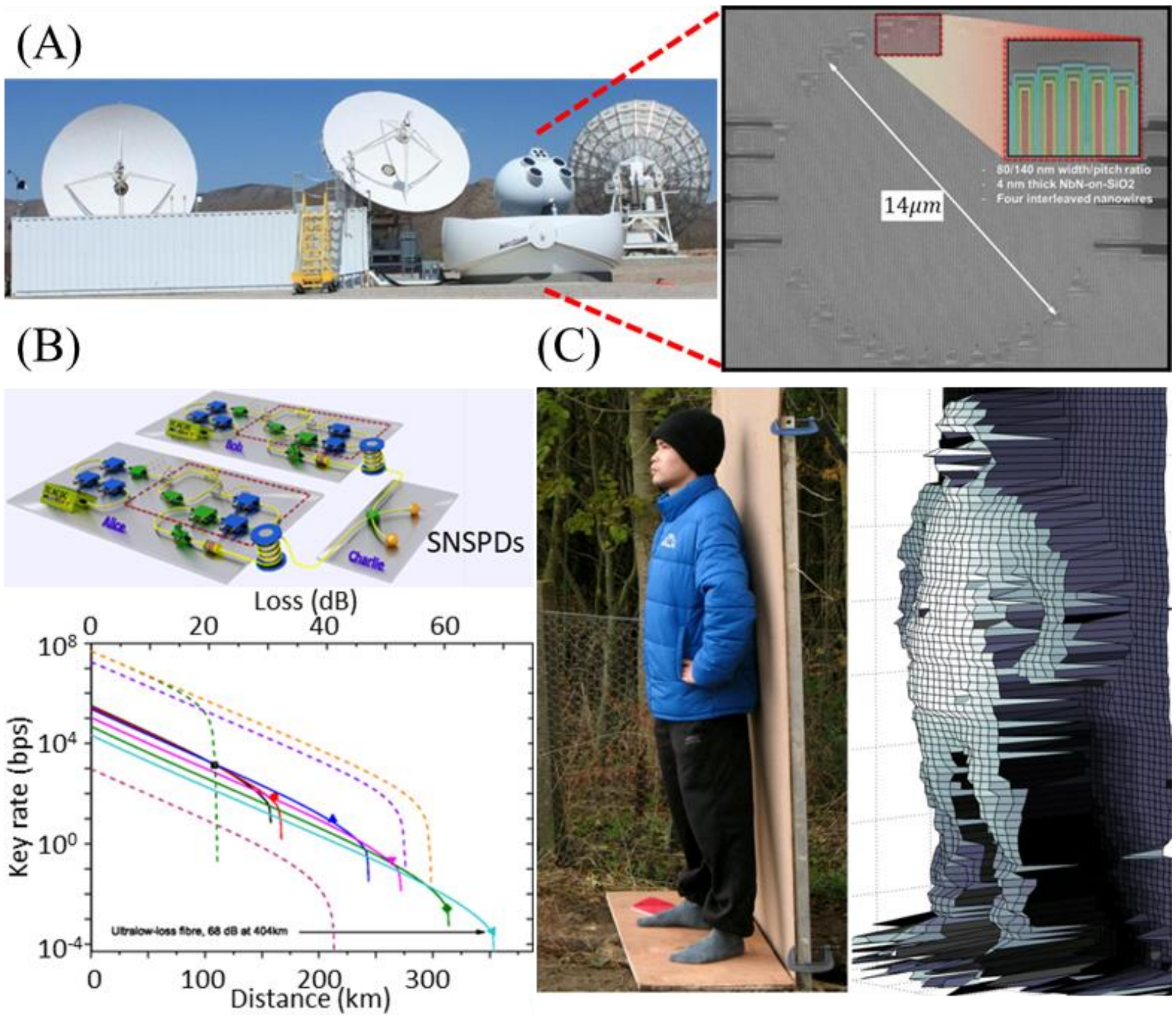

**Figure 1| SNSPD applications**. Examples for modern quantum and advanced technologies that are based on SNSPDs include: (**A**) Interstellar long-distance high-speed communication (device and system are highlighter). Adapted with permission.[37] 2015 © Springer Nature Switzerland AG; (**B**) Long-distance high-speed quantum key distribution. Reproduced with permission.[24] 2016 © American Physical Society; and (**C**) Long-distance imaging. Reproduced with permission.[30] 2013 © The Optical Society.

**Figure 2** illustrates schematically the system of SNSPDs. The core element of a SNSPD,

presented first by Gol'tsman et al. in 2001,[13] is a meandering superconducting nanowire, which is current biased, awaiting to convert the energy of an absorbed single photon to a measurable voltage. The output electronic signal from the meandering superconducting wire is amplified before it arrives to a counter (or an oscilloscope), while a bias tee sends some of the signal also to a shunt that is connected in parallel to the wire. The complete system includes also a single-photon source, which is usually an attenuated polarized laser light. Detailed discussion regarding system optimization can be found in a 2014 review by Dauler et al.[38]

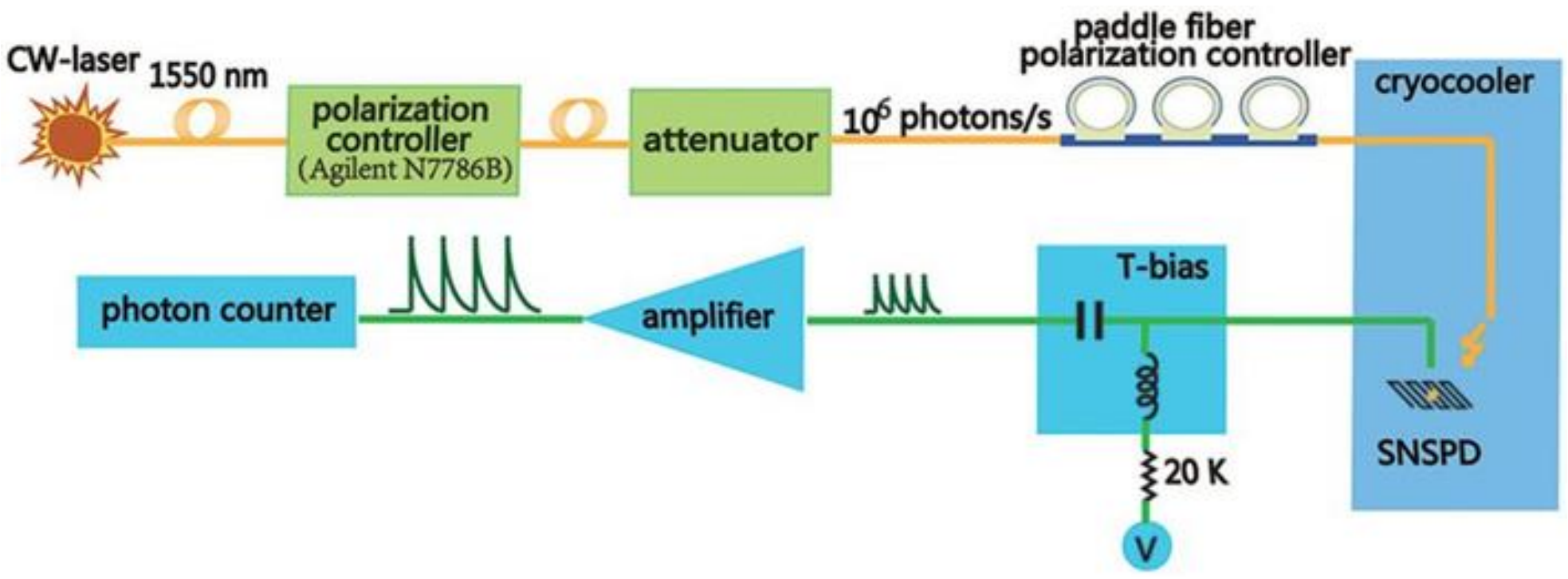

**Figure 2| A typical SNSPD system**. Light is transmitted from a single-photon source to a meandering superconducting nanowire that absorbs the photon and converts the signal from photonic to electronic. The electronic signal is amplified and measured with a counter (or an oscilloscope), while the superconducting wire is also connected in parallel to a shunt impedance that prevents the detector from latching. Reproduced with permission.[39] 2015 © Macmillan Publishers Limited.

Alongside the great progress in SNSPD applications and system optimization, there has been accumulated literature that concerns important advances related to the core element—the superconducting nanowire detector. Such advances include improvements in our understanding of the fundamental detection mechanisms as well as in the factors that affect the efficiency and timing properties. Despite these developments in our understanding of SNSPDs, which led to significant improvements in the SNSPD properties, there is a real need in organizing, analyzing and facilitating the accumulated data in an accessible manner. Such an analysis can help optimize the device properties by increasing the awareness of how they are affected by the choice of material, the device design and the operating conditions. In this review, we aim at bridging this gap. By surveying the existing literature, we present the mechanisms that govern device efficiency, dark-count rate, timing jitter and reset time of SNSPDs. We demonstrate how these four characteristics are affected by the material properties of the superconducting wires as well as by the operating parameters and by the design of the device. The origin of relevant material properties are also presented. Lastly, we summarize how to optimize the parameters that affect each of the four device performance properties, while we also cover several potential opportunities that have not yet been accomplished.

### 1.1 How to read this review?

The main objective of this review is to supply the readers with a set of tools for optimizing SNSPD efficiency and timing properties. We start with an introduction that covers the relevant fundamentals behind SNSPDs, which include both the basic detection mechanisms of SNSPDs and the properties of superconducting materials at the nanoscale. We should note that the manuscript is constructed so that readers not interested in the fundamentals of superconducting nanostructures or the basic mechanisms of SNSPDs can skip the corresponding sections and yet make benefit of our analysis of the effects of material parameters, device design and operating parameters on the SNSPD performance.

We then discuss in detail each of the four properties: detection efficiency, false counts, timing jitter and reset time. For each property, we first introduce the fundamental mechanisms that govern this property. We then continue and demonstrate on how the intrinsic properties--material properties (e.g. kinetic inductance, resistance, critical current and critical temperature, absorptance and homogeneity), device geometry and design as well as the system design--affect each of the four device-efficiency or timing characteristics. In the end of each chapter that concerns one of the four device characteristics, we present a summarizing paragraph that suggests how to improve the specific characteristic. We then complete the discussion by elaborating on how the extrinsic properties, which can be tuned during the experiment (bias current, temperature as well as wavelength and polarization of the light source) affect the different device characteristics. For each of the four device characteristics, we survey the evolution in performance as well as in the understanding of the governing mechanisms, with an emphasis on the state of the art. In addition to examples from the literature, we illustrate the different mechanisms visually, while we also facilitate the data regarding the effects of material properties on device characteristics in **Tables 1** and **2**. Lastly, we discuss within the text some of the open opportunities in the field that have still not been addressed.

We should note that the separation between efficiency and timing performance as well as between the four different characteristics is done here to assist us in supplying the readers with an organized overview from the technological-motivation perspective.

### 1.2 Detection mechanisms

To-date, the exact fundamental detection mechanisms of photons by SNSPDs is not yet completely understood. There are broadly two main competing mechanisms. The first model concerns a photon absorption in metals that increases the temperature locally by a progressing hot-spot.[13,40] The local temperature increase gives rise to an increase in resistance. When a superconductor is held near the phase transition, small changes in temperature give rise to large differences in resistance, allowing in turn single-photon sensitivity. In weak-link geometries, such as nanowires, the change in resistance is even larger and so is the resultant detection efficiency. Depending on the ratio between the photon wavelength ($\lambda$) and the binding energy of the Cooper pair (Δ), the hotspot diameter in <10-nm-thick SNSPDs typically ranges between a few nanometers to a few dozens of nanometers.[41,42] This diameter is usually smaller than the width of the wire (ca. 100 nm). Thus, in the hotspot model, after the formation and growth of the hotspot itself, the entire wire becomes resistive only because the current that flows in the area around the hotspot exceeds the critical value.[13,41] Specifically, Semenov et al.[41] showed a direct dependence of the detection of a photon with a certain wavelength on the energy gap already in 2001, suggesting the following limit on a detectable photon energy:

$$h\frac{c}{\lambda} > n\Delta N(0)k_{\mathrm{B}}T_{\mathrm{c}}Dd\tau_{\mathrm{th}} \quad \textbf{(1a)}$$

Here, $h, c, n, N(0), k_{\mathrm{B}}, D, d$ and $\tau_{\mathrm{th}}$ are respectively Planck's constant, speed of light, an energy-loss factor, the normal metal density of states at the fermi level, Boltzmann constant, normal state diffusivity, film thickness and electron thermalization time.

The authors[42] later added a correction by accounting for the influence of bias current ($I_{\mathrm{b}}$) on the hotspot size, i.e. defining the hotspot area as: $\mathrm{w} \cdot \left(1 - \frac{I_{\mathrm{b}}}{I_{\mathrm{c}}}\right)$, where w is the width of the nanowire and $I_{\mathrm{c}}$ is the wire critical current. Thus, assuming that $w \sim \sqrt{D\tau_{\mathrm{th}}}$ as well as $k_{\mathrm{B}}T_{\mathrm{c}} \approx \frac{\Delta}{\sqrt{\pi}}$, the cutoff energy to form a detectable hotspot becomes:

$$h\frac{c}{\lambda} = \frac{N_0\Delta^2 wd\sqrt{\pi D\tau_{\mathrm{th}}}}{\varsigma}\left(1 - \frac{I_{\mathrm{B}}}{I_{\mathrm{C}}}\right) \quad \textbf{(1b)}$$

Where ς is multiplication efficiency of quasiparticles.

The second model suggests that vortex-antivortex pairs govern the stability of the superconducting nanowire rather than the electron Cooper pairs. This model complies with the framework suggested by Berezinskii, Kosterlitz and Thouless (BKT)[43–45] regarding the phase transition mechanism in superconducting thin films.[46–49] There are two possible mechanisms to the vortex-assisted destruction of superconductivity. (i) The photon absorbed in the wire results in a vortex-antivortex pair. The current flow in the wire exerts a Lorentz force that acts on the vortex and the antivortex in opposite directions. When this force is large enough to break this vortex-antivortex pair, the wire transfers to the normal state with a measureable resistance. (ii) The second possible scenario is that the absorbed photon heats the wire and hence suppresses the energy barrier for a vortex to penetrate the wire. Consequently, the current flow helps the vortex cross the wire throughout its entire width, disrupting the superconductivity so that the wire becomes normal. Although the origin of these two scenarios are different in nature, it is challenging to distinguish between them experimentally from the photon detection only, because the resultant measureable current distribution after a single-photon detection event is identical.[47] Thus, other approaches are required, such as characterization of the device dark counts.[50]

The validity and limitations of the frameworks of the hotspot (typically, at high photon energies) and vortex-antivortex models (usually at low photon energies) in SNSPDs have been discussed in length by Engel et al.,[51] suggesting that currently, no single model that explains all the observations. However, we should note that the device material and geometry might dictate the governing mechanism, while different mechanisms can exist at the same detector, but at different photon energy and sometimes several mechanisms can take place even simultaneously. The different mechanisms are illustrated schematically in **Figure 3**, following Renema et al.[52]

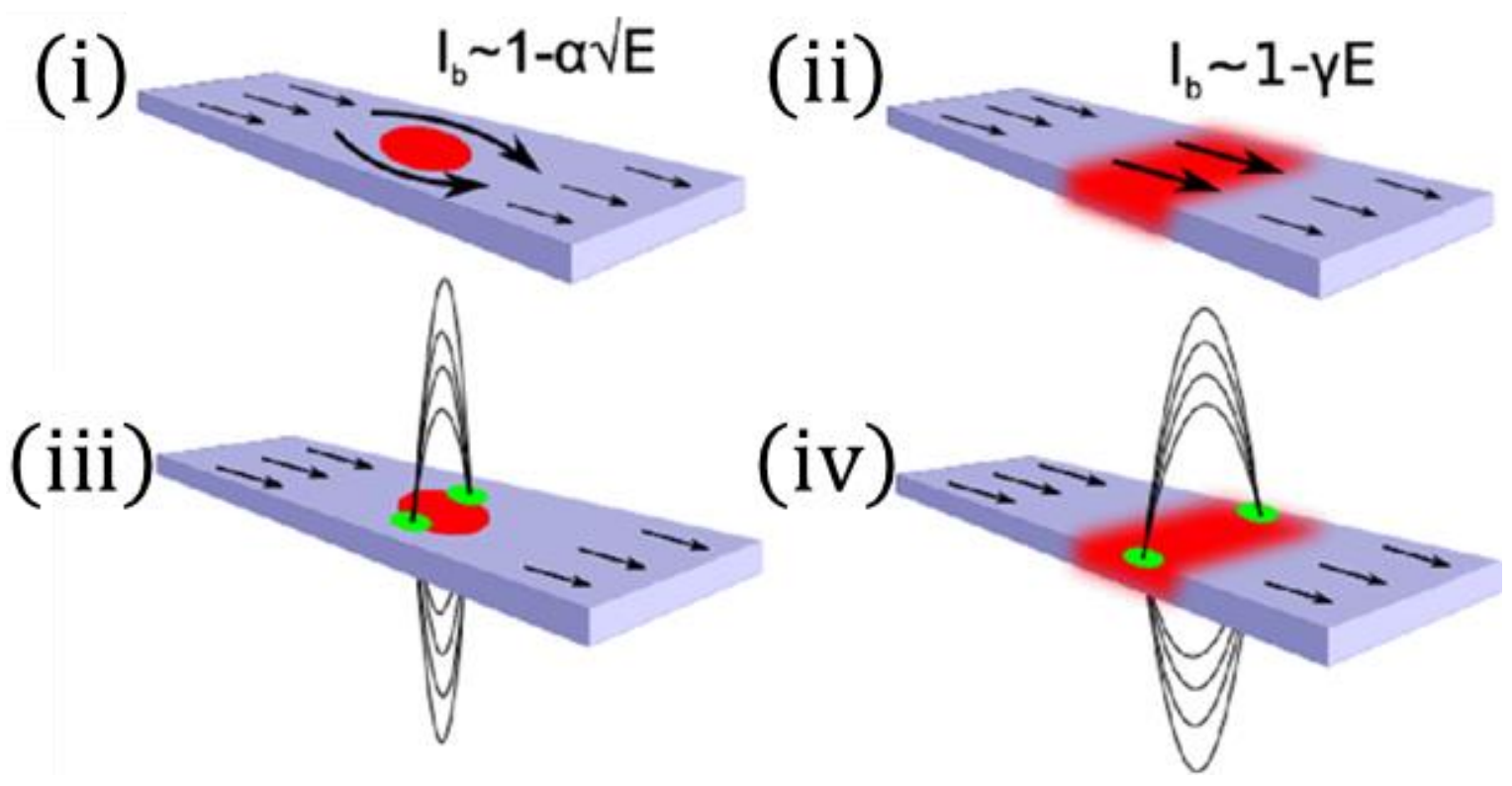

**Figure 3| Detection mechanisms in SNSPDs**. Competing detection mechanisms in SNSPDs include (i) hotspot formation by the breakage of Cooper pairs, leading to (ii) an increase in current density beyond its critical value; (iii) vortex crossing; and (iv) breakage of a vortex-antivortex pair. Reproduced with permission.[52] 2014 © American Physical Society.

## 1.3 Origin of device performance

The performance metrics of SNSPDs are divided into two categories from practical point of view: (i) efficiency (DE and DCR); and (ii) temporal characteristics (timing jitter and reset time). These characteristics depend on extrinsic and intrinsic sets of parameters as well as on the specific design of the device. Intrinsic parameters are those set by the material chemistry, structure, electronic properties, and geometry. These parameters are fixed for a given device. Extrinsic parameters are parameters that are tunable during a measurement, such as temperature, bias current as well as photon wavelength. These parameters are the degrees of freedom that one can play with when operating the device. In addition, system components and parameters, such as shunt impedance, electronic components (e.g. filters and amplifiers) are also affecting the device performance extrinsically. The dependence of device characteristics on some intrinsic and extrinsic properties is given in Table 2. We can now discuss the origin of these properties as well as how they affect the device performance.

## 1.4 Relationship between intrinsic properties in nano-superconducting structures

The individual intrinsic properties of bulk superconducting materials, such as critical temperature (as well as critical current and magnetic field), normal resistance and superconducting gap are intertwined and not completely independent.[53] Upon miniaturization of the superconducting structures, e.g. nanoscale films or wires, the superconducting properties are suppressed and the relationships between them changed (see **Figure 4**).[54,55] This size effect near the superconducting-to-insulating transition (SIT) is an active field of research, which has attracted even more attention after Kosterlitz and Thouless won the 2016 Nobel Prize for their theoretical work on superconductivity in thin films.[43–45] Thus, we give here only a brief summary of the parameters and relationships that are most relevant to SNSPDs both as thin films and as nanowires, which are affected by the SIT.

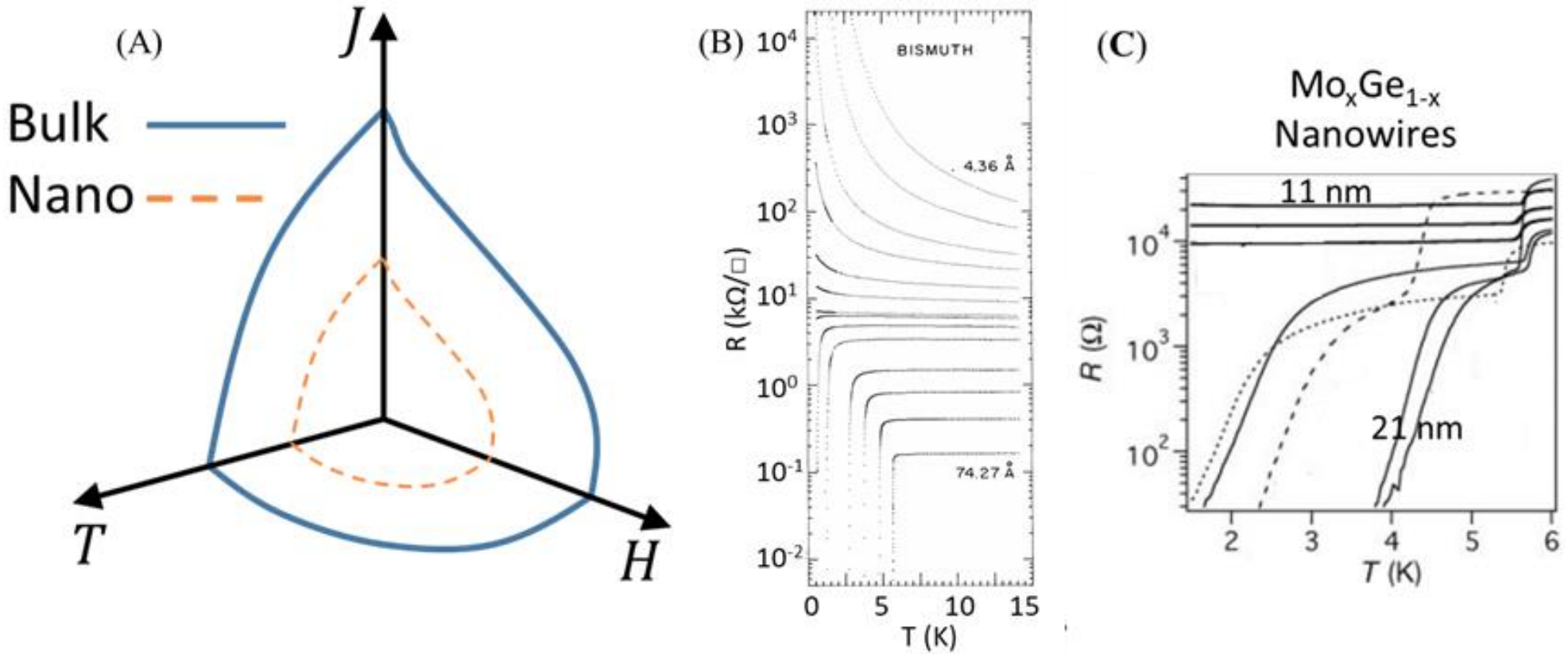

**Figure 4| Superconducting of superconducting properties at the nanoscale**. (**A**) Schematic illustration that demonstrates the suppression of superconducting properties upon miniaturization (within one eighth of the temperature-current density-magnetic field three-dimensional sphere). It is expected that below a certain size, the material loses completely its superconducting properties and becomes insulator, in a superconducting-to-insulator transition (SIT). (**B**) Upon miniaturization of thin superconducting bismuth films, the critical temperature decreases, while the normal-state resistance increases. Reproduced with permission.[54] 1989 © American Physical Society. (**C**) Similar size effect was reported for superconducting molybdenum germanium nanowires by Bezryadin, Lau and Tinkham. Reproduced with permission.[55] 2000 © Macmillan Publishers Limited.

Following Ambegaokar and Baratoff,[53,56] the multiplication between critical current and normal resistance ($R_{\mathrm{n}} = R_{\square} \cdot \frac{l}{w}$, $l$- length, w- width) in weak links (structures with suppressed superconductivity that are sandwiched between macroscopic electrodes) is approximately a constant, for a given temperature: $I_{\mathrm{c}} \cdot R_{\mathrm{n}} = \left(\frac{\pi\Delta}{2q_{\mathrm{e}}}\right)\tanh\left(\frac{\Delta}{2k_{\mathrm{B}}T}\right) \approx (T_{\mathrm{c}} - T) \cdot 635\,\frac{\mu V}{K}$, where $q_{\mathrm{e}}$ is electron charge. Likewise, You et al.[57] derived the dependence of kinetic inductance on material conductivity and energy gap, or equivalently, on the normal resistance and critical temperature:

$$L_{\mathrm{k}} = \frac{\hbar}{\pi\Delta\sigma_{\mathrm{n}}} \cdot \frac{l}{w \cdot d} = \frac{\hbar R_{\mathrm{n}}}{1.76\pi k_{\mathrm{B}} T_{\mathrm{c}}} \quad \mathbf{(2)}$$

Where $d$ is the wire thickness. Moreover, the relationship between the critical temperature, and the film thickness and sheet resistance has also been formulated empirically for a broad range of superconductors near the SIT limit: $T_{\mathrm{c}} = A \cdot R_{\square}^{-B}$ , where $A$ and $B$ are material-dependent constants that help classify the homogeneity of the superconductor (typically, $B \approx 1$ is the limit above which the materials are not crystalline, but amorphous).[58,59] This division is in agreement with accumulated data that suggest that the properties of amorphous and crystalline SNSPDs are different (with an emphasize on crystalline NbN and amorphous tungsten silicide, which are shown in **Figure 5**). There have been more traditional derivations for the dependence of $T_{\mathrm{c}}$ (and hence of $\Delta$) on the sheet resistance and thickness for thin films, depending on the dominating superconducting mechanism, which in turn, depends strongly on the material homogeneity. Nelson and Halperin[45] showed that under the vortex-antivortex dominancy of the BKT framework of thin-film superconductivity: $T_{\mathrm{c}} \approx T_{\mathrm{c}}^{0}\left(1 - \frac{0.17q_{\mathrm{e}}^{2}}{\sigma_{\mathrm{n}}\hbar}\right)$, where $T_{\mathrm{c}}^{0}$ is the bulk superconducting critical temperature, $\hbar$ is the reduced Planck constant and $\sigma_{\mathrm{n}}$ is the normal conductivity ($R_{\mathrm{n}} = 1/\sigma_{\mathrm{n}}$). McMillan[60] and Dynes[61] showed that when superconductivity is governed by phonon-mediated paired electrons (i.e. BCS framework) that interact with electric and magnetic impurities, the critical temperature becomes: $T_c = \frac{\Theta_D}{1.45}\exp\left[-\frac{1.04(1+\lambda^*)}{\lambda^* - \mu^*(1+0.62\lambda^*)}\right]$, where $\Theta_D$ is the Debye temperature, $\lambda^*$ and $\mu^*$ are coupling constants due to electric and magnetic impurities (respectively). Finkel'stein[62] showed that for homogeneous materials, such as amorphous molybdenum germanium, the critical temperature depends only on the sheet resistance, which is the parameter of disorder in the system: $\ln\left(\frac{T_c}{T_c^0}\right) = -\frac{e^2}{6\pi^2\hbar} g_1 R_\square \left(\ln\left(\frac{1}{T_c \tau_{mfp}}\right)\right)^3$, where $T_c$ and $T_c^0$ are the film and bulk critical temperature respectively, $g_1$ is an electron-electron interaction constant, $R_\square$ is the sheet resistance and $\tau_{mfp}$ is the electron mean-free-path time. Lastly, the absorption coefficient of the superconducting films or wires depends on their conductivity.[63,64]

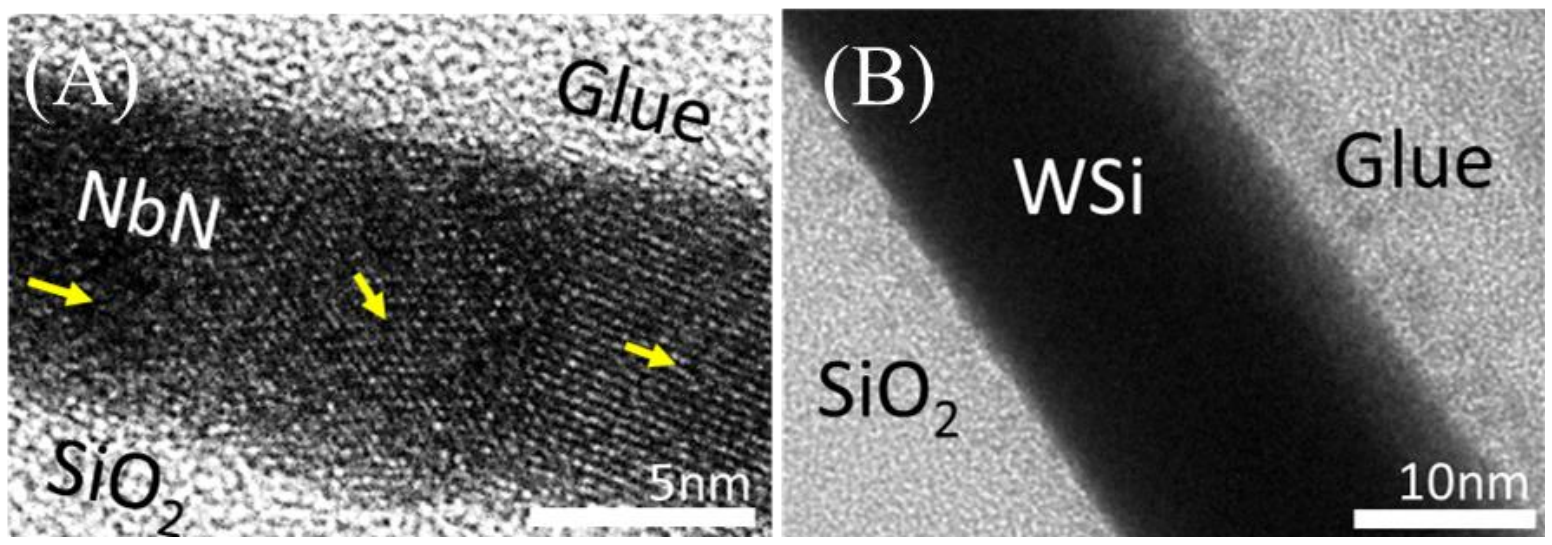

**Figure 5| Atomic structure of typical materials for SNSPDs**. (**A**) Atomic structure of a thin polycrystalline NbN film. (**B**) Amorphous structure of a thin tungsten silicide film. Reproduced with permission.[65] 2017 © IOP Publishing Ltd.

## 2. Intrinsic properties and functionality

We can now examine how different device characteristics are affected by the intrinsic properties. Although DE and DCR are not completely independent characteristics, while the same applies also to timing jitter and reset time, we will examine how intrinsic properties and device design affect each of these four characteristics individually.

## 2.1 Efficiency

### *2.1.1 Device Efficiency*

*Mechanism*

Device efficiency (DE) is the likelihood of detecting a photon, which is a multiplication of the probability of three independent sequential processes[16,66,67] as schematically illustrated in **Figure 6**. First, the photon has to arrive to the physical area of the device. The coupling of the device to the light source, which depends on the design of both the system and device, thus defines the first term of the probability ($\eta_{cpl}$). The second term ($\eta_{abs}$) reflects the probability of a photon that arrives to the region of the device to be absorbed in the superconducting wire. This term is the effective optical absorptance of the device ($\alpha$), which means it contains not only the material absorptance, but also the contribution of design factors, such as mirrors, fill factor, cavities and antennae, while $\alpha$ depends on the photon wavelength and polarization. The third part of the process is the quantum efficiency ($\eta_{qe}$). This intrinsic property defines the probability that a photon absorption event will produce a measurable signal. That is, $\eta_{qe}$ represents the likelihood that the energy transfer to the wire from photon that has been absorbed will heat the wire, break Cooper pairs (high photon energy) or move an unpaired vortex (lower photon energy) so that the wire will become resistive for a measurable time period. We should note that most studies discuss the efficiency at the communication-relevant wavelength, 1550 nm. The data presented in this review are usually concerns single-photon detection of this wavelength, unless noted otherwise. Schematic illustration of the how photons are absorbed and detected is given in Figure 6.

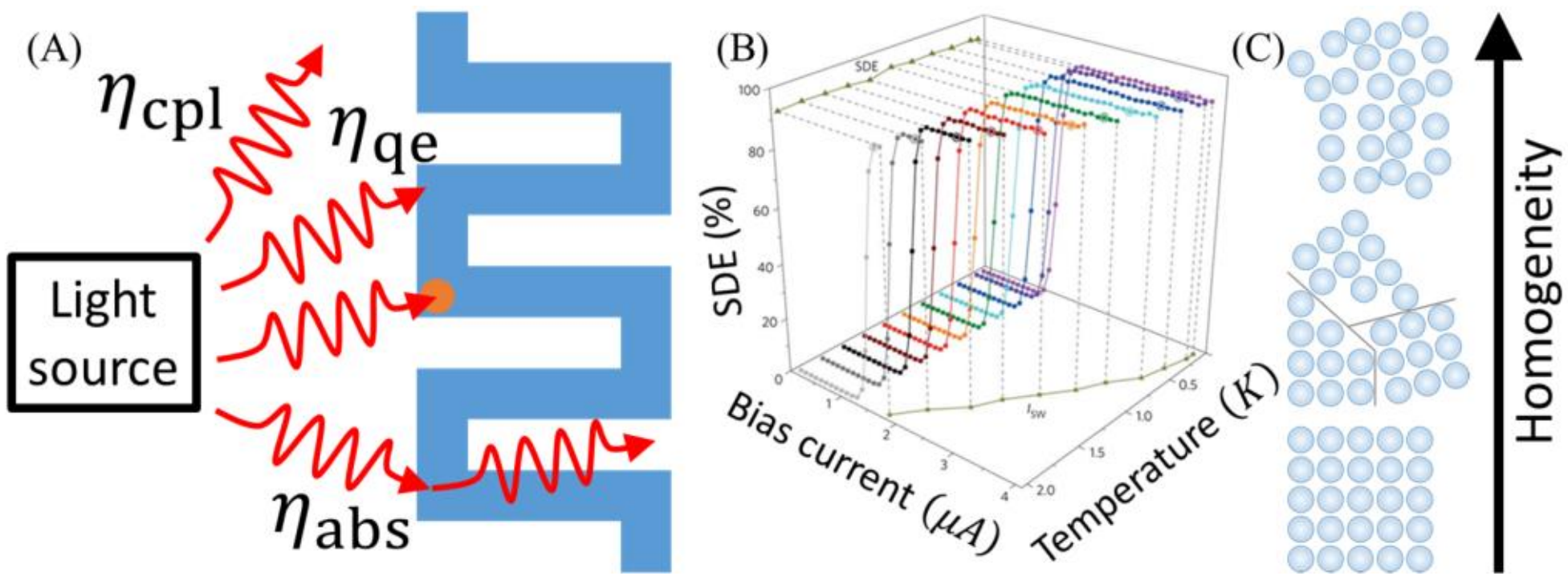

**Figure 6| Efficiency of photon-detection events in SNSPDs**. (**A**) Schematic illustration of the three-step photon-detection mechanism in SNSPDs— coupling to the optical system, spectral absorption, quantum efficiency. (**B**) Detection efficiency and its saturation as a function of extrinsic parameters for a $W_{0.55}Si_{0.45}$ SNSPD (reproduced with permission.[16] 2013 © Macmillan Publishers Limited), while the role of increasing material homogeneity (**C**) from crystalline to amorphous via polycrystalline (bottom-to-top) is also emphasized.

*Intrinsic and design solutions*

*Design-* $\eta_{cpl}$

To maximize DE, several optical designs have been introduced to increase both $\eta_{cpl}$ and $\eta_{abs}$. In most studies, the light is brought via an optical fiber to a chip that comprises the SNSPD,[16,17,68–80] some use manipulator to better align the fiber and the SNSPD,[76,77,81,82] while Miller's technique[83] for self-aligning transition edge sensors to optical by fibers by pre-patterning the chip has also been adopted for SNSPDs[16] (see details in Ref. 38). In these devices, large the active area[84,85] and fill factor[86] or optimization of the fill factor and the thickness[87] of the SNSPD play an important role in increasing DE. To improve the fiber-SNSPD coupling, Bachar et al.[72] processed a meander detector directly on the optical fiber (**Figure 7B**). Pernice et al.[88] demonstrated improved DE (91% compared to 3%) by integrating SNSPD and waveguide on the same silicon chip, while Khasminskaya et al.[89] integrated also the light source on the same chip **(Figure 7C).** Likewise, Najafi et al.[90,91] integrated the SNSPD with waveguides by flip-chip transferring the detector (**Figure 7D**). Lastly, several designs have focused on free-space coupling of single photons to the SNSPDs, e.g. for communication and imaging applications.[30,85,92–95]

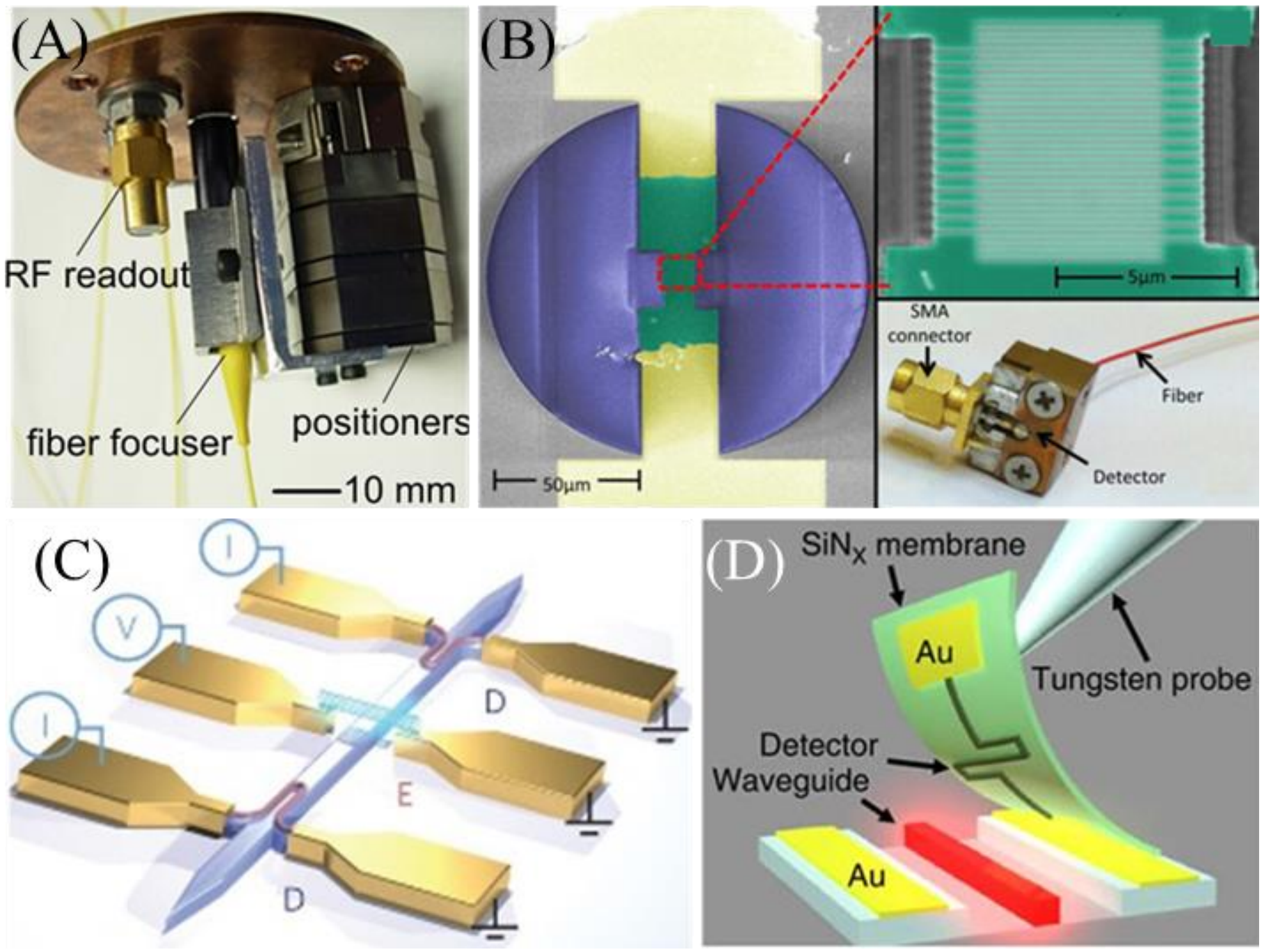

**Figure 7| Typical examples for system optical coupling of SNSPDs**. (**A**) The most common optical coupling of a meandering SNSPDs is aligning an optical fiber to the microns-area SNSPD that was processed on an independent chip. Reproduced with permission.[77] 2009 © The Optical Society. (**B**) A meandering SNSPD that was fabricated directly on the fiber. Reproduced with permission.[72] 2012 © AIP Publishing. (**C**) Waveguide-coupling of an on-chip integrated SNSPD. Reproduced with permission.[89] 2016 © Macmillan Publishers Limited. (**D**) Transferrable SNSPD that was processed on a membrane. Reproduced with permission.[90] 2015 © Macmillan Publishers Limited.

### *Intrinsic* $\eta_{qe}$

The accumulated literature on SNSPDs suggests that from the material perspective, there is a strong influence of material parameters on DE. For instance, Marsili et al.[16] demonstrated that SNSPDs based on amorphous materials might show efficiency (93% system efficiency tungsten silicide at 120 mK) higher than the previous crystalline SNSPDs. This high system efficiency has garnered much interest in the technological potential of SNSPDs and has initiated an ongoing race to higher efficiency at technologically optimal conditions. More recent works show promising increase of DE for other amorphous materials, mainly in molybdenum-based SNSPDs.[58,96–98] In addition to the design of the system, the high efficiency is attributed to 100% quantum efficiency, which was characterized by long-range saturation of the efficiency as a function of bias current (Figure 6).[99] This saturation behavior was then found to be more typical for amorphous materials.[16,75,97–101] Marsili et al.[16] also showed (see for example Figure 6B) that the saturation region decreases with increasing operating temperature. Given that bringing the device closer to or deeper in the saturation region is equivalent to improving the device detection performance, these results suggest that for a given temperature, reducing $\Delta_0$ (equivalent to reducing $T_c$, e.g. by changing the material) helps increase DE, in agreement with **Equations 1a-b**, where $\Delta_0$ is the Cooper-pair binding energy at $T = 0K$.

Dorenbos et al.[102] showed an increase in DE for SNSPDs based on the low-gap material, NbSi. On the other hand, the saturation region might shrink (see also Figure 6B), or even be eliminated, when $I_c$ is reduced due to design reasons, e.g. when the meandering wire allows significant current crowding.[103–105] Similarly, reduction of $I_c$ might decrease DE also when the wire is inhomogeneous and contains constrictions.[106] Such constrictions may originate along the width of the wire during fabrication by electron lithography (despite a transition of many researchers from a negative-tone process to positive-tone lithography[107]) or more severely, by focused ion beam.[108] Moreover, constrictions can arise also in the top (or bottom) face of the wire, e.g. due to substrate roughness (we should note that the substrate may affect DE also through the hotspot dynamics[109]). For instance, Atikian et al.[110] showed that SNSPDs on smooth (300-pm RMS roughness) diamond substrates contain no constrictions, while Rath et al.[111] used the diamond substrate for integrated nano-photonic circuits. Miki et al. demonstrated with both NbN[112] and NbTiN[113] SNSPDs that were grown epitaxially on MgO substrates that good lattice matching (along the ⟨100⟩ direction) also reduces constrictions and increases DE. Following the early work of Wang et al. in 1996,[114] Marsili et al.[115] showed that the epitaxial growth allows reduction of the substrate temperature during the sputtering, which is a necessity for integrated devices that require multiple fabrication steps (we should note that although most superconducting films are deposited via sputtering, molecular beam epitaxy[116] have also been demonstrated, while not much effort has been put in other deposition methods that are available for ultra-thin superconducting films, e.g. atomic layer deposition,[117] and pulsed laser deposition[118]). Likewise, the accumulated literature on NbN devices suggest that constrictions are greater when the grain size of the polycrystalline material is comparable to the width of the wire. That is, the grains of NbN films that are grown on MgO are comparable in size to the wire width as opposed to NbN that grows epitaxially on sapphire[59,119]--the first substrate used for SNSPDs[13](see detailed analysis by Espiau de Lamaestre et al.[120]). Another way to overcome the effect of grain size comparable to the wire width is to grow polycrystalline films with small grains, e.g. as in the case of amorphous substrates, such as $SiO_2$ (Figure 5)[65] or $Si_{1-x}N_x$,[91,121] both are preferable for integrated photonic circuits. The effects of substrate on NbN film properties were surveyed also by the Glasgow group in 2012,[122] in which other integration-relevant substrates, such as $LiNbO_3$ and GaAs were demonstrated useful for lower temperature deposition. Likewise, the Glasgow group[123] proposed another method to overcome the effects of substrate – using amorphous materials ($Mo_{1-x}Si_x$), in which there are no constrictions due to granularity or lattice matching. This approach can explain the previously reported relatively high detection efficiency in SNSPDs made of other amorphous materials, such as tungsten silicide.[16] Finally, the probability for constrictions increases also with increasing length of the SNSPD.[106] Thus, shorter wires are required, while waveguide coupling can compensate for the smaller active area.[88]

Recent works suggest that higher DE is obtained for SNSPDs made of superconductors with larger disorder (i.e. closer to the SIT). Thus, although amorphous materials allow for disorder naturally, crystalline materials with intentionally induced high disorder may also exhibit high efficiency. Indeed, Smirnov et al.[20] demonstrated 94% system efficiency for a crystalline NbN device that was designed with high $R_{\square}$. Nevertheless, Zadeh et al.[124] presented a crystalline NbTiN device with 92% system efficiency (for 1310 nm wavelength) and demonstrated that $R_{\square}$ should be optimized carefully, because higher sheet resistance requires thinner films,[59] which means that the optical absorptance might be affected as well.

*Design* $\eta_{abs}$

In addition to the effort put in increasing $\eta_{\mathrm{cpl}}$, there have been several device designs that address $\eta_{\mathrm{abs}}$, i.e. aiming at increasing the time that a photon stays near the device. Early works discussed the effects of fill factor[125–128] and film thickness.[80,113,129,130] Milostnaya et al.[131] then improved the DE thanks to the introduction of cavities, while Rosfjord et al.[127] demonstrated enhancement factor of 2.562 in DE by combining both cavities and anti-reflection coating (typically a 256-nm thick $SiO_2$ back layer is used for SNSPD systems with front illumination). Other methods to increase the optical absorption comprise the introduction of antennae including plasmonic antennae.[132–135] Examples for anti-reflecting coatings cavities and antennae are given in **Figure 8**.

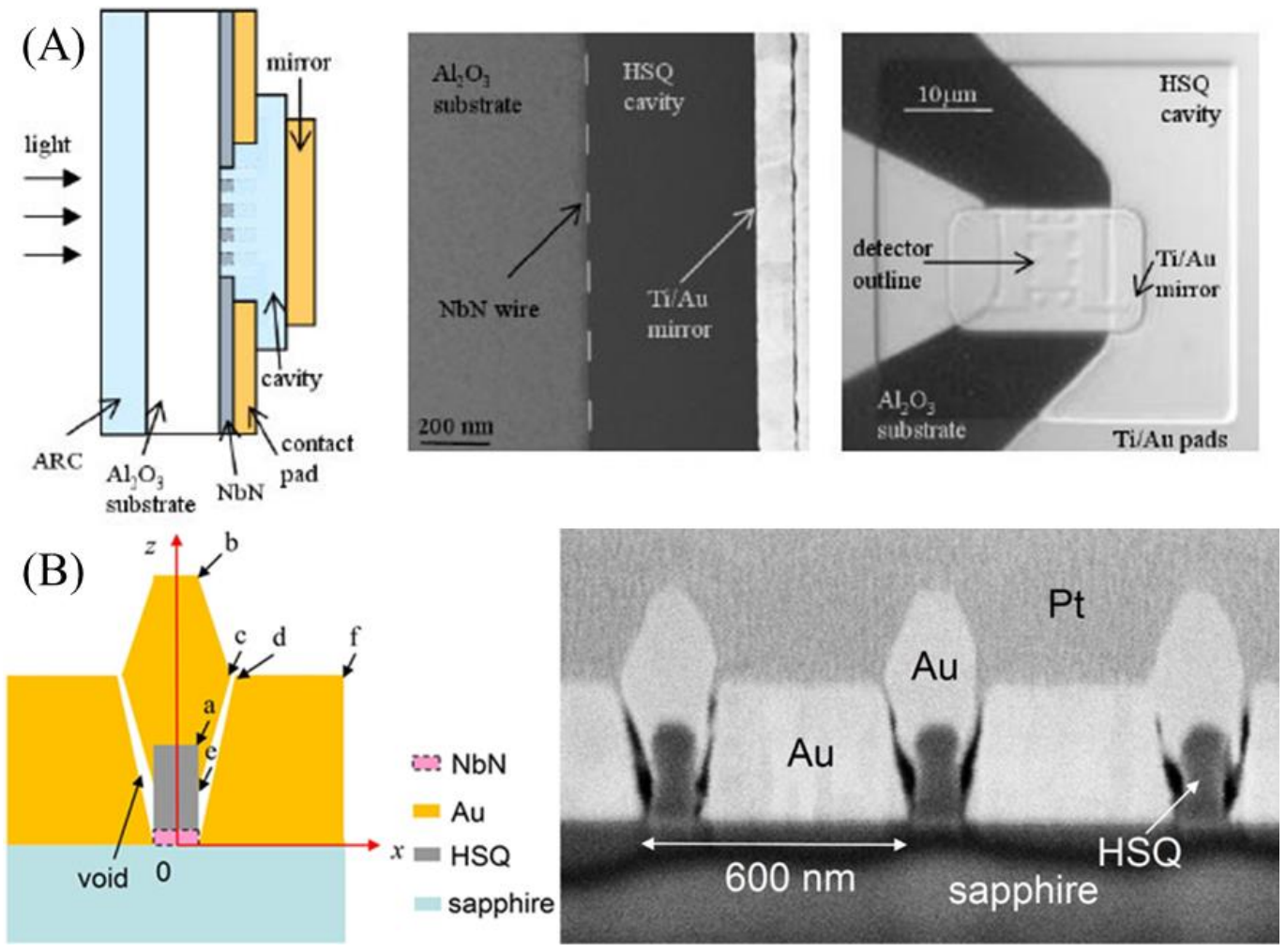

**Figure 8| Optical coupling of SNSPDs**. (**A**) Schematic illustration and electron micrographs of cavity and an anti-reflection mirror. Reproduced with permission.[127] 2006 © The Optical Society. (**B**) A SNSPD with optical antennae. Reproduced with permission.[132] 2001 © The Optical Society.

Another factor that must be regarded when discussing the photon absorption in the detector is the photon polarization and its interaction with the device. Being a wire, SNSPDs have one-dimensional symmetry that makes it sensitive to photon polarization. Typically, for a meandering SNSPD, the light-source polarization is tuned parallel to the SNSPD plane (transverse electric polarization, $T_{\mathrm{E}}$) to maximize absorption as illustrated e.g. in Figure 2. Anant et al.[130] showed already in 2008 that the photon absorption, and hence device efficiency, may vary substantially when changing the polarization when they measured 30% *contrast*, i.e. the difference divided by the sum of the parallel and perpendicular light absorption: $\frac{\eta_{\mathrm{abs}||}-\eta_{\mathrm{abs}\perp}}{\eta_{\mathrm{abs}||}+\eta_{\mathrm{abs}\perp}}$. An example for the efficiency dependence on polarization is given in **Figure 9**A.

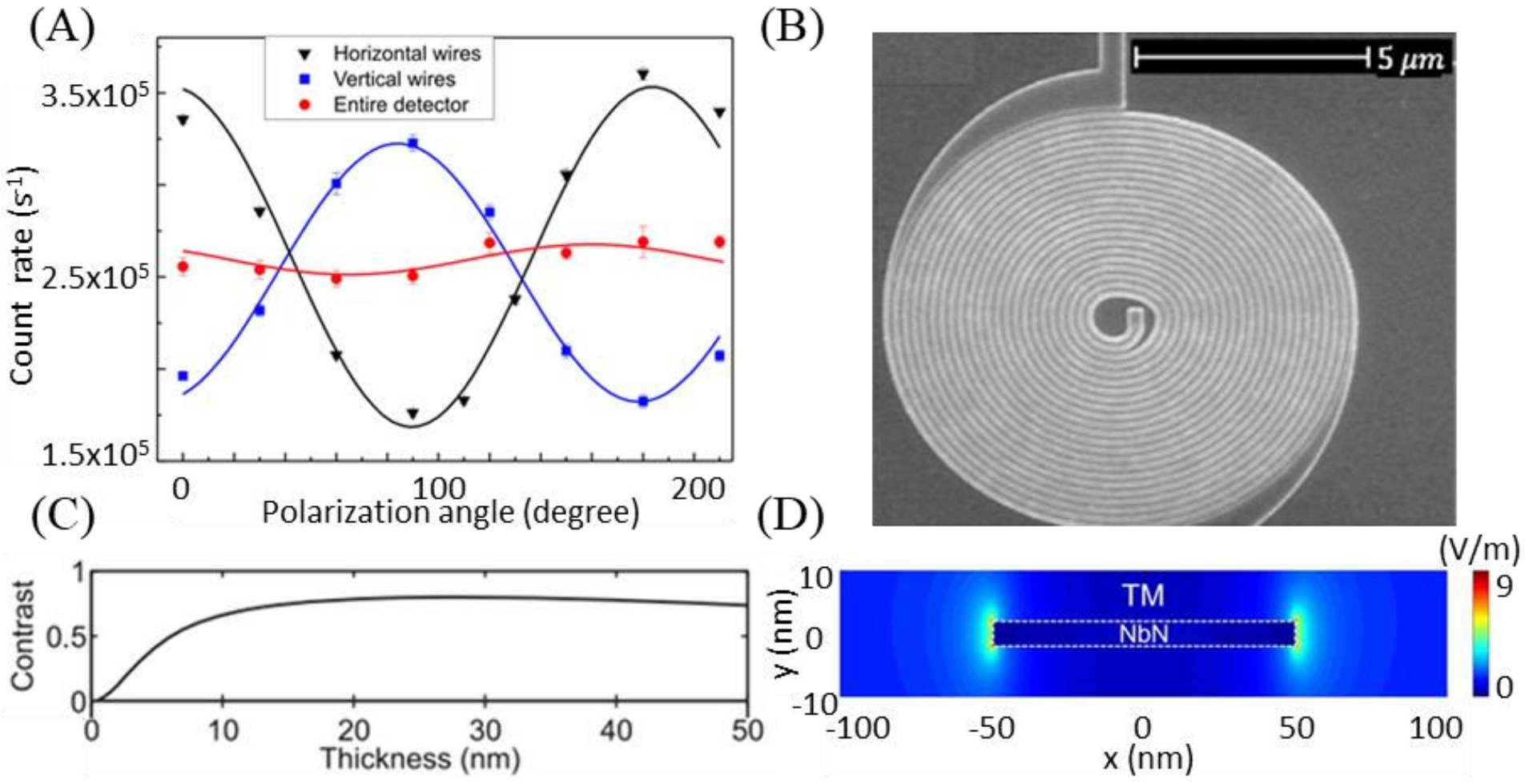

**Figure 9| Polarization effects on SNSPD absorption.** (**A**) Photon count rate of horizontal wires (black triangles) and vertical wires (blue squares) SNSPDs as a function of light polarization angle – 90° phase difference is observed. When the laser spot is focused on both detectors simultaneously (red circles) the sensitivity to polarization is reduced. Reproduced with permission.[136] 2008 © AIP Publishing LLC. (**B**) A typical NbTiN spiral SNSPD. Reproduced with permission.[136] 2008 © AIP Publishing LLC. (**C**) Polarization contrast (calculated) for a meandering NbN SNSPD increases with increasing wire thickness. Reproduced with permission.[87] 2009 © Cambridge University Press. (**D**) Distribution of the electric field amplitude around NbN nanowire for transverse magnetic polarized light (in the near-field framework). Reproduced with permission.[137] 2016 © Springer Nature Ltd.

Although the polarization of some of the light sources is linear and tunable, other sources relevant to systems that comprise SNSPDs have photon oriented in random polarization orientations. In these cases, there is a strong motivation to reduce the contrast so that the likelihood to detect a photon will be independent of its polarization. Reducing the contrast is usually done in two different strategies: (i) designing the SNSPD geometry and (ii) designing the dielectric medium around the SNSPD. By changing the usual meandering geometry with a spiral design (Figure 9B), Dorenbos et al.[136] reduced significantly the SNSPD sensitivity to photon polarization i.e. the contrast. These authors showed that the contrast observed in the spiral SNSPD is similar to placing two adjacent meandering SNSPDs in perpendicular orientation (see Figure 9A). Lately, Huang et al.[138] suggested that under illumination of photons with an arbitrary polarization, the average-photon detection efficiency of such a spiral design exceeds the efficiency of a meandering SNSPD with the same fill factor and material parameters. Spiral is not the only geometry that can reduce the contrast. Gu et al.[139] have recently proposed an alternative geometrical design to reduce the contrast – meandering the SNSPD in a nearly chaotic path, rather than the typical raster-like pattern. To the best of our knowledge, this promising design has not yet been tested experimentally.

Shape is not the only geometrical parameter that affects the contrast. Given the small dimensions of the SNSPDs, the polarization dependence of the light-matter interaction should be considered in the near-field region[137] so that the wire width (or fill factor) as well as thickness should both be taken into account. Figure 9C shows the contrast as a function of thickness. Here, Driessen et al.[87] found that the contrast increases with increases thickness (the contrast is saturating already below 10 nm for NbN).

In addition to the thickness contribution to the contrast, Driessen et al.[87] showed that the dielectric constant of the medium with which the light passes also affects the contrast. Calculating the contrast for a NbN SNSPD, they showed that the contrast decreases with increasing dielectric constant of the substrate. This contrast reduction did not arrive with no price – the difference between $\eta_{abs||}$ and $\eta_{abs\perp}$ indeed decreased, but the total absorption (and hence detection efficiency) also decreased significantly. The concept of using a dielectric medium was later adopted for helping overcome the localized (near-field) interaction of the photons with the asymmetric shape of the SNSPD that increases the sensitivity to a certain photon polarization (Figure 9D). Redaelli et al.[140] showed that covering the SNSPD area with a dielectric layer (for both filling the gaps and covering the wire) helps delocalize the photon-wire interaction, hence decreasing the polarization sensitivity of the detector without suppressing the detection efficiency. Likewise, Zheng et al.[137] showed that embedding the SNSPD within a dielectric medium can give rise to a combination of very low contrast (0.5%) and very high absorption (96%). The efficiency enhancement of such architecture can be attributed also to its cavity-like effect. Driessen et al.[87] showed that the maximum absorption, $A_{\max} = \frac{n_1}{n_1+n_3}$, is obtained when $R_{\square} = \frac{n_1 n_3}{n_1+n_3}$ where $R_{\square}$ is the superconducting wire sheet resistance and $n_1$,$n_2$ and $n_3$ are the refractive indices of the top layer, superconducting wire and the substrate respectively.

*DE optimization summary*

We can now summarize the factors that help increase the device efficiency. The main factor is the coupling of the SNSPD to the light source. Such coupling can be obtained by placing the device next to waveguides as well as adding cavities, antennae, wherever applicable. In the absence of waveguide coupling, anti-reflecting mirrors, high fill factor and large active area can also help enhance the light coupling. The choice of materials and the film thickness influence both on the device absorption. Moreover, the homogeneity of the material, i.e. lack of grains at all or at least at a size comparable to the SNSPD width, low-roughness substrate and a design that prevents current crowding and constrictions (e.g. short wires, mainly relevant to waveguide coupling) also play a major role in increasing DE. Polarization orientation of the photons can be optimized parallel to the SNSPD plane, while using an optimized dielectric medium can help both enhance the absorption as well as reducing the SNSPD sensitivity to photon polarization, i.e. reducing the contrast. Lastly, in addition to avoiding constrictions, the requirement for saturating DE as a function of $I_b$ suggest that materials with lower energy gap are preferable (also broadly leading to homogenous or amorphous materials).

*2.1.2 Dark counts and false counts*

Device efficiency is not the only factor that affects the performance of SNSPDs. High dark-count rate reduces our confidence that the measured signal is of a photon that was intentionally sent to the detector. Hence, high DCR limits realization of SNSPDs in communication and quantum key distribution technologies.[141,142]

*Mechanisms*

False counts are any measurable voltage pulses on the SNSPDs that do not originate from a photon that was sent to the device intentionally. Such false signals arise due to either straying photons or due to fluctuations from superconducting to normal state in a device that is held very close to the phase transformation. Thermal fluctuations,[50,143] fluctuations in the bias currents well as quantum fluctuations in the amplitude[144,145] and phase[55,146] of the superconducting order parameter, contribute all to false counts and are often called ‘dark counts’. Other fluctuations in the electronics can also contribute to false signals, but they can be distinguished from ‘true’ dark counts thanks to their time-evolution fingerprints[147] (yet, the rate of all false counts together, including those arise from straying photons, is usually called dark-count rate as in the current review). The dark-count and false-count mechanisms are illustrated schematically in **Figure 10.**

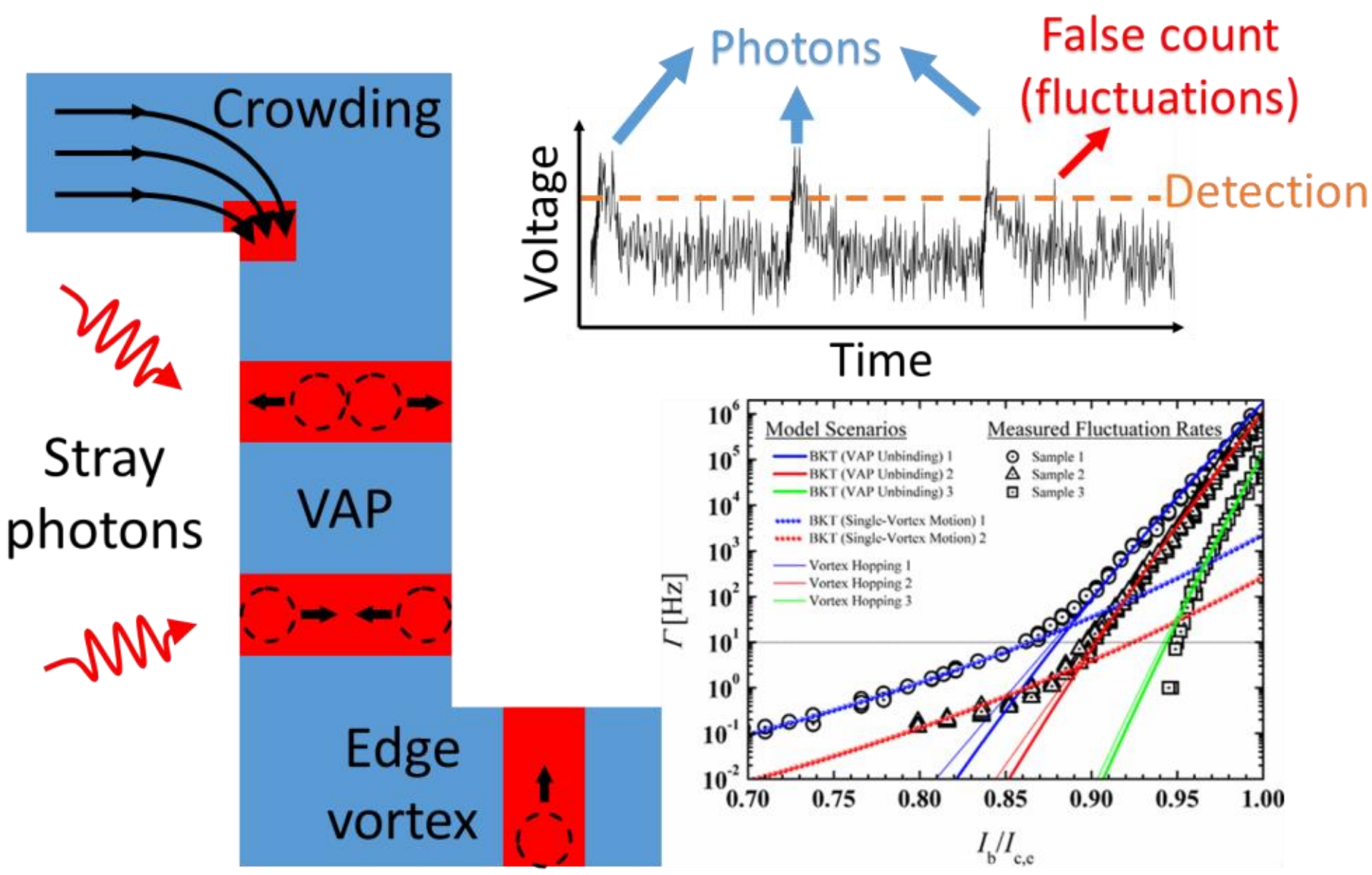

**Figure 10| Dark and false counts in SNSPDs**. Schematic illustration of the origin of false counts in SNSPDs, such as stray photons, current crowding, thermal-, electronic-, and vortex- fluctuations as well as low signal-to-noise ratio. Bottom right graph: fitting of the measured dark-count rate to the three mechanisms discussed in the text at different bias-current values. Reproduced with permission.[50] 2010 © American Physical Society.

Bartolf et al.[50] suggested that dark counts originate from three competing mechanisms, which are rather similar to the ‘true’ photon detection: (i) single-vortex motion; (ii) vortex-antivrotex unbinding; and (iii) vortex hopping. The dominancy of each of these mechanisms depend on experimental (i.e. extrinsic, and not intrinsic) parameters as demonstrated, e.g. in Figure 10.

*Design and material solutions*

Most solutions to overcome high DCR involve extrinsic parameters. Tunable parameters include temperature, wavelength and current bias. Likewise, hardware parameters include low-noise amplifiers and filters as well as good thermal and optical isolation. Yet, let us first discuss the parameters that involve design and materials.

*Intrinsic*

The demand for low sensitivity to fluctuations to reduce DCR dictates two main requirements from the material perspective. First, the material should be as homogenous as possible with respect to electron scattering (i.e. structure) as well as to the order parameter. For instance, Wollman et al. [22] demonstrated recently $10^{-4}$ $s^{-1}$ DCR for amorphous $Mo_xSi_{1-x}$ SNSPD that is operated at 4.2 K (though at the UV regime). Yet, there has been a recent effort to reduce DCR intrinsically also with crystalline materials, e.g. with $10^{-1}$ $s^{-1}$ for $MgB_2$.[148]

The binding superconducting energy gap also affects DCR. Specifically, Engel et al.[51] suggest that for a specific photon energy (wavelength), smaller energy gap increases the chance for fluctuations, giving rise to higher DCR. Consequently, higher $\Delta$ suggests also that higher $J_c$ and higher $T_c$ help reduce the DCR. Indeed, because thermal fluctuations ($\delta T$) are usually independent of the absolute temperature, higher critical temperatures allow smaller $\frac{\delta T}{T_c}$, which also help reduce the DCR for a given operating temperature. Similar considerations apply to electronic fluctuations ($\delta I$) and $J_c$. In addition, for this respect, superconductors that are closer to the SIT tend to have stronger fluctuations of the order parameter.[145,149] Thus, materials with larger $\Delta, T_c$ and $J_c$ (see **Table 3** for examples) are favorable for low DCR also from this perspective. Of course, these requirements are contradictory to the requirement for high DE. Thus, in addition to optimization of the intrinsic material properties and design, reducing the DCR while maintaining high DE moves the focus to the extrinsic properties.

*Design*

The likelihood for high DCR increases when the switching current becomes smaller than the critical current. Thus, design and fabrication reasons that give rise, e.g. to current crowding[103,104,150] and constrictions,[106] not only deteriorate DE but also increase DCR.[151]

Thermal phase slips as well as vortex-crossing events depend strongly on the SNSPD geometry. For instance, Inderbitzin et al.[152] demonstrated with a 100-nm thick Nb SNSPD that increasing film thickness helps avoid DCR, which is consistent with earlier demonstration of the Berggren group[147] that noise is reduced for thick and narrow wires. The idea of using narrow wires was supported also theoretically, by Engel et al.[51] Following Likharev's[153] analysis of superconducting weak links, Engel et al.[51] suggested that wires wider than ~4.4 times the coherence length of the Cooper pairs in the superconductor ($\xi$) allow for vortex crossing, hence giving rise to increase of the DCR. We should note that following this analysis, narrower wires behave as tunneling junctions. For a SNSPD with a typical geometry of ~5-nm thickness and ~100-nm width (much narrower SNSPDs have also been reported[147]), some materials (e.g. $\xi_{Nb} =$ 38nm[154]) will be considered weak links, while most materials used for SNSPDs will have a coherence length shorter than 4.4 times the coherence length (see Table 3 for comparison between coherence length and other material parameters of superconductors).

*DCR optimization summary*

We can now summarize the parameters leading to low dark-count rate. Low dark-count rate is obtained when the SNSPD is running far away from the fluctuation regime. Thus, homogeneous material (structure and superconducting order parameter) as well as a constriction-free and current-crowding free design are required. Moreover, narrow wires (<~4.4 times the coherence length) are also required, for decreasing the probability for vortex crossing. Because the intrinsic parameters that can help minimize fluctuations and hence the DCR (e.g. large $\Delta, T_c$ and $J_c$) are the opposite than those required for increasing DE, optimization of extrinsic properties (lower operating temperature and lower $I_b$, especially for devices with saturating DE) is usually more important for the DCR.

## 2.2 Timing

### *2.2.1 Timing jitter*

The fast timing jitter of SNSPDs is perhaps their most attractive property, outperforming competing technologies.[14,15] As opposed to a clear limit of e.g. DE=100%, it is not yet clear how small $\tau$ can be.[155–158] The mechanisms that dictate the timing jitter as well as the intrinsic and extrinsic properties that affect it are still not completely understood and have been examined recently very intensively.[155–158] Thus, optimizing $\tau$ and exploring its origin are still open tasks.

*Mechanism*

The uncertainty in time arrival for a detected photon is a combination of three different uncertainties.[16,157] The fundamental timing jitter of device ($\tau_d$) is usually believed to be affected by its geometry or design ($\tau_{dd}$)[157,158] as well as by intrinsic material properties ($\tau_{di}$), mainly the kinetic inductance,[18,159,160] electro-thermal properties[161] and material homogeneity.[158] The electronic noise or signal-to-noise ratio also contribute to the uncertainty in the timing of the signal ($\tau_e$). Lastly, another contribution to the jitter of the system, but which is not an inherent part of the SNSPD is the uncertainty in timing of the single-photon source ($\tau_s$).[159] Experimentally, $\tau$ is measured from the full-width-half-maximum (FWHM) of the histogram of counts as a function of time arrival of a large number of single-photon detection events (i.e. the instrument response function, IRF).[14] This histogram (the IRF) comprises the accumulated uncertainty in time arrival from the geometry, electronics and source. Thus, the complete timing jitter satisfies: $\tau = \sqrt{\sum \tau_i^2}$, where the $\tau_i$'s are the different components of the timing jitter (e.g. $\tau_d, \tau_e$ and $\tau_s$). A summary of these mechanisms can be found in a recent paper by Sidorova et al.,[162] while schematic illustrations of the jitter mechanisms are demonstrated in **Figure 11**.

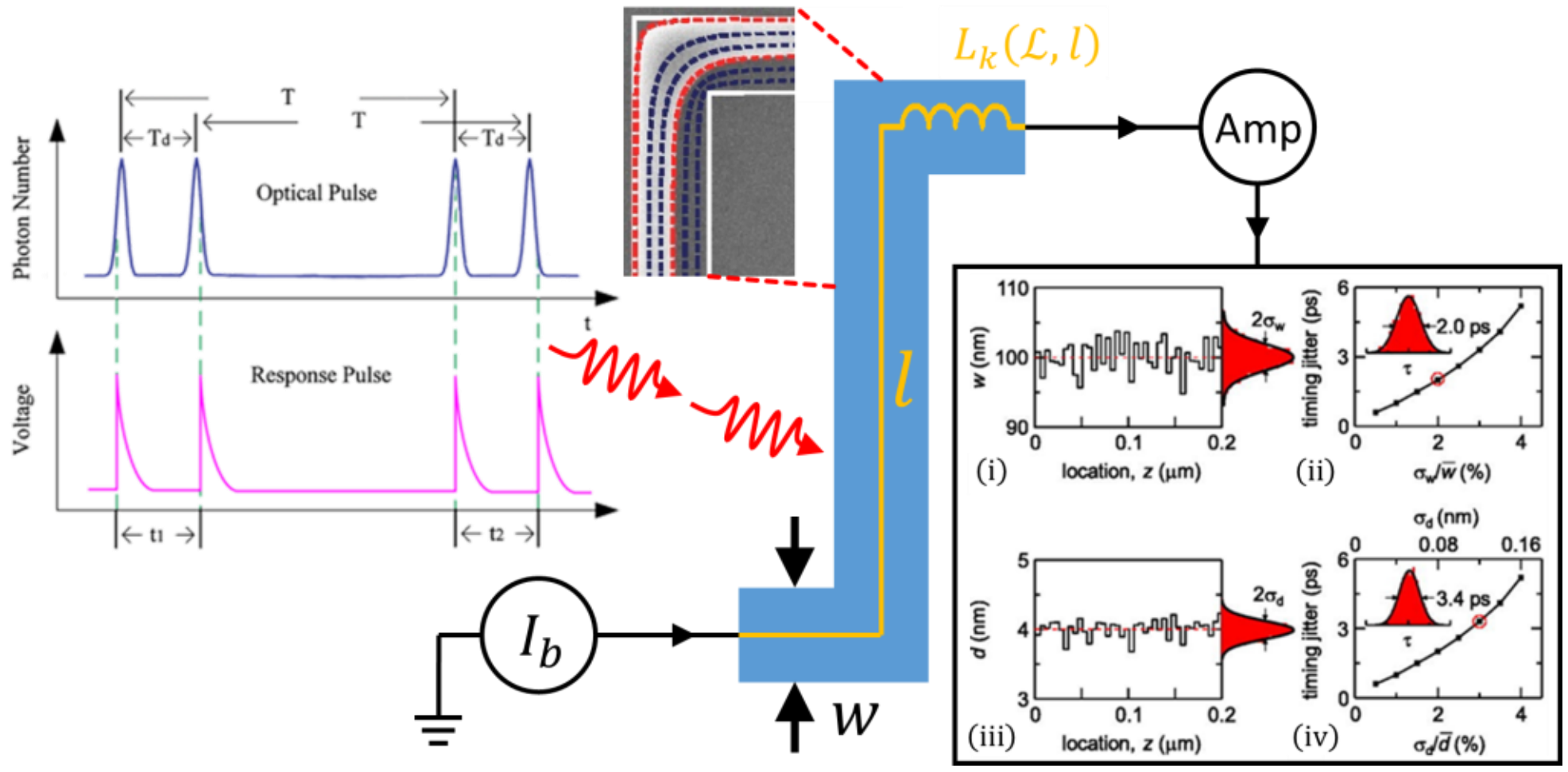

**Figure 11| Timing jitter in SNSPDs**. Illustration of the origin of uncertainty in arrival time (timing jitter) of photons detected with SNSPDs. Left-to right: illustration of the difference between two photon-detection events in time delay between the photon arrival (top) and the conversion of the photon to an electric signal (bottom), emphasizing the timing jitter definition. Reproduced with permission.[159] 2011 © Springer Nature Switzerland AG. Bias current, wire width and length, kinetic inductance and current crowding (Reproduced with permission.[151] 2012 © The Optical Society) are all intrinsic properties that affect the timing jitter, while timing uncertainty related to the light source and the noise arriving from the SNSPD amplifiers and electronics also influence the timing jitter. Finally, (i) constrictions in device width and (ii) thickness, giving rise to (iii-iv) respective increased timing jitter. Reproduced with permission.[158] 2017© AIP Publishing.

Recently, there have been several attempts to determine the fundamental limit of $\tau$. Wu et al.[156] suggested that the fundamental uncertainty in photon arrival corresponds to the time it takes for a vortex to cross the width of a SNSPD, limiting $\tau$ to roughly 1 ps. Likewise, Kozorezov et al.[163] suggested that in the case of amorphous superconductors, the fundamental timing jitter is dictated by Fano fluctuations, setting a limit of the same order of magnitude. These authors have also analyzed thoroughly the contribution of latency of the detector response very recently.[155] Lastly, the Berggren group[157] related the fundamental uncertainty in photon time arrival to the uncertainty in the specific position that the photon was absorbed along the superconducting nanowire, suggesting that the fundamental limit is intrinsic rather than electronic dependent.[164]

*Intrinsic and design solutions*

Addressing timing jitter reduction requires a combination of device design, material optimization as well as careful choice of the electronic system. Thus, complete overview of the parameters that govern the timing jitter should include both the following discussion as well as the sections that discuss system hardware.

*Intrinsic - material properties*

$R_{\square}$, $I_c$

Gol'tsman and co-authors[84,165–167] reported in earlier works sub-35 ps system jitter for a NbN-based SNSPDs. Following a previous analysis of Bertolini and Coche from 1968,[168] Zhao et al.[159] suggested in 2011 that increasing $I_\mathrm{c}$ as well as the normal resistance of the device ($R_\mathrm{n}$) results in an increase in the SNR and hence reducing $\tau$. Consequently, You et al.[169] examined thoroughly the behavior of jitter of a NbN SNSPD under various conditions, while they also introduced artificial noise to the system, allowing them to derive the governing mechanisms even more carefully. The outcome of this study, which showed τ=18 ps, confirmed that high $I_c$, which allows larger bias currents helps reduce the timing jitter (**Figure 12**).

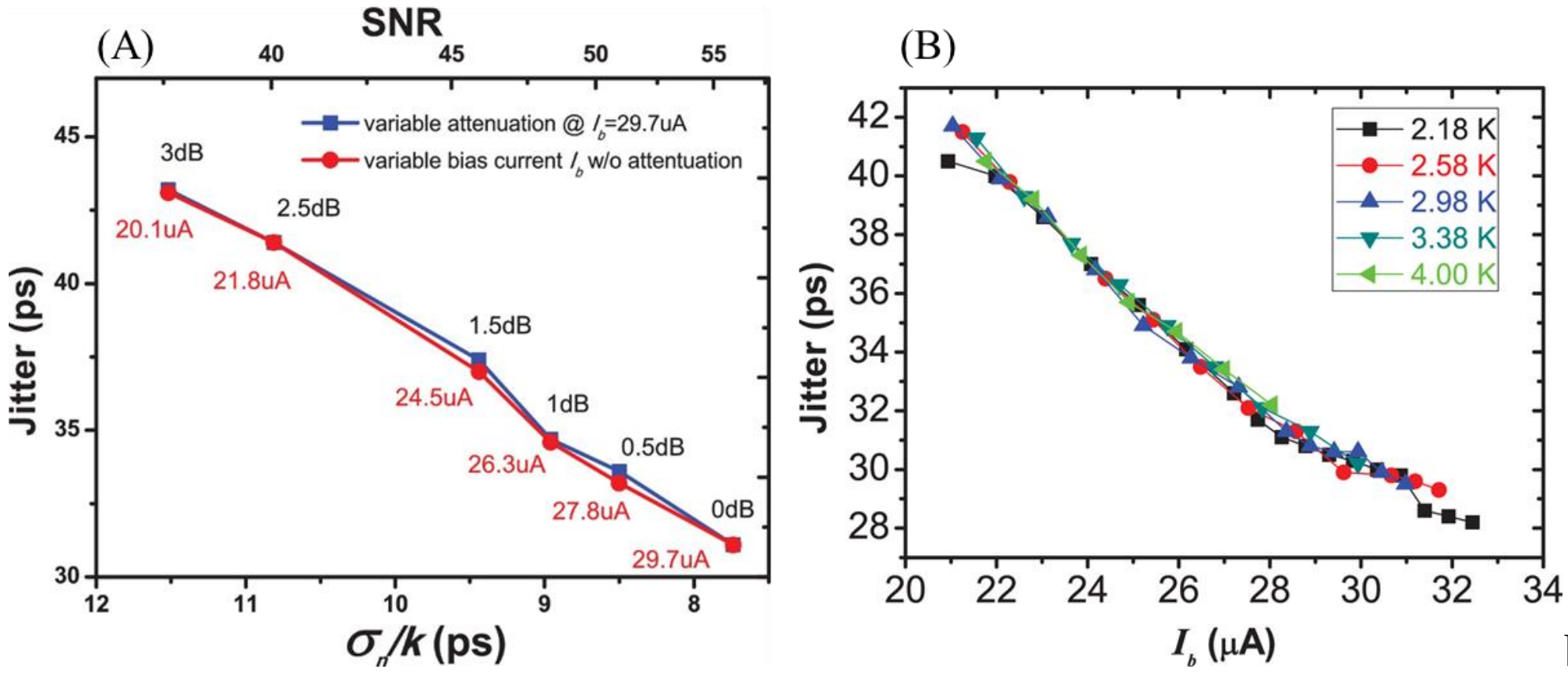

b

**Figure 12| Timing jitter in SNSPDs**. Timing jitter as a function of (**A**) signal-to-noise ratio and (**B**) bias current. The lack of dependence of $\tau$ on temperature (and hence on $I_c$) indicates that the absolute value of $I_b$ and not the normalized value $(I_{b_n})$ that affects the jitter. Reproduced with permission.[169] 2013 © AIP Publishing LLC.

$T_c$, $\Delta$

Both Δ and $T_c$ depend directly on the critical current density, $J_c$. Thus, as suggested, e.g. by Cheng et al.,[158] increasing $T_c$ (and Δ) are also expected to reduce the timing jitter. Indeed, as seen in **Figure 13**, when comparing timing jitter and $T_c$ of competitive SNSPDs made of different materials (albeit with a comparable design), we find that τ decreases for materials with lower $T_c$.

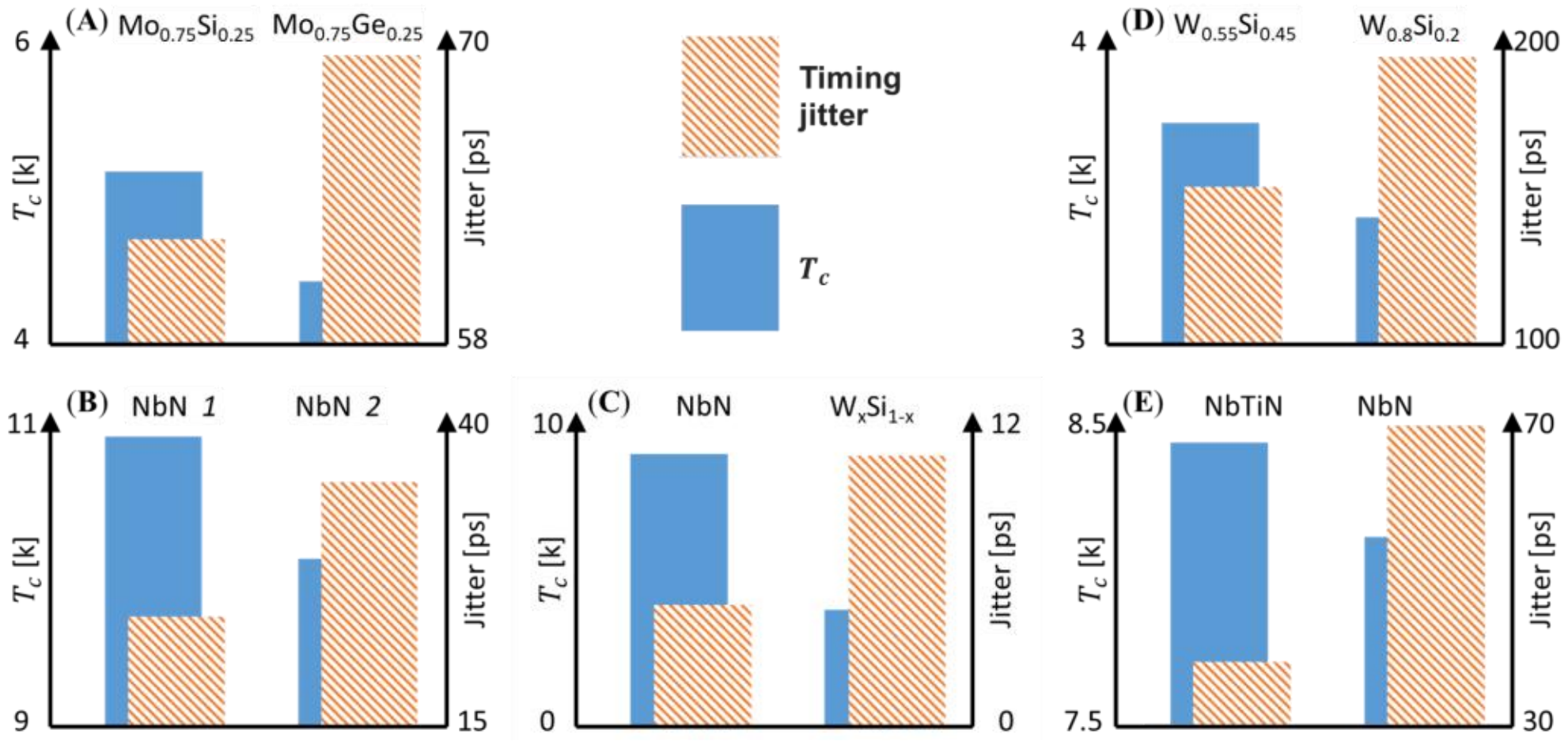

**Figure 13| Timing jitter vs. $T_c$.** $T_c$ and timing jitter of a selected set of comparable SNSPD devices. (**A**): Amorphous molybdenum silicide[22] and molybdenum germanium SNSPDs.[97]; (**B**): Two NbN SNSPDs of a similar design.[170] (**C**) Current records in timing jitter for NbN[18] and tungsten silicide[164] SNSPDs. (**D**) Two high-efficiency tungsten silicide SNSPDs.[16,100] (**E**) Comparable NbTiN and NbN SNSPDs.[171]

$L_k$

Kinetic inductance was found to have crucial effects on timing jitter.[157] Given the dependence of $L_\mathrm{k}$ on the sheet resistance (**Equation 2**), from the SNR perspective, higher $L_\mathrm{k}$ should help reduce τ. However, $L_\mathrm{k}$ determines also the speed at which the electric signal flows in the wire from the area of photon absorption to the electrodes. Thus, from this electromagnetic perspective, lower $L_\mathrm{k}$ is required to minimize τ.[157] Yet, the dominant contribution of $L_\mathrm{k}$ to the jitter is believed to arrive from the role it plays in the signal-transfer time rather than in the SNR. For instance, Kerman et al.[172] showed in 2006 that the rise time of SNSPD signals (i.e. the measurable voltage) depends on kinetic inductance: $t_\mathrm{rise} = \frac{L_\mathrm{k}}{Z_\mathrm{shunt}+R_\mathrm{n}}$, where $Z_\mathrm{shunt}$ is the impedance of the shunt that is connected in parallel to the SNSPD (this expression also strengthens the requirement for large $R_\mathrm{n}$, while yet maintaining $R_\mathrm{n} > Z_\mathrm{shunt}$). The kinetic inductance is an integration of the intrinsic sheet kinetic inductance (i.e. inductance per square of the materials, $L_{\mathrm{k}_\square}$) over the geometric path of the signal, which is defined by the length ($l$) and width ($w$) of the path (i.e. $\frac{l}{w}$= number of squares):

$$L_{\mathrm{k}} = L_{\mathrm{k}_\square} \cdot \frac{l}{w} \qquad (3)$$

Thus, the intrinsic kinetic inductivity ($L = L_{\mathrm{k}_\square} \cdot d$, $d$-thickness) defines how the signal rise time and hence the distribution multiple photon detection rise time events is skewed. Consequently, for a given geometry, smaller $L$ gives rise to smaller $\tau$. Indeed, when surveying recent literature for different materials and systems, Wu et al.[173] showed that decrease in rise time is translated directly to decrease in the timing jitter, while they also demonstrated 14.2 ps for their NbN SNSPD. We support their claim that the decrease in time rise can be translated to decrease in timing jitter by elucidating the fact that the distribution of the time arrival of the photons is also determined by the time rise. Thus, if the time rise increases, the distribution of photon time arrival will be skewed, leading to an increase in timing jitter. In **Figure 14**, we illustrate schematically this effect of the kinetic inductivity on the timing jitter. Here, when two photons are detected in different points along the SNSPD, the photon that was detected closer to the electrode ($h\nu_1$) will have a shorter path ($l_1 < l_2$) to go until the detected signal is generated in comparison to a photon that is detected further away from the electrode ($h\nu_2$). Because the effective kinetic inductance depends on the path (Equation 3) we obtain $L_{k_1} < L_{k_2}$. Now, we saw that the time rise is proportional to $L_k$. Consequently, the signal generated from the first photon will have a faster time rise than the second photon and will therefore be detected earlier ($\Delta t_1 < \Delta t_2$). The timing jitter is the uncertainty in time arrival, which means that fundamentally, the larger the difference in rise time between two photons that are detected along the wire, the higher the timing jitter. Hence, reducing the kinetic inductance will derive reduction in the rise time of each photon-detection event and therefore in the difference in the time it takes them to arrive to the detector, i.e. $\Delta t_2 - \Delta t_1$ is reduced. That is, decreasing the kinetic induction of the SNSPD helps reduces its rising time and consequentially its timing jitter (shortening the wire will have a similar effect to the decrease in kinetic inductance). Smirnov et al.[20] demonstrated recently further development of the dependence of rise time on kinetic inductance. Schuk et al.[174] suggested that the timing jitter of NbTiN, which has relatively low kinetic inductance (Figure 13), can be rather low, while the specific value they measured (51 ps) was relatively high due to the used electronics.

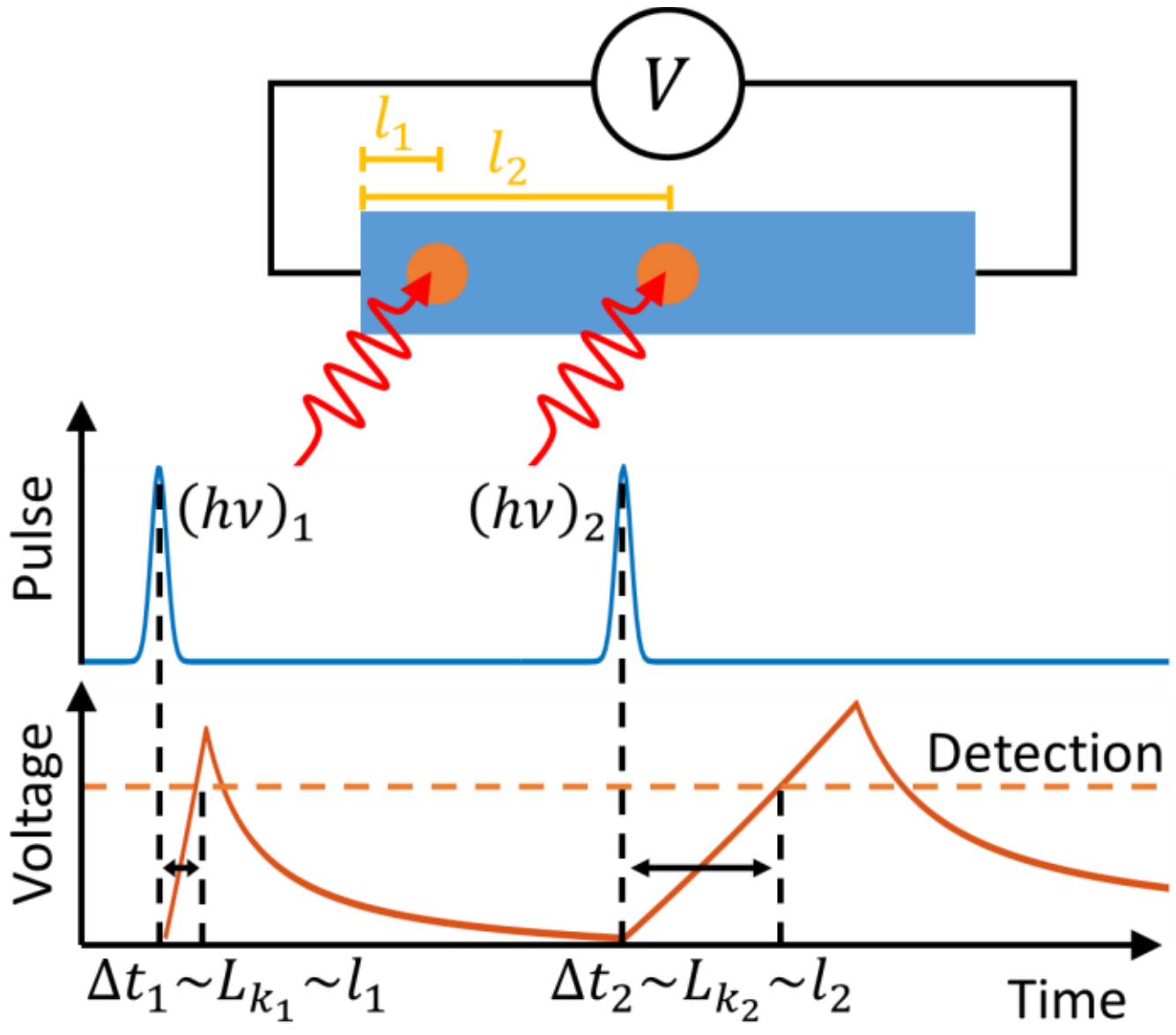

**Figure 14| Effects of kinetic inductance and device geometry on timing jitter**. Two identical photons that are absorbed in different places along the SNSPD (top) arrive at the electrodes, and hence producing a voltage signal, at different rise times ($\Delta t$), which depends on the distance between detection point and the SNSPD electrode ($l$). The consequential difference in rise time results in a difference in the time that passes between the moment that the photon was absorbed and the time at which the voltage pulse was detected. Thus, this difference in rise time, which increases for longer wires as well as for wires with higher kinetic inductance, adds to the uncertainty in the photon arrival time (i.e. timing jitter). A detailed discussion is given in the main text.

### *Homogeneity/amorphousness*

#### *Amorphous-material SNSPDs*

Empirically, amorphous materials exhibit a typical slower timing jitter than crystalline materials (Figure 13). It can be inferred that the slower timing jitter is due to the lower energy gap (and $T_c$) that such materials usually exhibit (Figure 13).[58,102] Marsili et al.[16] reported in 2013 τ=150 ps for their amorphous tungsten silicide ($W_{0.55}Si_{0.45}$) device, suggesting that the electronic noise in the system dominated this value. Yet, very recently, Korzh et al.[164] suggested that if the electronic noise of the system is reduced by using low-noise electronics and good impedance matching, while the geometric contribution to the timing jitter is also minimized, amorphous $W_xSi_{1-x}$ can exhibit τ as low as 10.3 ps. Verma et al.[97] measured in 2014 τ=69 ps for an amorphous molybdenum germanium ($Mo_xGe_{1-x}$) SNSPD. Korneeva et al.[96] reported in the same year τ=120 ps for molybdenum silicide ($Mo_xSi_{1-x}$) detectors, and Verma et al.[175] reduced τ in this material a year later to 76 ps, while Caloz et al.[98] showed τ=26 ps for $Mo_xSi_{1-x}$ SNSPDs.

*Crystalline-material SNSPDs*

In comparison to amorphous materials, SNSPDs based on crystalline materials demonstrate shorter timing jitter. There has been a continuous reduction of τ over the years for SNSPDs made of the common material NbN, since the Gol'tsman and co-authors[84,165–167] reported initially 35 ps. Recent studies showed sub-15 ps timing jitter for NbN SNSPDs,[169,173] while Korzh et al.[18] measured 2.7 ps, which is the lowest value reported in general. This value was measured though at λ=400 nm, while this SNSPD set also the record of τ=4.7 ps for λ=1550 nm. NbTiN demonstrate slightly larger τ, ranging from 68 ps[68] to values as low as 11.3 ps.[124] Lastly, Annunziata et al.[176] overcame processing challenges and measured $\tau < 100$ ps for a crystalline structure of metallic Nb SNSPDs.

*Hybrid amorphous-crystalline SNSPDs*

The above distinction between amorphous and crystalline SNSPDs raises a motivation to optimize the amorphousness of a superconductor. Following a recent analysis of the effects of amorphousness on superconducting thin films,[59] such optimization may open a new path for SNSPDs that benefit from the competitive detection properties of amorphous materials as well as from the superior timing performance of crystalline superconductors. It has been suggested recently[65] that using bilayers of amorphous and crystalline materials gives rise to hybridization of the superconducting properties via the proximity effect[177] (**Figure 15**). Indeed, SNSPDs made of a bilayer of 2-nm thick amorphous $W_xSi_{1-x}$ on 2-nm thick crystalline NbN demonstrated optimized detection performance with $\tau$=52 ps, much faster timing jitter than $W_xSi_{1-x}$ SNSPDs.[65] Further optimization of such hybrids is still under investigation.

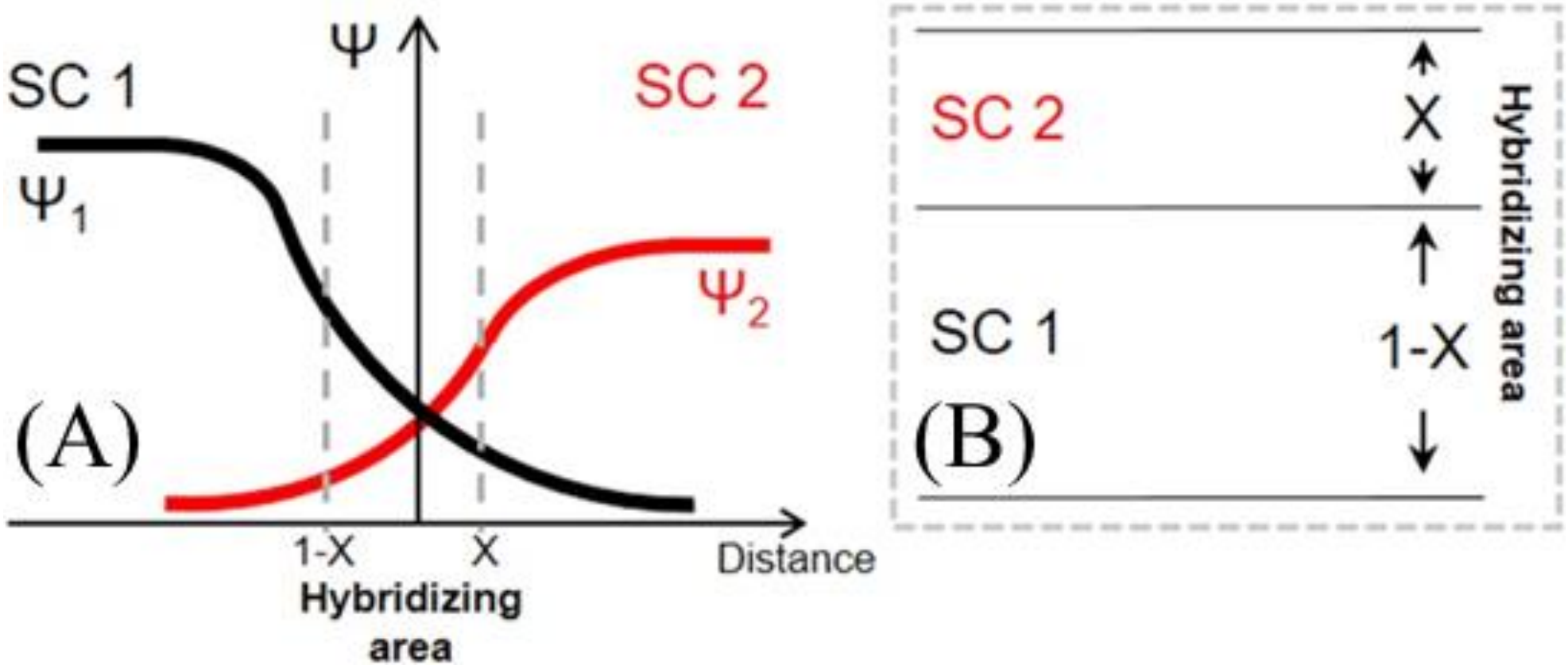

**Figure 15| Tunable functionality in superconducting nanostructures**. (**A**) The order parameter or electron wave function of two neighboring superconductors (SC1 and SC2) are mixed at the boundary between the two materials (within the dashed lines) due to the superconducting proximity effect.[177] (**B**) Processing bilayers with films thinner than these boundary gives rise to hybridization of functional properties. Reproduced with permission.[65] 2017. © IOP Publishing. Such hybridization method helped optimize device properties of a bilayer NbN-WxSi1-x SNSPD.[65]

*Material Summary*

Careful examination of the jitter reported for different materials allows us to examine the dominancy of the above parameters. For instance, Figure 13 shows a clear trend that suggests that from the material perspective, higher $T_c$ (and hence higher $J_c$ and $\Delta$) is the most promising parameter to reduce τ. This also means that from this perspective, the effect of $L$ is secondary. That is, optimization of $L$ is important for the jitter only after the material was chosen, and this optimization is usually done by device design rather than by changing the material. We should note though, that because $T_c$ scales roughly inverse to the homogeneity or amorphousness of the material, it is impossible currently to distinguish which of these two properties dominates τ.

*Device design*

*Electronics*

To fulfill the performance potential of the SNSPDs, there is a strong desire to minimize τ closer to the intrinsic limit $\tau_d$. Thus, the signal-to-noise ratio has to be reduced to allow $\tau_e \ll \tau_d$. In his 2013 paper on $W_xSi_{1-x}$ SNSPDs, Marsili et al.[16] the authors used cryogenic pre-amplifiers to reduce the timing jitter τ=96 ps from 121 ps with room-temperature electronics. Likewise, Zadeh et al.[124] who used the same method to obtain τ=48.8 ps for NbTiN detectors. To-date, the systems with the reported shortest timing jitter for all materials involve cryogenic components, such as pre-amplifiers, filters and bias teas.[18]

*Device Design*

*Constrictions*

The device geometry affects strongly several factors important to the jitter, such as bias and switching currents, $L_{\mathrm{k}}$ and $R_{\mathrm{n}}$. The effects of geometry on timing jitter became clearer when devices that involve SNSPDs connected in parallel (superconducting nanowire avalanche photodetectors, SNAPs) and other large-area devices demonstrated high SNR but high timing jitter.[147,178–180] Similarly to DE and DCR, constrictions in the wires (narrower areas due to inhomogeneity in fabrication, due to substrates effects, due to granular structure or due to current crowding) deteriorate also timing performance.[155,158,181–183] Pictorial illustration of the effects of constrictions and inhomogeneities on timing jitter is given in Figure 11.

*Dimensions – length and width*

The dimensions of the wires also have a strong effect on the timing jitter. Longer and narrower wires comprise higher probability for inhomogeneities and constrictions.[147,155,158,178–183] These dimensions affect also the timing jitter more directly. Wu et al.[156] suggested that wider wires allow larger variation in the time that a vortex crosses the wire, i.e. increasing the uncertainty in of the photon arrival time. Likely, very recent works suggest that magnetic effects, e.g. due to the meandering geometry, also contributes to the timing jitter.[184] Additionally, the Berggren group suggested[93] that the increase in $L_{\mathrm{k}}$ (**Equation 3**) due to an increase in the device length plays a significant role in the timing jitter, Using this hypotheses, recent efforts have led to sub-5 ps for a short and linear NbN SPSND[18] **(Figure 16)** and 10.3 ps for $W_xSi_{1-x}$ SNSPDs.

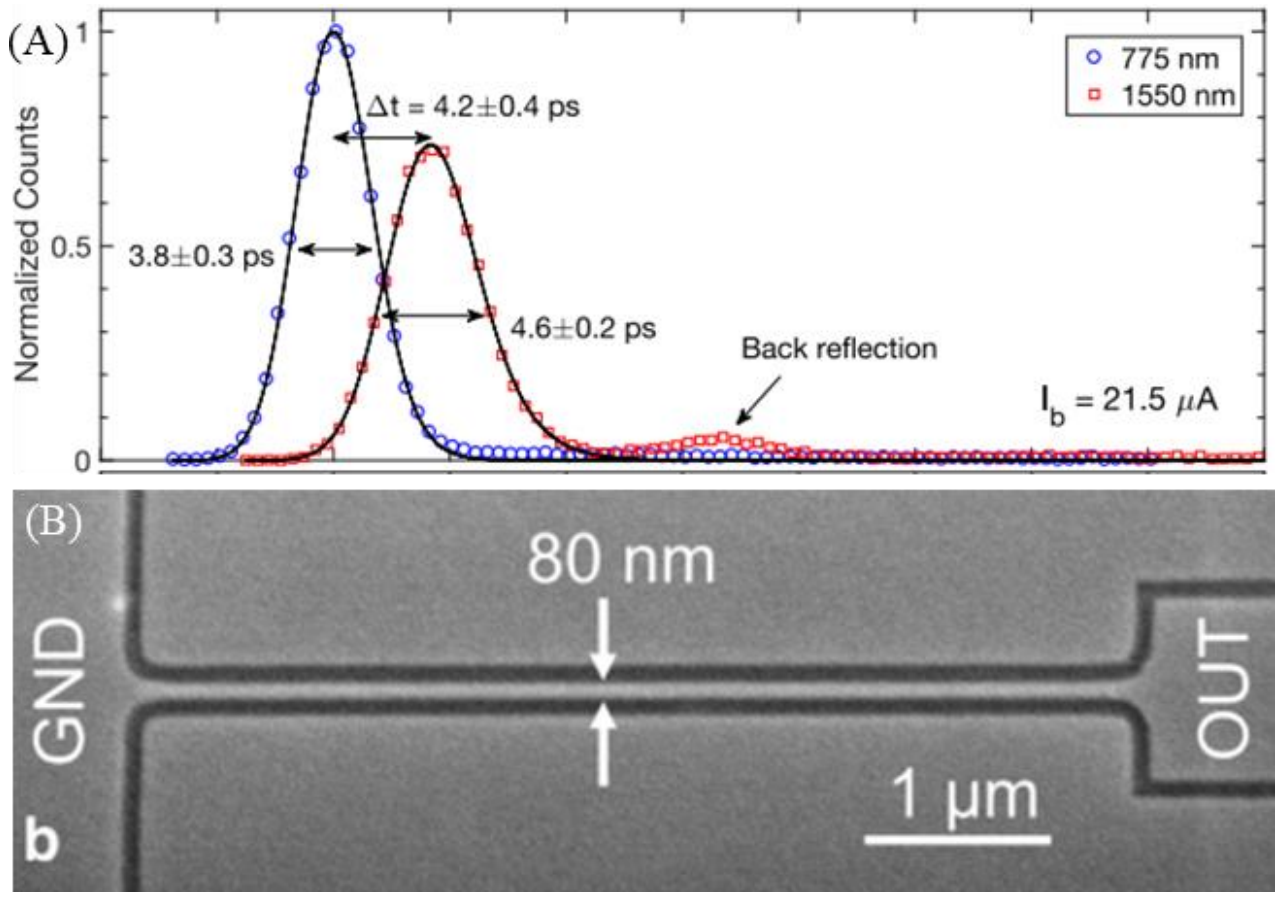

**Figure 16| Sub-5 ps timing jitter in a linear short-wire geometry NbN SNSPD**. Reproduced with permission.[18] 2018 © arXiv.

We should note that contradictory to the require for short length to reduce $L_{\mathrm{k}}$, the necessity in larger $R_{\mathrm{n}}$ (for higher SNR and shorter rise times, $t_{\mathrm{rise}}$) implies that larger wires are preferable. That is, similarly to Equation 3, the device normal resistance equals to the sheet resistance times the number of squares: $R_{\mathrm{n}} = R_{\square} \cdot \frac{l}{w}$. Hence, the geometrical requirements that stem from large $R_{\mathrm{n}}$ contradict those associated with small $L_{\mathrm{k}}$. Practically, $R_{\square}$ can be controlled rather easily, e.g. by changing the film thickness or the disorder in the superconductor.[59] Thus, the geometrical considerations related to minimization of τ can be governed by $L_{\mathrm{k}}$, as demonstrated by the above examples (including Figure 13).[18,93]

*Measurement scheme*

Although short wires are favorable for reducing τ via the reduction of kinetic inductance, such devices may be less attractive from the device efficiency perspective. One way to overcome this challenge is to increase the DE, e.g. by coupling the SNSPD to waveguides.[88,111,174,185,186] However, there are other solutions that do not require a special design, but rather, they involve a different measurement scheme. Such readout schemes can be implemented also to short wires, reducing further τ. Specifically, Calandry et al.[157] measured the two electrodes of the wires separately, in a differential measurement reduced the timing jitter in a NbN SNSPD from 35 ps to 29 ps in a simple wire and to 12 ps for wires with a different geometric design ($\tau_e < 7$). Likewise, Shcheslavskiy et al.[187] developed a different readout scheme to measure low jitter (τ=7.9 ps) in a NbN SNSPD, while they also surveyed jitter measurements from other devices that have been done with this readout scheme.

*Jitter optimization summary*

We can now sum up the above conclusions regarding the design of a SNSPD system with minimized timing jitter: (i) choosing a material with high $T_c$ (or hybridizing a bilayer of two materials); (ii) growing a film with high sheet resistance (thin or disordered); (iii) designing a short device for low $L_k$ and low constriction (waveguide coupling can help compensate for the DE), while avoiding current crowding; (iv) processing a narrow wire for low vortex-crossing uncertainty (as long as no constrictions); (v) choosing electronic and readout components with low noise, if possible, at cryogenic temperatures; and (vi) measuring the differential signal between the two electrodes of the wire.

*2.2.2 Reset time*

*Introduction*

From an application perspective, reset time defines the frequency band of a single device. That is, the shorter the reset time, the larger is the amount of data (number of photons) that can be detected in a given time. We should note that while the jitter dictates a fundamental limit for the system, having a system with a large number of devices might overcome limitations that presumably arise from the device reset time. For instance, a system can be designed so that when one SNSPD detects a photon and the device has not yet reset, the next photon will be directed to a different SNSPD, which is ready to detect it.[125,161,172,188–190] Yet, for many applications as well as for the sake of improving a specific device performance, it is important to understand the parameters that affect the reset time as well as to discuss methods to minimize it.

*Mechanism and Definitions*

When a SNSPD detects a single-photon, it becomes normal. At this stage, the current flows through the shunt and the device cools down to become superconducting again. Thus, after a photon-detection event there is a certain time period at which the device is not ready to detect a second photon as it was ready to detect the first photon. We can distinguish between two sequential processes that take place during this time. Initially, when the SNSPD is normal, a second photon cannot be detected at all. However, when the superconducting wire is then cooled down, the current gradually flows from the shunt back to the detector. During this time, the SNSPD starts being able to detect another photon, albeit with limited efficiency. The DE increases with time until the entire current flows again through the SNSPD only and the system is ready to detect another photon, exactly as it was ready to detect the first one.[191,192] Following Migdal et al.,[191] the first step, at which no second photon can be detected at all, is called *dead time* $(t_{\mathrm{dead}})$, while the second step, at which the DE is gradually recovered is called *reset time* $(t_{\mathrm{reset}})$. These processes are illustrated in **Figure 17**. We should note that the shunt and electronic components play a significant role in the reset time. That is, the role of the shunt is to allow fast reset time, while yet successfully absorbing the inductance energy $\left(\frac{1}{2}L_{\mathrm{k}}I_{\mathrm{b}}^{2}\right)$ from the wire and let it cool back to the superconducting state without latching.[193] Other electronic components, such as amplifiers and filter may also contribute to this tradeoff.

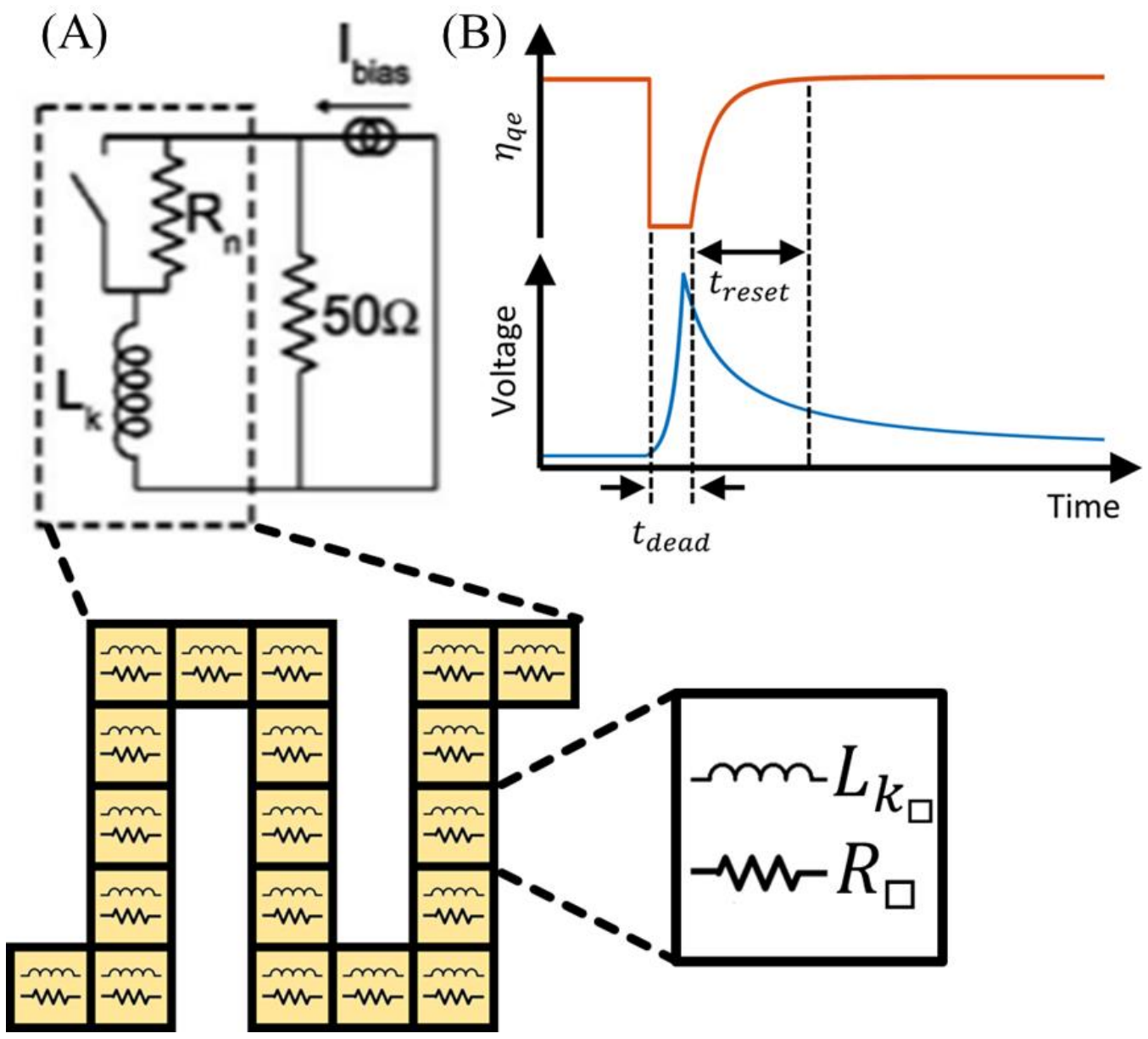

**Figure 17| Reset time.** (**A**) Equivalent circuit of the inductance and switchable resistance in SNSPD connected in parallel to a shunt resistor, emphasizing the geometric and device parameters that affect the reset time. Reproduced with permission.[172] 2006 © AIP Publishing LLC. (**B**) Efficiency (top) and voltage (bottom) time evolution upon a photon detection, at which initially another arriving photon cannot be detected ($t_{dead}$), while the efficiency then increases gradually ($t_{reset}$) until the device has completely recovered and is ready to detect the next photon.

From a practical measurable device characteristic point of view, reset time is measured as the time at which the voltage between the two electrodes decays[172] (Figure 17). Typically, this curve of voltage vs. time can be simplify fit to an exponent for obtaining the device reset time.[19,76,111,188,194] Kerman et al.[172] showed that the fall time is:

$$t_{\text{reset}} = \frac{L_{\text{k}}}{Z_{\text{shunt}}} \quad \textbf{(4)}$$

(which is reminiscent of the above expression that describes the rise time), while later works suggest that the $Z_{\text{shunt}}$ should be replaced by the impedance of the electronics, which may include e.g. the capacitance and resistance of the amplifiers.[19,65] There is no agreement what parts of this exponential decay should exactly be included as the reset time. For instance, it has been suggested that $t_{\text{reset}}$ is measured for the time at which the signal dropped to $\frac{1}{e}$ of its peak value.[97,99,188,195] On the other hand, other works[16,96] suggest that the reset time is the time at which the signal drops from 90% of the peak value to 10%, which is more consistent with the usual method of measuring phase transition in nano superconducting structures.[59,196] We illustrated schematically the mechanisms that are involved in setting the reset time in Figure 17.

*Intrinsic and design solutions*

The methods to address reset time reduction usually involve tunability of material and design parameters that affect $L_{\text{k}}$, while $R_{\text{n}}$ is important when comparing to the shunt, for avoiding latching. Electronic components, such as shunt resistor and effective shunt impedance are also important. Finally, geometric design that allows further reduction of $L_{\text{k}}$ is also a common method to reduce $t_{\text{reset}}$.

*Intrinsic – material properties*

The main material property that affects the reset time is inductivity. Typically, crystalline materials, which have lower $L$, demonstrate faster reset time than amorphous SNSPDs[16,22,175] (though low-$L$ amorphous materials, such as $Mo_xSi_{1-x}$ also exhibit low reset time[96]). For instance, Gol'tsman[13] reported ca. 100-ps reset time already in the first NbN SNSPD paper. Typical NbN devices with larger active area or higher DE as well as with different device design demonstrated reset time of a few nanoseconds, ranging from 100 ps for special designs to a few dozens of nanoseconds[170,197] or even 100 ns for an extreme device size in diameter.[178] NbTiN has smaller inductivity than NbN,[113] giving rise presumably to shorter $t_{\text{reset}}$. However, the reset time of NbTiN SNSPDs in the literature is slightly slower than that of NbN SNSPDs, ranging from 6.58 ns[198] to 10 ns,[199] while Schuck et al.[174] demonstrated in 2013 1.2-ns reset time for a short

SNSPD that is coupled to a waveguide. We should note that the faster reset time in NbN devices stems most probably from the fact that much more effort has been put into optimizing NbN devices rather than to improve NbTiN. Indeed, Yang et al.[171] examined carefully the different in performance between NbTiN and NbN SNSPDs of a similar design and found that the reset time of the NbTiN SNSPD was 17 ns, faster than 32-ns measured for a NbN SNSPD. The authors suggest that this difference is due to a two-fold difference between the kinetic inductance of the NbTiN device and that of the NbN SNSPD. Note though that there was a two-fold difference also in the normal resistance (NbN is larger) as well as in the critical current (NbTiN is larger) in these devices.

Another parameter that one should bear in mind is the normal resistance of the wire. $R_\mathrm{n}$, which affects the reset time only indirectly.[172] That is, the device resistance does not play a direct role in the reset time, but it does affect the shunt resistance, which in turn affects the reset time directly (the resistance affects also the reset time due to the relationship between resistance and inductance, as was given in Equation 3). The shunt is chosen to be larger than the device resistance, allowing for cooling of the wire after a photon-detection event. **Equation 4** suggests that the larger the shunt impedance, the shorter the reset time is. Thus, presumably, larger $R_\mathrm{n}$ values are preferable. However, large resistance gives rise to an increase in accumulated heat in the wire during a photon-detection event. When the heat energy exceeds the limit that the shunt can help divert (with respect to the hotspot energy), the wire latches after a single-photon detection event and never returns to its superconducting state, and hence cannot function as a photon detector. Therefore, the resistance should be optimized so that it is the largest possible, yet without letting the device to latch.[200] From the material perspective, the optimization is done for the sheet resistance, which is the material intrinsic property. However, the device resistance is determined also by the design, which we will discuss next.

*Device design*

There are three main methods to design SNSPDs with reduced reset time. These methods concern (i) changing the resistance of the wire or the shunt that is connected in parallel to the wire, (ii) changing the device geometry as well as (iii) processing higher hierarchy of SNSPD structures. The first method is to play around with the shunt, i.e. to increase its impedance, without affecting the device performance (without decreasing DE).[193] Yang et al.[161] showed that adding a resistance in series to a NbN SNSPD (effectively increasing $R_{\mathrm{n}}$), the reset time can be reduced by about one half. Moreover, Kerman et al.[201] increased the load (shunt) resistance to reduce $t_{\mathrm{reset}}$ from 7.2 ns to 2.4 ns.

Optimizing the geometry with respect to the reset time is obtained mainly by minimizing the inductance, i.e. shortening the wire (see Equation 3). In the first NbN SNSPD paper by Gol'tsman et al.,[13] the 100-ps reset time can be ascribed to the short length of the wire. The same reset time was demonstrated a bit later[165] also to a larger-area device, though with a small fill factor and hence a medium-size length of ca. 40 μm. The problem with short detectors and small effective areas is that the DE is also low. Thus, a combination of small wires that are coupled efficiently to the light source, e.g. with waveguides, cavities or antennae, allows improvement in reset time without decreasing DE. For instance, in 2012, Pernice et al.[88] coupled in 2012 a short wire (10 μm) to a waveguide, demonstrating 450-ps reset time (and 91% on-chip DE for 1550-nm wavelength). Likewise, in 2016, Vetter et al.[19] coupled a SNSPD with cavity, allowing them to measure 119-ps reset time for a short wire (and 30% on-chip DE at 1550 nm). In addition, the current record of 1.2-ns reset time for an NbTiN SNSPD was measured in 2013[174] also for a relatively short device that was coupled to a waveguide (< 67.7% DE at 1542 nm).

The last design aspect relevant to the minimization is hierarchical SNSPD structures. Following the suggestion of Kerman et al.,[172] Dauler et al.[190] proposed in 2007 to connect a number of wires in parallel to each other, where each wire has a kinetic inductance: $L_{\mathrm{k}}^{\mathrm{wire}}$ (see **Figure 18A-B**). The inductance of the entire device in such a geometry is: $L_{\mathrm{k}}^{\mathrm{tot}} = \frac{L_{\mathrm{k}}^{\mathrm{wire}}}{N}$. That is, the kinetic inductance of this device is smaller than the inductance of a single wire $\left(L_{\mathrm{k}}^{\mathrm{wire}}\right)$ by a factor of $N$ (Figure 18C). Based on similar arguments, at about the same time, Ejrnaes et al.[202] demonstrated a device, in which all wires are biased close to their switching-current value, so that the entire current that flows in the device is much larger than the typical bias current in a single-wire device (~5 times larger). When a photon hits one of these parallel wires, it becomes resistive, so that the other wires, which are still superconducting, serve as a shunt to this resistive wire and absorbs all of its excessive current. Consequently, after the photon was absorbed in one of the wires, the current that was redistributed in the parallel wires exceeds the switching values also in all of these other wires. This avalanche effect gives rise not only to a faster process (due to lower kinetic inductance), but also to a higher single-to-noise ratio, as was demonstrated, e.g. by the MIT group.[179] This superconducting nanowire avalanche single-photon detectors (SNAP)[202] exhibited 12.5-ns reset time for a large-area device . Cheng et al.[194] presented later a large-area SNAP with 890-ps reset time (with saturating DE vs. $I_{\mathrm{b}}$ behavior, but the low system absorption set the DE on 20% at 1550 nm). Another method to form avalanche detection, albeit with a larger reset time, is by connecting parallel SNSPDs not electrically, but thermally. The NIST[203] group demonstrated layered architecture of two SNSPDs with a sandwiched layer that allows electric isolation, but yet conduct thermally. Here, a photon-induced hotspot in one wire grows also to the neighboring wire, so that the two wires become normal together, similarly to SNAP. The idea of Dauler et al.[190] was implemented not only in SNAPs, but also in a slightly modified design that allows fast reset time as well as photon counting[188,189] as well as photon allocation. Implementing this idea, Tarkhov et al.[21] demonstrated a device with $t_{\mathrm{reset}}$=50 ps for a photon counter (up to 12 photons) that can detect single photons with a potential for ca. 300-nm spatial resolution (at λ=950 nm) (Figure 18D). Verma et al.[204] demonstrated a similar design for a four-pixel array of SNSPDs made of $W_xSi_{1-x}$, with 24-ns reset time, while Zhao et al.[205] measured sub-4 ns reset time for an equivalent NbN-based system with integrated on-chip inductors and resistors.

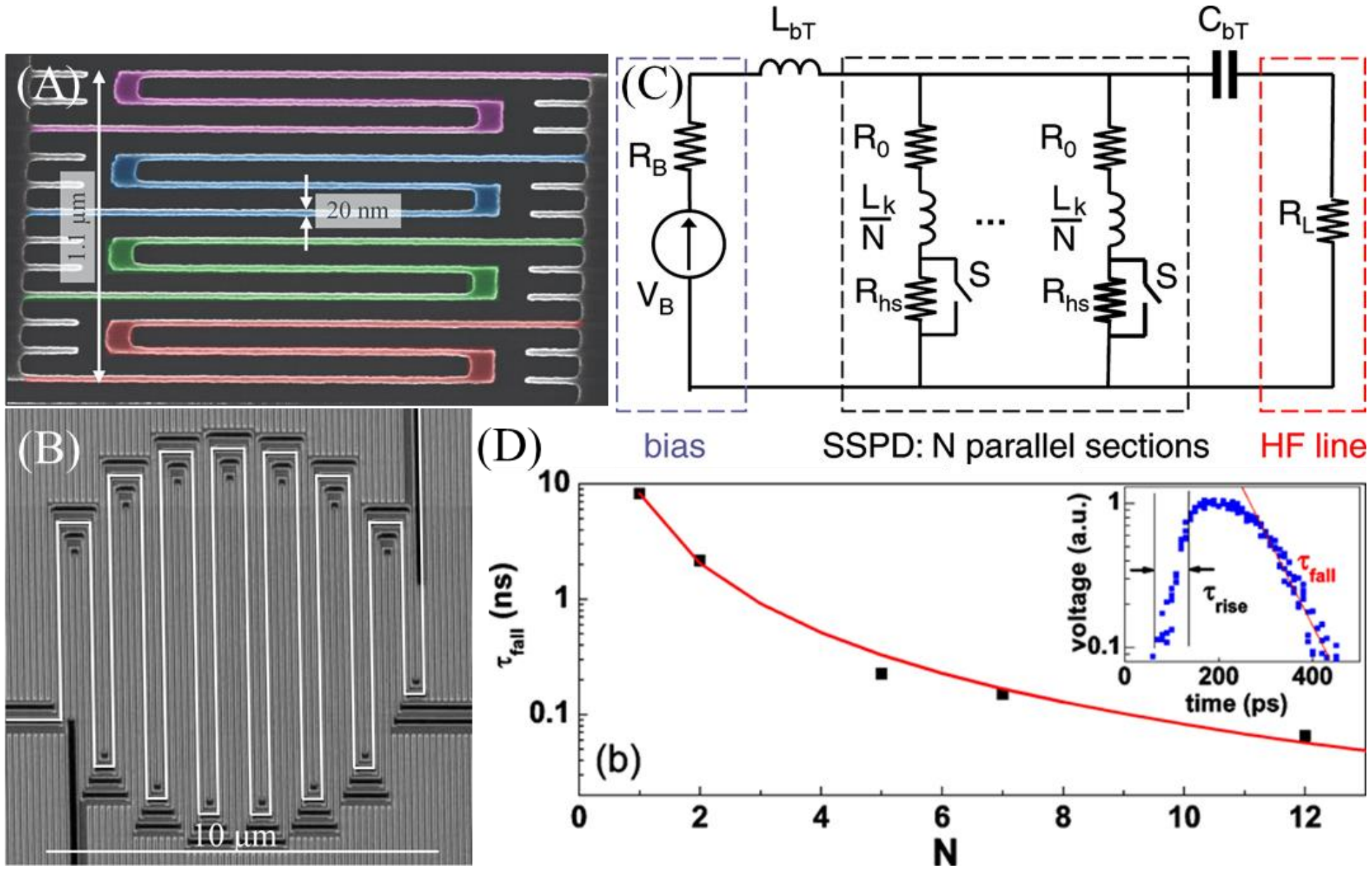

**Figure 18| Parallel-wire design of SNSPDs**. Scanning electron micrograph (false colors) of SNSPDs with parallel-wire design: (**A**) a superconducting nanowire avalanche photo-detector (SNAP). Reproduced with permission.[206] 2012 © AIP Publishing LLC. (**B**) an interleaved photon counter. Reproduced with permission.[207] 2008 © arXiv. (**C**) Equivalent circuit and (**D**) improved reset time with increasing number of parallel-connected wires. Reproduced with permission.[21] 2008 © AIP Publishing LLC.

*Reset-time optimization summary*

Summarizing the requirements for short reset time in SNSPDs suggests that the main factor that governs this property is the device inductance. Hence, choosing a low-$L$ material, processing a short wire (and compensating for the coupling by integrating, e.g. s waveguide) and connecting several wires in parallel contribute all to faster reset times.

## 3. Extrinsic/experimental parameters influence

After having discussed how the device characteristics are affected by intrinsic properties, we can now examine the effects of the extrinsic properties, with a focus on current bias, operating temperature and photon wavelength. The working conditions of these extrinsic properties are set by the intrinsic behavior. For instance, the critical current and critical temperature will dictate, respectively the operating current and temperature range as illustrated in Figure 4A.

### 3.1 Current bias

Modulating the bias currents helps us bring the system closer or further away to the phase transition. Thus, increasing the bias current increases also the likelihood to detect a photon (i.e. higher DE, see Figure 6B), but also increasing the chance of detecting false signals, due to increase in fluctuations (i.e. higher DCR). Bartolf et al.[50] suggested that the origin of dark counts stems from three competing mechanisms, depending on the bias current, which is normalized by the switching current (an intrinsic property): $I_{\mathrm{b_n}} = I_{\mathrm{b}}/I_{\mathrm{s}}$. The dominant mechanism for low bias current values is single-vortex motion. For larger $\mathrm{I_b}$, the dark counts are mostly associated with vortex-antivrotex unbinding and vortex hopping.

Practically, Marsili et al.[16] used tungsten silicide-based SNSPDs to demonstrated that nearly ideal DE is obtained when devices that operate at a saturation region of the DE as a function of current bias. Here, biasing the wire at the plateau region at a value much lower than the switching current allows us to operate the device at its highest DE, while yet being less sensitive to the undesirable fluctuations that are the origin of the DCR. Similar behavior has been obtained for NbN[20] and NbTiN[124] SNSPDs as well as for NbN-$W_5Si_3$ hybrid devices.[65]

The current bias affects also the resultant signal, so that the voltage of a photon-detection event is higher with increasing bias current. Consequently, increasing the current bias helps improve signal-to-noise ratio (SNR) and hence reducing the timing jitter.[157,169,175]

### 3.2 Operating temperature (and a word on high-$T_c$ superconducting nanowires)

In the framework of phonon-mediated electron pairing (i.e. BCS superconductivity), the superconducting properties depend on the electron-electron binding energy, $\Delta$, which in turn depends on the temperature. Thus, under this framework, the temperature affects SNSPD-related parameters, such as critical current. The interplay between current and temperature (as well as magnetic field) is illustrated schematically in Figure 4A, while this interplay is even more pronounced in thin films, closer to the SIT (Figure 4B-C). Hence, the operating temperature affects SNSPDs similarly to the effects of bias current (e.g. see increase in DE saturation region with decreasing temperature in Figure 6B).

The operating temperature is a balance between the requirement to reduce thermal-related fluctuations and increasing the working temperature range with respect to $T_c$ and the requirement for cost reduction of the cryocoolers (for deeper insights on cryocoolers see e.g. reviews by Radebaugh[208,209]). There is a big jump in cost for refrigerating systems that operate around the He lambda point (2.1768 K), which practically suggests a ~ 2.5-K limit. Such temperature are obtained by e.g. Gifford-McMahon refrigerators (the current commercially available state-of-the-art closed-cycle cooler for SNSPDs) that have been adopted for SNSPDs by the NIST group,[210] while pulse tube can obtain slightly higher temperatures around this range. The next big step is at the 4.2-K temperature of liquid helium, which is typically too high for conventional SNSPDs, even for NbN-based devices,[211–213] while miniaturized 4-k closed-cycle coolers have been developed for space applications and demonstrated for SNSPDs by Gemmell et al.[214]). Examining NbN-based SNSPDs, Yamashita et al.[215] demonstrated a constant decrease in system detection efficiency from 15% at 16 mK through ca. 10% at 2.5 K, which then dropped already to 2% at 4 K. These authors associated their observation to the decrease in $I_c$ with increasing temperatures, while the work was backed-up with a recent study by Zhang et al.[197] on a 90%-SDE SNSPD. Lower temperatures are also favorable for reducing noise stemming from thermal fluctuations in the superconducting state, i.e. thermal phase slips (in the BCS framework) or vortex-related fluctuations (in the BKT framework).[212] Phenomenological analysis of these fluctuations were also discussed in the literature.[197,216] Lowering the operating temperature may also assist in reducing some of the electronic noise in the low-temperature components and in the bias current that flows in the SNSPD. Lastly, lowering the temperature-related fluctuation by reducing the temperature as well as facilitating good thermal isolation of the device has a strong influence on the reduction of dark-count rate.[197,212] In this respect, using low-temperature electronics, Shibata et al.[17,217,218] demonstrated that reducing the temperature helps reduce false counts that arrive from black-body radiation.

Similar to other superconducting technologies, there is a continuous effort to develop SNSPDs based on higher critical temperatures. For instance, there have been several reports on $MgB_2$ SNSPDs,[148,219,220] working also at 4.2 K and higher with low dark-count rate. We should note that SNSPDs based on high-$T_c$ superconductors, such as the cuprates, have not been successfully implemented as SNSPDs, perhaps the high energy gap is too large to allow single-photon detection. Yet, nanowires made of high-$T_c$ superconductors have already been implemented as sensitive photon detectors (not single-photons).[221] Recently, by looking at the dark counts of YBCO nanowires Ejrnaes et al.[222] concluded that such structures might potentially lead to SNSPD.

### 3.3 Wavelength

A major advantage of SNSPDs over competing technologies is their sensitivity to photons with energy lower than that of visible light, mainly in the near-IR range, at the communication-relevant wavelengths. For a given device, photons with shorter wavelength are detected more efficiently.[22,174,223] That is, photons with higher energy destroy superconductivity more easily and hence are detectable with higher probability. This effect is attributed to the fact that higher input energy breaks more Cooper pairs (with a binding energy $\Delta$), giving rise to a larger hotspot and hence increase DE, as seen e.g. in Equations 1a-b. Therefore, typically, for a given wavelength (and a given device design), materials with lower $\Delta$ (or lower $T_c$) exhibit higher DE.[224] We suggest that also the reverse conclusion is correct, i.e. materials with lower $T_c$ (or lower $\Delta$) are better for detectors of longer-wavelength photons. This conclusion is supported, e.g. by the observation of Baek et al.[99] of saturating DE for amorphous tungsten silicide SNSPDs at $\lambda$=1850 nm. However, there have been several successful attempts to increase the operating wavelength to 5 μm[223] and even to 10.6 μm.[225] also by narrowing the SNSPD width. Extending the range to shorter wavelengths is usually less complex, as was demonstrated e.g. for single photon detection at the soft x-ray range (around 10 keV) made of TaN and Nb SNSPDs[152,226] as well as for high performance amorphous SNSPDs at the UV range (useful e.g. for characterizing trapped ions).[22]

Detection of other high energy particles, such as alpha and beta particles, was also demonstrated with SNSPDs e.g. by a NbTiN device presented by Azouz et al.[227] Likewise, SNSPDs have also been utilized in detecting neutral single molecules and nano particles, including e.g. hemoglobin, insulin tryptophan, gramicidin, polypeptide and myoglobin.[228]

Another effect of wavelength that should be discussed is the spectral absorption of the material (or the device). This effect is more trivial than the effect of wavelength on superconductivity. The intrinsic material absorption at the required operating wavelength plays usually an important role in optimizing the thickness of the film, from which a SNSPD is processed as well as the fill factor.[130] The optical system, which includes e.g. cavities, antennae and antireflective surfaces, is also optimized for the operating wavelength (Figure 8).[127,132] Moreover, operating at wavelength that are not suitable for common optical fibers or silicon-based waveguides, require more complex systems and is therefore less common in the literature. Yet, we should note that pushing the SNSPDs to work at lower energies, suitable e.g. to the 3-5 μm[223] and 8-12 μm[225] atmospheric windows, may increase the attractiveness of SNSPDs.

We should note that the wavelength affects also properties other than efficiency. For instance, the recent record for the SNSPD timing jitter was measured at wavelengths shorter than the typical 1550 nm.[18] The decrease in timing jitter with increasing wavelength may be attribute e.g. to a hotspot that propagates faster or to the higher signal-to-noise ratio, although the exact mechanisms remain unexplored.

## 4. Outlook

SNSPDs are one of the most mature technologies available for quantum applications today in the single-photon arena and can already be found as a shelf product for detecting single photons.

The recent improvements in SNSPD performance as efficient and fast detectors in the communication-relevant wavelength is attractive for a large variety of applications in space technologies, nearly commercial quantum systems and in academic research of quantum optics. Yet, there are still several rather big challenges that are awaiting to be addressed: what is the fundamental origin of the timing jitter and how can we optimize SNSPDs accordingly? Can the properties of SNSPDs be optimized so that all performance properties will be maximized in one device? Can the advantageous properties of SNSPDs be available also for longer wavelength in the mid-IR, enabling free-space communication at the IR atmospheric windows at 3-5 $\mu m$ and at 8-12 $\mu m$? Can SNSPDs fabrication be scalable, allowing large-scale devices and large number of adjacent devices that maintain the superior properties? What are the avenues leading for such scalability – new materials? New design? New fabrication processes? New systems? Can SNSPDs be attractive for a broader user community, e.g. by increasing the operating temperature or by simplifying the optical coupling? Can the fabrication process of SNSPDs allow easier integration with other optical components and systems?

The origin of the SNSPD advantageous properties and the methods to improve them by means of material, geometrical design and system apparatus are discussed thoroughly in the text and are summarized in the end of each Section, while a more general summary is presented in Table 2. Likewise, the fundamentals of the challenges specified in this Section are also discussed thoroughly in the current paper. Thus, we hope that this review can be helpful not only for understanding SNSPDs but also for finding a path to address these intriguing challenges.

## 5. Tables

**Table 1.** SNSPD characteristics of typical high-performance devices for different materials.

| *Material* | *Thickness [nm]* | *Intrinsic* | *SDE [%]* | *DCR [cps]* | *Jitter [ps]* | *reset time [ns]* | *Reference and comments* |
|---|---|---|---|---|---|---|---|
| *NbN* | 5 | | 7 (QE) | $10^1$ | | | [229] Optical lithography |
| | 7 | $R_{\square} = 520\ \Omega$ $RRR = 0.64$ | 94 | | | | [20] |
| | 6 | $R_{\square} = 630\ \Omega$ $RRR = 0.74$ | 40-10 (QE) | | | | [230] $\lambda = 2.5 - 3\ \mu m$ |
| | 7 | $\rho^{@20K} = 293\ \mu\Omega \cdot cm$ | | $10^2$ | 68-110 | 32 | [171] |
| | 7 | | 90.2-92.1 | $10^1$ | 79 | 48.5 | [197] |
| | 5-6 | | | | 14.2 | | [173] |
| | 8 | | 20 | $10^2$ | | | [231] |
| | 7 | $RRR$=0.8 | | | 2.7-4.6 | | [18] $\lambda = 400\ nm - 1.55\ \mu m$ |
| | 6.5 | | 65 | $10^2$ | 81 | 100 | [178] Diameter: $100\ \mu m$ |
| | 6.5 | | 70 | $10^2$ | 176 | 5.6 | [188] |
| | 3.5 | | 12 | $10^{-1}$ | | | [218] |
| | 7 | | 85 | $10^{-1}$ | | | [78] |
| | | | | | | | |
| *NbTiN* | 9 | | 86 | | 10.91 | | [124] Cavity |
| | 5 | $\rho^{@20K} = 156\ \mu\Omega \cdot cm$ | 77.5 | $10^2$ | 38-46 | 17 | [171] |
| | 5 | $R_n > 1000\ \Omega$ $L_k = 342\ nH$ | 74 | $10^2$ | 68 | | [68,199] |
| | 6 | | 23 (DE) | $10^3$ | 60 | 6.58 | [198] |
| | 8.4 | | 91.5-93.3 | $10^2$ | 49 | 20 | [124] |
| | 8 | $R_{\square} = 231\ \Omega$ $L_{k_{\square}} = 0.144\ nH$ | 68 | $10^{-1}$ | 53 | 1.2 | [174] Waveguide |
| | | | | | | | |
| *NbSi* | 10 | $R_{\square} = 262\ \Omega$ $L_k = 50\ nH$ | 0.1 | $10^4$ | | 1 | [102] $\lambda = 1700\ nm$ |
| | | | | | | | |
| *TaN* | 3.9 | $R_{\square} = 590\ \Omega$ | 0.01 | $10^5$ | | 5 | [232] |
| | 5 | $R_{\square} = 500\ \Omega$ | 20 | | | | [233] $\lambda = 700\ nm$ |
| | | | | | | | |
| *Nb* | 7.5 | $R_{\square}^{@10K} = 105\ \Omega$ | 6 | | 100 | 6 | [176] |
| | 7.5 | $R_\square = 110\ \Omega$ | 5 | | | 4.6 | [193] |
| *Nb** | 4.5 | $R_\square = 30 - 70\ \Omega$ | 26 | $10^0$ | | 2-30 | [234] $\lambda = 404$ nm |
| | | | | | | | |
| *WSi* | 4-5 | $\rho^{@300\mathrm{K}} = 200\ \mu\Omega\cdot\mathrm{cm}$ | 93 | $10^3$ | 150 | 40 | [16] |
| | | | | | 10.3 | | [164] 160 nm long |
| | 4.5 | | 20 | $10^2$ | | 14 | [99] |
| | 4.6 | | 78(70) | $10^2(1)$ | 191(225) | | [100] |
| | 5 | | 88 | | 60 | | [38] |
| | | | | | | | |
| *MoSi* | 10 | | 76(69) | $10^{-4}$ | 56(62) | 90(11) | [22] $\lambda = 250 - 370$ nm |
| | 6.6 | $L_{k_\square} = 196$ pH | 87.1 | $10^2$ | 76 | 35 | [97,175] |
| | 7 | $RRR = 0.94$ | 80 | $10^3$ | 26 | | [98] |
| | 10 | | 5 | $10^1$ | 51 | | [235] |
| | 4 | $R_\square = 460\ \Omega$ <br> $L_{k_\square} = 150$ pH | 18 | $10^1$ | 120 | 6 | [96] |
| | | | 20 | $10^2$ | | | [236] |
| | | | | | | | |
| *MoGe* | 7.5 | | 25-30 | $10^2$ | 69 | 9 | [97] |
| | | | | | | | |
| *MoN* | 3.6 | $R_\square = 425\ \Omega$ <br> $L_k = 250$ nH | | | | 5 | [237] |
| | | | | | | | |
| *NbN/αW₅Si₃* | 2/2 | $R_\square = 447\ \Omega$ | 96 (QE) | $10^3$ | 52 | 5 | [65] |
| | | | | | | | |
| *MgB₂* | 15 | | | $10^{-1}$ | | | [148] $@T = 5$ K |
| | 10 | $\rho^{@300\mathrm{K}} = 300\ \mu\Omega\cdot\mathrm{cm}$ | $10^{-3}$ | | | 2 | [219] |
| | 10 | | $10^{-7}$ | | | 5 | [220] |

**Table 2.** Summary of the dependence of performance metrics on material properties, device design and operation conditions.

| *Performance metric* | | *Intrinsic* | | | | | | | | | *Extrinsic* | | | |
|---|---|---|---|---|---|---|---|---|---|---|---|---|---|---|
| | | *Geometry* | | | *Material* | | | | | | *Operation* | | | *Other* |
| | | Length | Width | Thickness | $T_c$ | $I_c$ | $\Delta$ | $L_k$ | $R_\square$ | Homogeneity | Temp. | $I_b$ | Wave-length | |
| *Efficiency* | DE | Long [238] | Narrow [147,223] | Optimization between: disorder (thin[20]) and absorptance [124] | Low [16] | High [16] | Low [102] | - | High [20] | Homogeneous [58] | Low [197,215] | High [16,99] | Short [174,223] | Waveguides,[88] cavities,[131] ARC,[127] antennae,[132] constriction free[106] |
| | DCR | Short [197] | Narrow [147] | Thick [152] | High [50,143] | High [151] | High [239] | - | - | Homogeneous [22] | Low [50,143] | Low [50] | Indirect ($I_c$[240]) | Constriction and current crowding free,[151] low-temperature filters[218] |
| *Time* | Jitter | Short [18] | Narrow [156] | Indirect effect ($L_k$[241], $I_c$[169]) | High (Fig. 12) | High [169] | High [158] | Low [157,158] | High (SNR[159]) | Optimization between: constrictions (homogeneous[158]) and intrinsic speed (crystalline, empiric[65]) | Low [169] | High [169,175] | Short [18] | Low noise,[169] differential measurement[157] |
| | Reset time | Short [19,88] | Indirect effect (narrow for high $Z$[172]) | Indirect effect (thin for high $Z$[172]) | Indirect effect ($I_c$[193]) | High [193] | Indirect effect ($Z$ [161]) | Low [172] | Optimized with the shunt[161] | Crystalline (empiric) | Low [242] | Low [193] | Short [22] | Parallel design (SNAP[202]/ME[190]) |

**Table 3.** Superconducting properties of common superconducting materials.

| *Material* | $T_c[K]$ | $J_c\left[\frac{MA}{cm^2}\right]$ | $2\Delta[meV]$ | $\xi[nm]$ **Thin film* |
|---|---|---|---|---|
| *NbN* | 16 [243] | 2-4 [115] | 4.9 [244] | 6.5* [245] |
| *NbTiN* | 16 [246] | 8 [113] | 4.5 [246] | 170 [246] |
| *Nb* | 9.2 [247] | 2-7 [248] | 3 [249] | 38 [154] |
| *WSi* | 5 [250] | 0.8 [58] | 1.52 [58] | 7-7.4* [251] |
| *MoSi* | 7.5 [252] | 1.3 [96] | 2.28 [58] | 7-14* [253] |
| *MoGe* | 7.4 [58] | 1.2 [58] | 2.2 [58] | 5 [254] |
| $MgB_2$ | 39 [255] | 1.1 [219] | 4-14 [256] | 5.2 [257] |

## 6. Acknowledgments

We would like acknowledges financial support from the Zuckerman STEM Leadership Program, the Horev Fellowship for Leadership in Science and Technology, supported by the Taub Foundation, the Israel Science Foundation (ISF) grant # 1602/17 as well as the Eliyahu Pen Research Fund.

## 7. References

[1] M. F. Riedel, D. Binosi, R. Thew, T. Calarco, *Quantum Sci. Technol.* **2017**, *2*, 030501.

[2] M. Mohseni, P. Read, H. Neven, S. Boixo, V. Denchev, R. Babbush, A. Fowler, V. Smelyanskiy, J. Martinis, *Nature* **2017**, *543*, 171.

[3] E. Gibney, *Nature* **2017**, *545*, 16.

[4] J. I. Cirac, H. J. Kimble, *Nat. Photonics* **2017**, *11*, 18.

[5] M. Sharma, *China J.* **2018**, *80*, 24.

[6] K. Singh, K. L. Chuan, A. Ekert, C. C. Theng, J. Hogan, E. Tan, In *50 Years of Science in Singapore*; WORLD SCIENTIFIC, 2017; pp. 361–387.

[7] A. Herman, *Am. Aff. J.* **2008**, *2*.

[8] T. E. Northup, R. Blatt, *Nat. Photonics* **2014**, *8*, 356.

[9] I. S. Osborne, *Science (80-. ).* **2018**, *361*, 38.

[10] A. Aspuru-Guzik, P. Walther, *Nat. Phys.* **2012**, *8*, 285.

[11] J. P. Kraack, *Nat. Phys.* **2018**, *14*, 776.

[12] I. Aharonovich, D. Englund, M. Toth, *Nat. Photonics* **2016**, *10*, 631.

[13] G. N. Gol'tsman, O. Okunev, G. Chulkova, A. Lipatov, A. Semenov, K. Smirnov, B. Voronov, A. Dzardanov, C. Williams, R. Sobolewski, *Appl. Phys. Lett.* **2001**, *79*, 705.

[14] R. H. Hadfield, *Nat. Photonics* **2009**, *3*, 696.

[15] M. D. Eisaman, J. Fan, A. Migdall, S. V. Polyakov, *Rev. Sci. Instrum.* **2011**, *82*, 071101.

[16] F. Marsili, V. B. Verma, J. A. Stern, S. Harrington, A. E. Lita, T. Gerrits, I. Vayshenker, B. Baek, M. D. Shaw, R. P. Mirin, S. W. Nam, *Nat. Photonics* **2013**, *7*, 210.

[17] H. Shibata, K. Shimizu, H. Takesue, Y. Tokura, *Opt. Lett.* **2015**, *40*, 3428.

[18] B. A. Korzh, Q.-Y. Zhao, S. Frasca, J. P. Allmaras, T. M. Autry, E. A. Bersin, M. Colangelo, G. M. Crouch, A. E. Dane, T. Gerrits, F. Marsili, G. Moody, E. Ramirez, J. D. Rezac, M. J. Stevens, E. E. Wollman, D. Zhu, P. D. Hale, K. L. Silverman, R. P. Mirin, S. W. Nam, M. D. Shaw, K. K.

Berggren, *arXiv Prepr. arXiv1804.06839* **2018**, 1.

[19] A. Vetter, S. Ferrari, P. Rath, R. Alaee, O. Kahl, V. Kovalyuk, S. Diewald, G. N. Goltsman, A. Korneev, C. Rockstuhl, W. H. P. Pernice, *Nano Lett.* **2016**, *16*, 7085.

[20] K. Smirnov, A. Divochiy, Y. Vakhtomin, P. Morozov, P. Zolotov, A. Antipov, V. Seleznev, *Supercond. Sci. Technol.* **2018**, *31*, 035011.

[21] M. Tarkhov, J. Claudon, J. P. Poizat, A. Korneev, A. Divochiy, O. Minaeva, V. Seleznev, N. Kaurova, B. Voronov, A. V Semenov, G. Gol'tsman, *Appl. Phys. Lett.* **2008**, *92*, 241112.

[22] E. E. Wollman, V. B. Verma, A. D. Beyer, R. M. Briggs, F. Marsili, J. P. Allmaras, A. E. Lita, R. P. Mirin, S. W. Nam, M. D. Shaw, *Opt. Express* **2017**, *25*, 26792.

[23] Z. Zhang, C. Chen, Q. Zhuang, F. N. C. Wong, J. H. Shapiro, *Quantum Sci. Technol.* **2017**, *3*, 025007.

[24] H. L. Yin, T. Y. Chen, Z. W. Yu, H. Liu, L. X. You, Y. H. Zhou, S. J. Chen, Y. Mao, M. Q. Huang, W. J. Zhang, H. Chen, M. J. Li, D. Nolan, F. Zhou, X. Jiang, Z. Wang, Q. Zhang, X. Bin Wang, J. W. Pan, *Phys. Rev. Lett.* **2016**, *117*, 190501.

[25] A. V. Glejm, A. A. Anisimov, L. N. Asnis, Y. B. Vakhtomin, A. V. Divochiy, V. I. Egorov, V. V. Kovalyuk, A. A. Korneev, S. M. Kynev, Y. V. Nazarov, R. V. Ozhegov, A. V. Rupasov, K. V. Smirnov, M. A. Smirnov, G. N. Goltsman, S. A. Kozlov, *Bull. Russ. Acad. Sci. Phys.* **2014**, *78*, 171.

[26] J. Yin, Y. Cao, Y. H. Li, S. K. Liao, L. Zhang, J. G. Ren, W. Q. Cai, W. Y. Liu, B. Li, H. Dai, G. B. Li, Q. M. Lu, Y. H. Gong, Y. Xu, S. L. Li, F. Z. Li, Y. Y. Yin, Z. Q. Jiang, M. Li, J. J. Jia, G. Ren, D. He, Y. L. Zhou, X. X. Zhang, N. Wang, X. Chang, Z. C. Zhu, N. Le Liu, Y. A. Chen, C. Y. Lu, R. Shu, C. Z. Peng, J. Y. Wang, J. W. Pan, *Science (80-. ).* **2017**, *356*, 1140.

[27] L. Xue, Z. Li, L. Zhang, D. Zhai, Y. Li, S. Zhang, M. Li, L. Kang, J. Chen, P. Wu, Y. Xiong, *Opt.*

*Lett.* **2016**, *41*, 3848.

[28] D. M. Boroson, B. S. Robinson, D. V Murphy, D. A. Burianek, F. Khatri, J. M. Kovalik, Z. Sodnik, D. M. Cornwell, *Free. Laser Commun. Atmos. Propag. XXVI* **2014**, *8971*, 89710S.

[29] C. M. Natarajan, L. Zhang, H. Coldenstrodt-Ronge, G. Donati, S. N. Dorenbos, V. Zwiller, I. A. Walmsley, R. H. Hadfield, *Opt. Express* **2013**, *21*, 893.

[30] A. McCarthy, N. J. Krichel, N. R. Gemmell, X. Ren, M. G. Tanner, S. N. Dorenbos, V. Zwiller, R. H. Hadfield, G. S. Buller, *Opt. Express* **2013**, *21*, 8904.

[31] L. K. Shalm, E. Meyer-Scott, B. G. Christensen, P. Bierhorst, M. A. Wayne, M. J. Stevens, T. Gerrits, S. Glancy, D. R. Hamel, M. S. Allman, K. J. Coakley, S. D. Dyer, C. Hodge, A. E. Lita, V. B. Verma, C. Lambrocco, E. Tortorici, A. L. Migdall, Y. Zhang, D. R. Kumor, W. H. Farr, F. Marsili, M. D. Shaw, J. A. Stern, C. Abellán, W. Amaya, V. Pruneri, T. Jennewein, M. W. Mitchell, P. G. Kwiat, J. C. Bienfang, R. P. Mirin, E. Knill, S. W. Nam, *Phys. Rev. Lett.* **2015**, *115*, 250402.

[32] R. Riedinger, S. Hong, R. A. Norte, J. A. Slater, J. Shang, A. G. Krause, V. Anant, M. Aspelmeyer, S. Gröblacher, *Nature* **2016**, *530*, 313.

[33] J. Zhang, N. Boiadjieva, G. Chulkova, H. Deslandes, G. N. Gol’tsman, A. Korneev, P. Kouminov, M. Leibowitz, W. Lo, R. Malinsky, O. Okunev, A. Pearlman, W. Slysz, K. Smirnov, C. Tsao, A. Verevkin, B. Voronov, K. Wilsher, R. Sobolewski, *Electron. Lett.* **2003**, *39*, 1086.

[34] M. G. Tanner, S. D. Dyer, B. Baek, R. H. Hadfield, S. Woo Nam, *Appl. Phys. Lett.* **2011**, *99*, 201110.

[35] C. M. Natarajan, M. G. Tanner, R. H. Hadfield, *Supercond. Sci. Technol.* **2012**, *25*, 063001.

[36] T. Yamashita, S. Miki, H. Terai, *IEICE Trans. Electron.* **2017**, *100*, 274.

[37] D. M. Boroson, B. S. Robinson, *Lunar Atmos. Dust Environ. Explor. Mission* **2015**, 115.

[38] E. A. Dauler, M. E. Grein, A. J. Kerman, F. Marsili, S. Miki, S. W. Nam, M. D. Shaw, H. Terai, V. B. Verma, T. Yamashita, *Opt. Eng.* **2014**, *53*, 081907.

[39] Q. Guo, H. Li, L. You, W. Zhang, L. Zhang, Z. Wang, X. Xie, M. Qi, *Sci. Rep.* **2015**, *5*, 9616.

[40] A. M. Kadin, M. W. Johnson, *Appl. Phys. Lett.* **1996**, *69*, 3938.

[41] A. D. Semenov, G. N. Gol'tsman, A. A. Korneev, *Phys. C* **2001**, *351*, 349.

[42] A. Semenov, A. Engel, H. W. Hübers, K. Il'in, M. Siegel, *Eur. Phys. J. B* **2005**, *47*, 495.

[43] J. M. Kosterlitz, D. J. Thouless, *J. Phys. C* **1973**, *6*, 1181.

[44] V. L. Berezinskii, *Sov. Phys. JETP* **1971**, *32*, 493.

[45] B. I. Halperin, D. R. Nelson, *J. Low Temp. Phys.* **1979**, *36*, 599.

[46] A. M. Kadin, M. Leung, A. D. Smith, J. M. Murduck, *Appl. Phys. Lett.* **1990**, *57*, 2847.

[47] A. M. Kadin, M. Leung, A. D. Smith, *Phys. Rev. Lett.* **1990**, *65*, 3193.

[48] A. N. Zotova, D. Y. Vodolazov, *Phys. Rev. B* **2012**, *85*, 024509.

[49] L. N. Bulaevskii, M. J. Graf, C. D. Batista, V. G. Kogan, *Phys. Rev. B - Condens. Matter Mater. Phys.* **2011**, *83*, 144526.

[50] H. Bartolf, A. Engel, A. Schilling, K. Il'In, M. Siegel, H. W. Hübers, A. Semenov, *Phys. Rev. B* **2010**, *81*, 024502.

[51] A. Engel, J. J. Renema, K. Il'In, A. Semenov, *Supercond. Sci. Technol.* **2015**, *28*, 114003.

[52] J. J. Renema, R. Gaudio, Q. Wang, Z. Zhou, A. Gaggero, F. Mattioli, R. Leoni, D. Sahin, M. J. A. De Dood, A. Fiore, M. P. Van Exter, *Phys. Rev. Lett.* **2014**, *112*, 117604.

[53] M. Tinkham, *Introduction to superconductivity*; Courier Corporation, 2004.

[54] D. B. Haviland, Y. Liu, A. M. Goldman, *Phys. Rev. Lett.* **1989**, *62*, 2180.

[55] A. Bezryadin, C. N. Lau, M. Tinkham, *Nature* **2000**, *404*, 971.

[56] V. Ambegaokar, A. Baratoff, *Phys. Rev. Lett.* **1963**, *11*, 104.

[57] L. You, X. Shen, X. Yang, *Chinese Sci. Bull.* **2010**, *55*, 441.

[58] A. E. Lita, V. B. Verma, R. D. Horansky, J. M. Shainline, R. P. Mirin, S. Nam, *MRS Proc.* **2015**, *1807*, 1.

[59] Y. Ivry, C. S. Kim, A. E. Dane, D. De Fazio, A. N. McCaughan, K. A. Sunter, Q. Zhao, K. K. Berggren, *Phys. Rev. B* **2014**, *90*, 214515.

[60] W. L. McMillan, *Phys. Rev.* **1968**, *167*, 331.

[61] R. C. Dynes, *Solid State Commun.* **1972**, *10*, 615.

[62] A. M. Finkel'stein, *Phys. B Phys. Condens. Matter* **1994**, *197*, 636.

[63] M. Tinkham, *Phys. Rev.* **1956**, *104*, 845.

[64] R. E. Glover, M. Tinkham, *Phys. Rev.* **1957**, *108*, 243.

[65] Y. Ivry, J. J. Surick, M. Barzilay, C. S. Kim, F. Najafi, E. Kalfon-Cohen, A. D. Dane, K. K. Berggren, *Nanotechnology* **2017**, *28*, 435205.

[66] M. Hofherr, D. Rall, K. Ilin, M. Siegel, A. Semenov, H. W. Hübers, N. A. Gippius, *J. Appl. Phys.* **2010**, *108*, 014507.

[67] F. Najafi, F. Marsili, V. B. Verma, Q. Zhao, M. D. Shaw, K. K. Berggren, S. W. Nam, In *Superconducting Devices in Quantum Optics*; Springer, Cham, 2016; pp. 3–30.

[68] S. Miki, T. Yamashita, H. Terai, Z. Wang, *Opt. Express* **2013**, *21*, 10208.

[69] X. Yang, H. Li, W. Zhang, L. You, L. Zhang, X. Liu, Z. Wang, W. Peng, X. Xie, M. Jiang, *Opt.*

*Express* **2014**, *22*, 16267.

[70] L. Zhang, M. Gu, T. Jia, R. Xu, C. Wan, L. Kang, J. Chen, P. Wu, *IEEE Photonics J.* **2014**, *6*, 1.

[71] Z. Yan, M. K. Akhlaghi, J. L. Orgiazzi, A. Hamed Majedi, *J. Mod. Opt.* **2009**, *56*, 380.

[72] G. Bachar, I. Baskin, O. Shtempluck, E. Buks, *Appl. Phys. Lett.* **2012**, *101*, 262601.

[73] G. Fujii, D. Fukuda, T. Numata, A. Yoshizawa, H. Tsuchida, S. Inoue, T. Zama, In *Lecture Notes of the Institute for Computer Sciences, Social-Informatics and Telecommunications Engineering*; Springer, Berlin, Heidelberg, 2010; Vol. 36 LNICST, pp. 220–224.

[74] M. E. Grein, M. Willis, A. Kerman, E. Dauler, B. Romkey, D. Rosenberg, J. Yoon, R. Molnar, B. S. Robinson, D. Murphy, D. M. Boroson, In *CLEO: 2014*; OSA: Washington, D.C., 2014; p. SM4J.5.

[75] H. Le Jeannic, V. B. Verma, A. Cavaillès, F. Marsili, M. D. Shaw, K. Huang, O. Morin, S. W. Nam, J. Laurat, *Opt. Lett.* **2016**, *41*, 5341.

[76] R. Cheng, X. Guo, X. Ma, L. Fan, K. Y. Fong, M. Poot, H. X. Tang, *Opt. Express* **2016**, *24*, 27070.

[77] X. Hu, T. Zhong, J. E. White, E. A. Dauler, F. Najafi, C. H. Herder, F. N. C. Wong, K. K. Berggren, *Opt. Lett.* **2009**, *34*, 3607.

[78] W. J. Zhang, X. Y. Yang, H. Li, L. X. You, C. L. Lv, L. Zhang, C. J. Zhang, X. Y. Liu, Z. Wang, X. M. Xie, *Supercond. Sci. Technol.* **2018**, *31*, 035012.

[79] S. Miki, T. Yamashita, M. Fujiwara, M. Sasaki, Z. Wang, *IEEE Trans. Appl. Supercond.* **2011**, *21*, 332.

[80] X. Hu, C. W. Holzwarth, D. Masciarelli, E. A. Dauler, K. K. Berggren, *IEEE Trans. Appl. Supercond.* **2009**, *19*, 336.

[81] F. Stellari, A. J. Weger, S. Kim, D. Maliuk, P. Song, H. A. Ainspan, Y. Kwark, C. W. Baks, U. Kindereit, V. Anant, others, In *ISTFA 2013: Proceedings from the 39th International Symposium for Testing and Failure Analysis*; 2013; p. 182.

[82] X. Hu, F. Najafi, J. Mower, F. Bellei, X. Mao, P. Kharel, Y. Ivry, A. McCaughan, L. L. Cheong, K. K. Berggren, D. R. Englund, In *Frontiers in Optics 2013*; OSA: Washington, D.C., 2013; p. FW1C.5.

[83] A. J. Miller, A. E. Lita, B. Calkins, I. Vayshenker, S. M. Gruber, S. W. Nam, *Opt. Express* **2011**, *19*, 9102.

[84] A. Verevkin, J. Zhang, R. Sobolewski, A. Lipatov, O. Okunev, G. Chulkova, A. Korneev, K. Smirnov, G. N. Gol'tsman, A. Semenov, G. N. Gol'tsman, A. Semenov, G. N. Gol'tsman, A. Semenov, *Appl. Phys. Lett.* **2002**, *80*, 4687.

[85] M. S. Allman, V. B. Verma, M. Stevens, T. Gerrits, R. D. Horansky, A. E. Lita, F. Marsili, A. Beyer, M. D. Shaw, D. Kumor, R. Mirin, S. W. Nam, *Appl. Phys. Lett.* **2015**, *106*, 192601.

[86] J. K. W. Yang, A. J. Kerman, E. A. Dauler, B. Cord, V. Anant, R. J. Molnar, K. K. Berggren, *IEEE Trans. Appl. Supercond.* **2009**, *19*, 318.

[87] E. F. C. Driessen, F. R. Braakman, E. M. Reiger, S. N. Dorenbos, V. Zwiller, M. J. A. de Dood, *Eur. Phys. J. Appl. Phys.* **2009**, *47*, 10701.

[88] W. H. P. Pernice, C. Schuck, O. Minaeva, M. Li, G. N. Goltsman, A. V. Sergienko, H. X. Tang, *Nat. Commun.* **2012**, *3*, 1325.

[89] S. Khasminskaya, F. Pyatkov, K. Słowik, S. Ferrari, O. Kahl, V. Kovalyuk, P. Rath, A. Vetter, F. Hennrich, M. M. Kappes, G. Gol'tsman, A. Korneev, C. Rockstuhl, R. Krupke, W. H. P. Pernice, *Nat. Photonics* **2016**, *10*, 727.

[90] F. Najafi, J. Mower, N. C. Harris, F. Bellei, A. Dane, C. Lee, X. Hu, P. Kharel, F. Marsili, S.

Assefa, K. K. Berggren, D. Englund, *Nat. Commun.* **2015**, *6*, 5873.

[91] F. Najafi, J. Mower, X. Hu, F. Bellei, P. Kharel, A. Dane, Y. Ivry, L. L. Cheong, K. Sunter, D. Englund, K. K. Berggren, *Opt. Soc. Am.* **2013**, QF1A.6.

[92] S. Chen, D. Liu, W. Zhang, L. You, Y. He, W. Zhang, X. Yang, G. Wu, M. Ren, H. Zeng, Z. Wang, X. Xie, M. Jiang, *Appl. Opt.* **2013**, *52*, 3241.

[93] Q. Y. Zhao, D. Zhu, N. Calandri, A. E. Dane, A. N. McCaughan, F. Bellei, H. Z. Wang, D. F. Santavicca, K. K. Berggren, *Nat. Photonics* **2017**, *11*, 247.

[94] H. Zhou, Y. He, L. You, S. Chen, W. Zhang, J. Wu, Z. Wang, X. Xie, *Opt. Express* **2015**, *23*, 14603.

[95] F. Bellei, A. P. Cartwright, A. N. McCaughan, A. E. Dane, F. Najafi, Q. Zhao, K. K. Berggren, *Opt. Express* **2016**, *24*, 3248.

[96] Y. P. Korneeva, M. Y. Mikhailov, Y. P. Pershin, N. N. Manova, A. V Divochiy, Y. B. Vakhtomin, A. A. Korneev, K. V Smirnov, A. G. Sivakov, A. Y. Devizenko, G. N. Goltsman, *Supercond. Sci. Technol.* **2014**, *27*, 095012.

[97] V. B. Verma, A. E. Lita, M. R. Vissers, F. Marsili, D. P. Pappas, R. P. Mirin, S. W. Nam, *Appl. Phys. Lett.* **2014**, *105*, 022602.

[98] M. Caloz, M. Perrenoud, C. Autebert, B. Korzh, M. Weiss, C. Schönenberger, R. J. Warburton, H. Zbinden, F. Bussières, *Appl. Phys. Lett.* **2018**, *112*.

[99] B. Baek, A. E. Lita, V. Verma, S. W. Nam, *Appl. Phys. Lett.* **2011**, *98*, 251105.

[100] V. B. Verma, B. Korzh, F. Bussières, R. D. Horansky, A. E. Lita, F. Marsili, M. D. Shaw, H. Zbinden, R. P. Mirin, S. W. Nam, *Appl. Phys. Lett.* **2014**, *105*, 122601.

[101] E. E. Wollman, V. Verma, R. M. Briggs, A. D. Beyer, R. Mirin, S. W. Nam, F. Marsili, M. D.

Shaw, In *Conference on Lasers and Electro-Optics*; 2016; p. FW4C.4.

[102] S. N. Dorenbos, P. Forn-Díaz, T. Fuse, A. H. Verbruggen, T. Zijlstra, T. M. Klapwijk, V. Zwiller, *Appl. Phys. Lett.* **2011**, *98*, 251102.

[103] J. R. Clem, K. K. Berggren, *Phys. Rev. B - Condens. Matter Mater. Phys.* **2011**, *84*, 174510.

[104] H. L. Hortensius, E. F. C. Driessen, T. M. Klapwijk, K. K. Berggren, J. R. Clem, *Appl. Phys. Lett.* **2012**, *100*, 182602.

[105] D. Henrich, P. Reichensperger, M. Hofherr, J. M. Meckbach, K. Il'In, M. Siegel, A. Semenov, A. Zotova, D. Y. Vodolazov, *Phys. Rev. B* **2012**, *86*, 144504.

[106] A. J. Kerman, E. A. Dauler, J. K. W. Yang, K. M. Rosfjord, V. Anant, K. K. Berggren, G. N. Gol'tsman, B. M. Voronov, *Appl. Phys. Lett.* **2007**, *90*, 101110.

[107] J. K. W. Yang, E. Dauler, A. Ferri, A. Pearlman, A. Verevkin, G. Gol'tsman, B. Voronov, R. Sobolewski, W. E. Keicher, K. K. Berggren, *IEEE Trans. Appiled Supercond.* **2005**, *15*, 626.

[108] W. Li, J. C. Fenton, P. A. Warburton, *IEEE Trans. Appl. Supercond.* **2009**, *19*, 2819.

[109] L. Zhang, L. You, X. Yang, J. Wu, C. Lv, Q. Guo, W. Zhang, H. Li, W. Peng, Z. Wang, X. Xie, *Sci. Rep.* **2018**, *8*, 1486.

[110] H. A. Atikian, A. Eftekharian, A. Jafari Salim, M. J. Burek, J. T. Choy, A. Hamed Majedi, M. Lončar, *Appl. Phys. Lett.* **2014**, *104*, 122602.

[111] P. Rath, O. Kahl, S. Ferrari, F. Sproll, G. Lewes-Malandrakis, D. Brink, K. Ilin, M. Siegel, C. Nebel, W. Pernice, *Light Sci. Appl.* **2015**, *4*, e338.

[112] S. Miki, M. Fujiwara, M. Sasaki, Z. Wang, *IEEE Trans. Appl. Supercond.* **2007**, *17*, 285.

[113] S. Miki, M. Takeda, M. Fujiwara, M. Sasaki, A. Otomo, Z. Wang, *Appl. Phys. Express* **2009**, *2*, 075002.

[114] Z. Wang, A. Kawakami, Y. Uzawa, B. Komiyama, *J. Appl. Phys.* **1996**, *79*, 7837.

[115] F. Marsili, D. Bitauld, A. Fiore, A. Gaggero, F. Mattioli, R. Leoni, M. Benkahoul, F. LÚvy, *Opt. Express* **2008**, *16*, 3191.

[116] S. Z. Lin, O. Ayala-Valenzuela, R. D. McDonald, L. N. Bulaevskii, T. G. Holesinger, F. Ronning, N. R. Weisse-Bernstein, T. L. Williamson, A. H. Mueller, M. A. Hoffbauer, M. W. Rabin, M. J. Graf, *Phys. Rev. B* **2013**, *87*, 184507.

[117] M. Ziegler, L. Fritzsch, J. Day, S. Linzen, S. Anders, J. Toussaint, H. G. Meyer, *Supercond. Sci. Technol.* **2013**, *26*, 025008.

[118] W. N. Kang, H.-J. Kim, E.-M. Choi, C. U. Jung, S.-I. Lee, *Science (80-. ).* **2001**, *292*, 1521.

[119] A. Semenov, B. Günther, U. Böttger, H. W. Hübers, H. Bartolf, A. Engel, A. Schilling, K. Ilin, M. Siegel, R. Schneider, D. Gerthsen, N. A. Gippius, *Phys. Rev. B* **2009**, *80*, 054510.

[120] R. Espiau De Lamaëstre, P. Odier, J. C. Villégier, *Appl. Phys. Lett.* **2007**, *91*, 232501.

[121] C. Schuck, W. H. P. Pernice, H. X. Tang, *2013 IEEE Photonics Conf. IPC 2013* **2013**, 370.

[122] M. G. Tanner, L. S. E. Alvarez, W. Jiang, R. J. Warburton, Z. H. Barber, R. H. Hadfield, *Nanotechnology* **2012**, *23*, 505201.

[123] D. Bosworth, S. L. Sahonta, R. H. Hadfield, Z. H. Barber, *AIP Adv.* **2015**, *5*, 087106.

[124] I. E. Zadeh, J. W. N. Los, R. B. M. Gourgues, V. Steinmetz, G. Bulgarini, S. M. Dobrovolskiy, V. Zwiller, S. N. Dorenbos, *APL Photonics* **2017**, *2*, 111301.

[125] A. J. Kerman, B. S. Robinson, R. J. Barron, D. O. Caplan, M. L. Stevens, J. J. Carney, S. A. Hamilton, W. E. Keicher, E. A. Dauler, J. K. W. Yang, K. Rosfjord, V. Anant, K. K. Berggren, In *2006 Digest of the LEOS Summer Topical Meetings*; IEEE; pp. 9–10.

[126] G. N. Goltsman, K. Smirnov, P. Kouminov, B. Voronov, N. Kaurova, V. Drakinsky, J. Zhang, A.

Verevkin, G. N. Gol'tsman, K. Smirnov, P. Kouminov, B. Voronov, N. Kaurova, V. Drakinsky, J. Zhang, A. Verevkin, R. Sobolewski, G. N. Gol, K. Smirnov, P. Kouminov, B. Voronov, N. Kaurova, V. Drakinsky, J. Zhang, A. Verevkin, G. N. Gol'tsman, K. Smirnov, P. Kouminov, B. Voronov, N. Kaurova, V. Drakinsky, J. Zhang, A. Verevkin, R. Sobolewski, *IEEE Trans. Appl. Supercond.* **2003**, *13*, 192.

[127] K. M. Rosfjord, J. K. W. Yang, E. A. Dauler, A. J. Kerman, V. Anant, B. M. Voronov, G. N. Gol'tsman, K. K. Berggren, *Opt. Express* **2006**, *14*, 527.

[128] L. B. Zhang, L. Kang, J. Chen, P. H. Wu, In *34th International Conference on Infrared, Millimeter, and Terahertz Waves, IRMMW-THz 2009*; IEEE, 2009; pp. 1–2.

[129] A. Semenov, P. Haas, H. W. Hubers, K. Ilin, M. Siegel, A. Kirste, D. Drung, T. Schurig, A. Engel, *J. Mod. Opt.* **2009**, *56*, 345.

[130] V. Anant, A. J. Kerman, E. A. Dauler, K. W. Joel, K. M. Rosfjord, K. K. K. Berggren, J. K. W. Yang, K. M. Rosfjord, K. K. K. Berggren, *Opt. Express* **2008**, *16*, 10750.

[131] I. Milostnaya, A. Korneev, I. Rubtsova, V. Seleznev, O. Minaeva, G. Chulkova, O. Okunev, B. Voronov, K. Smirnov, G. Gol'Tsman, W. Slysz, M. Wegrzecki, M. Guziewicz, J. Bar, M. Gorska, A. Pearlman, J. Kitaygorsky, A. Cross, R. Sobolewski, *J. Phys. Conf. Ser.* **2006**, *43*, 1334.

[132] X. Hu, E. A. Dauler, R. J. Molnar, K. K. Berggren, *Opt. Express* **2011**, *19*, 17.

[133] C. Zhang, R. Z. Jiao, *Chinese Phys. B* **2012**, *21*, 120306.

[134] R. M. Heath, M. G. Tanner, T. D. Drysdale, S. Miki, V. Giannini, S. A. Maier, R. H. Hadfield, *Nano Lett.* **2015**, *15*, 819.

[135] M. Csete, Á. Sipos, A. Szalai, F. Najafi, G. Szabó, K. K. Berggren, *Sci. Rep.* **2013**, *3*, 2406.

[136] S. N. Dorenbos, E. M. Reiger, N. Akopian, U. Perinetti, V. Zwiller, T. Zijlstra, T. M. Klapwijk,

*Appl. Phys. Lett.* **2008**, *93*, 161102.

[137] F. Zheng, R. Xu, G. Zhu, B. Jin, L. Kang, W. Xu, J. Chen, P. Wu, *Sci. Rep.* **2016**, *6*, 22710.

[138] J. Huang, W. J. Zhang, L. X. You, X. Y. Liu, Q. Guo, Y. Wang, L. Zhang, X. Y. Yang, H. Li, Z. Wang, X. M. Xie, *Supercond. Sci. Technol.* **2017**, *30*, 074004.

[139] C. Gu, Y. Cheng, X. Zhu, X. Hu, *Adv. Photonics 2015* **2015**, JM3A.10.

[140] L. Redaelli, V. Zwiller, E. Monroy, J. M. Gérard, *Supercond. Sci. Technol.* **2017**, *30*, 035005.

[141] S. Miki, M. Fujiwara, R.-B. Jin, T. Yamamoto, M. Sasaki, In *Superconducting Devices in Quantum Optics*; Springer, Cham, 2016; pp. 107–135.

[142] H. Takesue, S. W. Nam, Q. Zhang, R. H. Hadfield, T. Honjo, K. Tamaki, Y. Yamamoto, *Nat. Photonics* **2007**, *1*, 343.

[143] M. Hofherr, D. Rall, K. Il'In, A. Semenov, H. W. Hübers, M. Siegel, *J. Low Temp. Phys.* **2012**, *167*, 822.

[144] A. Engel, A. D. Semenov, H. W. Hübers, K. Il'in, M. Siegel, *Phys. C* **2006**, *444*, 12.

[145] B. Sacépé, C. Chapelier, T. I. Baturina, V. M. Vinokur, M. R. Baklanov, M. Sanquer, *Phys. Rev. Lett.* **2008**, *101*, 157006.

[146] A. Bezryadin, *J. Phys. Condens. Matter* **2008**, *20*, 43202.

[147] F. Marsili, F. Najafi, E. Dauler, F. Bellei, X. Hu, M. Csete, R. J. Molnar, K. K. Berggren, *Nano Lett.* **2011**, *11*, 2048.

[148] A. E. Velasco, D. P. Cunnane, S. Frasca, T. Melbourne, N. Acharya, R. Briggs, A. D. Beyer, M. D. Shaw, B. S. Karasik, M. A. Wolak, V. B. Verma, A. E. Lita, H. Shibata, M. Ohkubo, N. Zen, M. Ukibe, X. Xi, F. Marsili, In *Optics InfoBase Conference Papers*; OSA: Washington, D.C., 2017; Vol. Part F42-C, p. FF1E.7.

[149] Y. Dubi, Y. Meir, Y. Avishai, *Nature* **2007**, *449*, 876.

[150] K. Ilin, D. Henrich, Y. Luck, Y. Liang, M. Siegel, D. Y. Vodolazov, *Phys. Rev. B* **2014**, *89*, 184511.

[151] K. A. Mohsen, A. Haig, E. Amin, L. Marko, M. A. Hamed, *Phys. Rev. B* **2012**, *86*, 23610.

[152] K. Inderbitzin, A. Engel, A. Schilling, K. Ilin, M. Siegel, *Appl. Phys. Lett.* **2012**, *101*, 162601.

[153] K. K. Likharev, *Rev. Mod. Phys.* **1979**, *51*, 101.

[154] B. W. Maxfield, W. L. McLean, *Phys. Rev.* **1965**, *139*, A1515.

[155] J. P. Allmaras, A. G. Kozorezov, B. A. Korzh, M. D. Shaw, *arXiv Prepr. arXiv1805.00130* **2018**.

[156] H. Wu, C. Gu, Y. Cheng, X. Hu, *Appl. Phys. Lett.* **2017**, *111*, 062603.

[157] N. Calandri, Q.-Y. Zhao, D. Zhu, A. Dane, K. K. Berggren, *Appl. Phys. Lett.* **2016**, *109*, 152601.

[158] Y. Cheng, C. Gu, X. Hu, *Appl. Phys. Lett.* **2017**, *111*, 062604.

[159] Q. Zhao, L. Zhang, T. Jia, L. Kang, W. Xu, J. Chen, P. Wu, *Appl. Phys. B Lasers Opt.* **2011**, *104*, 673.

[160] K. V. Smirnov, A. V. Divochiy, Y. B. Vakhtomin, M. V. Sidorova, U. V. Karpova, P. V. Morozov, V. A. Seleznev, A. N. Zotova, D. Y. Vodolazov, *Appl. Phys. Lett.* **2016**, *109*, 052601.

[161] J. K. W. Yang, A. J. Kerman, E. A. Dauler, V. Anant, K. M. Rosfjord, K. K. Berggren, *IEEE Trans. Appl. Supercond.* **2007**, *17*, 581.

[162] M. Sidorova, A. Semenov, H. W. Hübers, I. Charaev, A. Kuzmin, S. Doerner, M. Siegel, *Phys. Rev. B* **2017**, *96*.

[163] A. G. Kozorezov, C. Lambert, F. Marsili, M. J. Stevens, V. B. Verma, J. P. Allmaras, M. D. Shaw, R. P. Mirin, S. W. Nam, *Phys. Rev. B* **2017**, *96*, 054507.

[164] B. Korzh, Q.-Y. Zhao, S. Frasca, D. Zhu, E. Ramirez, E. Bersin, M. Colangelo, A. E. Dane, A. D. Beyer, J. Allmaras, E. E. Wollman, K. K. Berggren, M. D. Shaw, M. D. Shaw, In *Conference on Lasers and Electro-Optics*; OSA: Washington, D.C., 2018; p. FW3F.3.

[165] J. Zhang, W. Slysz, A. Verevkin, O. Okunev, G. Chulkova, A. Korneev, A. Lipatov, G. N. Gol'tsman, R. Sobolewski, *IEEE Trans. Appl. Supercond.* **2003**, *13*, 180.

[166] R. Sobolewski, A. Verevkin, G. N. Gol'tsman, A. Lipatov, K. Wilsher, *IEEE Trans. Appl. Supercond.* **2003**, *13*, 1151.

[167] A. Korneev, A. Lipatov, O. Okunev, G. Chulkova, K. Smirnov, G. Gol'tsman, J. Zhang, W. Slysz, A. Verevkin, R. Sobolewski, *Microelectron. Eng.* **2003**, *69*, 274.

[168] G. Bertolini, A. Coche, *Semiconductor Detectors*; North Holland Publishing Co., 1968.

[169] L. You, X. Yang, Y. He, W. Zhang, D. Liu, W. Zhang, L. Zhang, L. Zhang, X. Liu, S. Chen, Z. Wang, X. Xie, *AIP Adv.* **2013**, *3*, 072135.

[170] F. Najafi, A. Dane, F. Bellei, Q. Zhao, K. A. Sunter, A. N. McCaughan, K. K. Berggren, *IEEE J. Sel. Top. Quantum Electron.* **2015**, *21*, 1.

[171] X. Yang, L. You, L. Zhang, C. Lv, H. Li, X. Liu, H. Zhou, Z. Wang, *IEEE Trans. Appl. Supercond.* **2017**, *28*, 1.

[172] A. J. Kerman, E. A. Dauler, W. E. Keicher, J. K. W. Yang, K. K. Berggren, G. Gol'tsman, B. Voronov, *Appl. Phys. Lett.* **2006**, *88*, 111116.

[173] J. Wu, L. You, S. Chen, H. Li, Y. He, C. Lv, Z. Wang, X. Xie, *Appl. Opt.* **2017**, *56*, 2195.

[174] C. Schuck, W. H. P. Pernice, H. X. Tang, *Sci. Rep.* **2013**, *3*, 1893.

[175] V. B. Verma, B. Korzh, F. Bussières, R. D. Horansky, S. D. Dyer, A. E. Lita, I. Vayshenker, F. Marsili, M. D. Shaw, H. Zbinden, R. P. Mirin, S. W. Nam, *Opt. Express* **2015**, *23*, 33792.

[176] A. J. Annunziata, D. F. Santavicca, J. D. Chudow, L. Frunzio, M. J. Rooks, A. Frydman, D. E. Prober, *IEEE Trans. Appl. Supercond.* **2009**, *19*, 327.

[177] L. N. Cooper, *Phys. Rev. Lett.* **1961**, *6*, 689.

[178] C. L. Lv, H. Zhou, H. Li, L. X. You, X. Y. Liu, Y. Wang, W. J. Zhang, S. J. Chen, Z. Wang, X. M. Xie, *Supercond. Sci. Technol.* **2017**, *30*, 115018.

[179] Q. Zhao, A. N. McCaughan, A. E. Dane, F. Najafi, F. Bellei, D. De Fazio, K. A. Sunter, Y. Ivry, K. K. Berggren, *Opt. Express* **2014**, *22*, 24574.

[180] M. Ejrnaes, A. Casaburi, R. Cristiano, O. Quaranta, S. Marchetti, N. Martucciello, S. Pagano, A. Gaggero, F. Mattioli, R. Leoni, P. Cavalier, J. C. VilĺÍgier, *Appl. Phys. Lett.* **2009**, *95*.

[181] Q. Wang, J. J. Renema, A. Engel, M. P. van Exter, M. J. A. de Dood, *Opt. Express* **2015**, *23*, 24873.

[182] R. Gaudio, K. P. M. Op ’T Hoog, Z. Zhou, D. Sahin, A. Fiore, *Appl. Phys. Lett.* **2014**, *105*, 222602.

[183] H. L. Hortensius, E. F. C. Driessen, T. M. Klapwijk, *IEEE Trans. Appl. Supercond.* **2013**, *23*, 2200705.

[184] M. Sidorova, A. Semenov, A. Kuzmin, I. Charaev, S. Doerner, M. Siegel, *IEEE Trans. Appl. Supercond.* **2018**, *28*.

[185] O. Kahl, S. Ferrari, V. Kovalyuk, G. N. Goltsman, A. Korneev, W. H. P. Pernice, *Sci. Rep.* **2015**, *5*, 10941.

[186] M. K. Akhlaghi, E. Schelew, J. F. Young, *Nat. Commun.* **2015**, *6*, 8233.

[187] V. Shcheslavskiy, P. Morozov, A. Divochiy, Y. Vakhtomin, K. Smirnov, W. Becker, *Rev. Sci. Instrum.* **2016**, *87*, 053117.

[188] J. Huang, W. Zhang, L. You, C. Zhang, C. Lv, Y. Wang, X. Liu, H. Li, Z. Wang, *Supercond. Sci. Technol.* **2018**, *31*, 074001.

[189] E. A. Dauler, A. J. Kerman, B. S. Robinson, J. K. W. Yang, B. Voronov, G. Goltsman, S. A. Hamilton, K. K. Berggren, *J. Mod. Opt.* **2009**, *56*, 364.

[190] E. A. Dauler, B. S. Robinson, A. J. Kerman, J. K. W. Yang, K. M. Rosfjord, V. Anant, B. Voronov, G. Gol'tsman, K. K. Berggren, *IEEE Trans. Appl. Supercond.* **2007**, *17*, 279.

[191] A. Migdall, S. V Polyakov, J. Fan, J. C. Bienfang, *Exp. Methods Phys. Sci.* **2013**, *45*, 64.

[192] L. Costrell, *Accurate Determination of the Deadtime and Recovery Characteristics of Geig er-Muller Counters*; 1949; Vol. 42.

[193] A. J. Annunziata, O. Quaranta, D. F. Santavicca, A. Casaburi, L. Frunzio, M. Ejrnaes, M. J. Rooks, R. Cristiano, S. Pagano, A. Frydman, D. E. Prober, *J. Appl. Phys.* **2010**, *108*, 084507.

[194] R. Cheng, M. Poot, X. Guo, L. Fan, H. X. Tang, *IEEE Trans. Appl. Supercond.* **2017**, *27*, 1.

[195] G. Goltsman, A. Korneev, A. Divochiy, O. Minaeva, M. Tarkhov, N. Kaurova, V. Seleznev, B. Voronov, O. Okunev, A. Antipov, K. Smirnov, Y. Vachtomin, I. Milostnaya, G. Chulkova, *J. Mod. Opt.* **2009**, *56*, 1670.

[196] T. I. Baturina, S. V. Postolova, A. Y. Mironov, A. Glatz, M. R. Baklanov, V. M. Vinokur, *EPL* **2012**, *97*, 17012.

[197] W. J. Zhang, L. X. You, H. Li, J. Huang, C. L. Lv, L. Zhang, X. Y. Liu, J. J. Wu, Z. Wang, X. M. Xie, *Sci. China Physics, Mech. Astron.* **2017**, *60*, 120314.

[198] M. G. Tanner, C. M. Natarajan, V. K. Pottapenjara, J. A. O'Connor, R. J. Warburton, R. H. Hadfield, B. Baek, S. Nam, S. N. Dorenbos, E. B. Urea, T. Zijlstra, T. M. Klapwijk, V. Zwiller, *Appl. Phys. Lett.* **2010**, *96*, 221109.

[199] S. N. Dorenbos, E. M. Reiger, U. Perinetti, V. Zwiller, T. Zijlstra, T. M. Klapwijk, *Appl. Phys. Lett.* **2008**, *93*, 131101.

[200] R. H. Hadfield, A. J. Miller, S. W. Nam, R. L. Kautz, R. E. Schwall, *Appl. Phys. Lett.* **2005**, *87*, 1.

[201] A. J. Kerman, D. Rosenberg, R. J. Molnar, E. A. Dauler, *J. Appl. Phys.* **2013**, *113*, 144511.

[202] M. Ejrnaes, R. Cristiano, O. Quaranta, S. Pagano, A. Gaggero, F. Mattioli, R. Leoni, B. Voronov, G. Gol'Tsman, *Appl. Phys. Lett.* **2007**, *91*, 262509.

[203] V. B. Verma, A. E. Lita, M. J. Stevens, R. P. Mirin, S. W. Nam, *Appl. Phys. Lett.* **2016**, *108*, 131108.

[204] V. B. Verma, R. Horansky, F. Marsili, J. A. Stern, M. D. Shaw, A. E. Lita, R. P. Mirin, S. W. Nam, *Appl. Phys. Lett.* **2014**, *104*, 051115.

[205] Q. Zhao, A. McCaughan, F. Bellei, F. Najafi, D. De Fazio, A. Dane, Y. Ivry, K. K. Berggren, *Appl. Phys. Lett.* **2013**, *103*, 142602.

[206] F. Najafi, F. Marsili, E. Dauler, R. J. Molnar, K. K. Berggren, *Appl. Phys. Lett.* **2012**, *100*, 152602.

[207] E. A. Dauler, A. J. Kerman, B. S. Robinson, J. K. W. Yang, B. Voronov, G. Gol'tsman, S. A. Hamilton, K. K. Berggren, *arXiv Prepr. arXiv0805.2397* **2008**.

[208] R. Radebaugh, *Proc. IEEE* **2004**, *92*, 1719.

[209] R. Radebaugh, *J. Phys. Condens. Matter* **2009**, *21*, 164219.

[210] R. H. Hadfield, M. J. Stevens, S. S. Gruber, A. J. Miller, R. E. Schwall, R. P. Mirin, S. W. Nam, *Opt. Express* **2005**, *13*, 10846.

[211] A. Engel, K. Inderbitzin, A. Schilling, R. Lusche, A. Semenov, H. W. Hübers, D. Henrich, M. Hofherr, K. Il'In, M. Siegel, *IEEE Trans. Appl. Supercond.* **2013**, *23*, 2300505.

[212] T. Jia, C. Wan, L. Zhao, Y. Zhou, Q. Zhao, M. Gu, X. Jia, L. Zhang, B. Jin, J. Chen, L. Kang,

*Chinese Sci. Bull.* **2014**, *59*, 3549.

[213] S. Miki, M. Fujiwara, M. Sasaki, Z. Wang, *IEEE Trans. Appl. Supercond.* **2009**, *19*, 332.

[214] N. R. Gemmell, M. Hills, T. Bradshaw, T. Rawlings, B. Green, R. M. Heath, K. Tsimvrakidis, S. Dobrovolskiy, V. Zwiller, S. N. Dorenbos, M. Crook, R. H. Hadfield, *Supercond. Sci. Technol.* **2017**, *30*, 11LT01.

[215] T. Yamashita, S. Miki, W. Qiu, M. Fujiwara, M. Sasaki, Z. Wang, *Appl. Phys. Express* **2010**, *3*, 102502.

[216] T. Yamashita, S. Miki, K. Makise, W. Qiu, H. Terai, M. Fujiwara, M. Sasaki, Z. Wang, *Appl. Phys. Lett.* **2011**, *99*, 161105.

[217] H. Shibata, K. Fukao, N. Kirigane, S. Karimoto, H. Yamamoto, *IEEE Trans. Appl. Supercond.* **2017**, *27*, 1.

[218] H. Shibata, K. Shimizu, H. Takesue, Y. Tokura, *Appl. Phys. Express* **2013**, *6*, 072801.

[219] H. Shibata, H. Takesue, T. Honjo, T. Akazaki, Y. Tokura, *Appl. Phys. Lett.* **2010**, *97*, 212504.

[220] H. Shibata, T. Akazaki, Y. Tokura, *Appl. Phys. Express* **2013**, *6*, 6.

[221] R. Arpaia, M. Ejrnaes, L. Parlato, F. Tafuri, R. Cristiano, D. Golubev, R. Sobolewski, T. Bauch, F. Lombardi, G. P. Pepe, *Phys. C* **2015**, *509*, 16.

[222] M. Ejrnaes, L. Parlato, R. Arpaia, T. Bauch, F. Lombardi, R. Cristiano, F. Tafuri, G. P. Pepe, *Supercond. Sci. Technol.* **2017**, *30*, 12LT02.

[223] F. Marsili, F. Bellei, F. Najafi, A. E. Dane, E. A. Dauler, R. J. Molnar, K. K. Berggren, *Nano Lett.* **2012**, *12*, 4799.

[224] *Comment: A similar argument regarding the effects of photon energy on the detection via the breakage of Cooper pairs can be rephrased also by revolving the detection mechanism to vortices,*

*when considering that higher input energy gives rise to higher probability*.

[225] A. Korneev, Y. Korneeva, I. Florya, B. Voronov, G. Goltsman, *Phys. Procedia* **2012**, *36*, 72.

[226] K. Inderbitzin, A. Engel, A. Schilling, *IEEE Trans. Appl. Supercond.* **2013**, *23*, 2200505.

[227] H. Azzouz, S. N. Dorenbos, D. De Vries, E. B. Ureña, V. Zwiller, *AIP Adv.* **2012**, *2*, 032124.

[228] M. Marksteiner, A. Divochiy, M. Sclafani, P. Haslinger, H. Ulbricht, A. Korneev, A. Semenov, G. Gol'tsman, M. Arndt, *Nanotechnology* **2009**, *20*, 455501.

[229] N. V Minaev, M. A. Tarkhov, D. S. Dudova, P. S. Timashev, B. N. Chichkov, V. N. Bagratashvili, *Laser Phys. Lett.* **2018**, *15*, 026002.

[230] P. I. Zolotov, A. V Divochiy, Y. B. Vakhtomin, P. V Morozov, V. A. Seleznev, K. V Smirnov, *J. Phys. Conf. Ser.* **2017**, *917*, 062037.

[231] Q. Guo, L. You, H. Li, W. Zhang, L. Zhang, X. Liu, X. Yang, S. Chen, Z. Wang, X. Xie, *IEEE Trans. Appl. Supercond.* **2017**, *27*, 1.

[232] A. Engel, A. Aeschbacher, K. Inderbitzin, A. Schilling, K. Il'In, M. Hofherr, M. Siegel, A. Semenov, H. W. Hübers, *Appl. Phys. Lett.* **2012**, *100*, 062601.

[233] K. Il'In, M. Hofherr, D. Rall, M. Siegel, A. Semenov, A. Engel, K. Inderbitzin, A. Aeschbacher, A. Schilling, *J. Low Temp. Phys.* **2012**, *167*, 809.

[234] T. Jia, L. Kang, L. Zhang, Q. Zhao, M. Gu, J. Qiu, J. Chen, B. Jin, *Appl. Phys. B* **2014**, *116*, 991.

[235] J. Li, R. A. Kirkwood, L. J. Baker, D. Bosworth, K. Erotokritou, A. Banerjee, R. M. Heath, C. M. Natarajan, Z. H. Barber, M. Sorel, R. H. Hadfield, *Opt. Express* **2016**, *24*, 13931.

[236] M. Caloz, B. Korzh, N. Timoney, M. Weiss, S. Gariglio, R. J. Warburton, C. Schönenberger, J. Renema, H. Zbinden, F. Bussières, *Appl. Phys. Lett.* **2017**, *110*, 083106.

[237] Y. Korneeva, I. Florya, S. Vdovichev, M. Moshkova, N. Simonov, N. Kaurova, A. Korneev, G.

Goltsman, *IEEE Trans. Appl. Supercond.* **2017**, *27*, 1.

[238] C. Delacour, J. Claudon, J. P. Poizat, B. Pannetier, V. Bouchiat, R. E. De Lamaestre, J. C. Villegier, M. Tarkhov, A. Korneev, B. Voronov, G. Gol'Tsman, *Appl. Phys. Lett.* **2007**, *90*, 191116.

[239] A. Engel, J. J. Renema, K. Il'In, A. Semenov, *Supercond. Sci. Technol.* **2015**, *28*, 114003.

[240] S. Chen, L. You, W. Zhang, X. Yang, H. Li, L. Zhang, Z. Wang, X. Xie, *Opt. Express* **2015**, *23*, 10786.

[241] R. Meservey, P. M. Tedrow, *J. Appl. Phys.* **1969**, *40*, 2028.

[242] A. J. Kerman, J. K. W. Yang, R. J. Molnar, E. A. Dauler, K. K. Berggren, *Phys. Rev. B* **2009**, *79*, 100509.

[243] B. T. Matthias, T. H. Geballe, V. B. Compton, *Rev. Mod. Phys.* **1963**, *35*, 1.

[244] E. A. Antonova, D. R. Dzhuraev, G. P. Motulevich, V. A. Sukhov, *Zhurnal Ehksperimental'noj i Teor. Fiz.* **1981**, *80*, 2426.

[245] D. Hazra, N. Tsavdaris, S. Jebari, A. Grimm, F. Blanchet, F. Mercier, E. Blanquet, C. Chapelier, M. Hofheinz, *Supercond. Sci. Technol.* **2016**, *29*, 105011.

[246] T. Hong, K. Choi, K. I. Sim, T. Ha, B. C. Park, H. Yamamori, J. H. Kim, *J. Appl. Phys.* **2013**, *114*, 243905.

[247] D. K. Finnemore, T. F. Stromberg, C. A. Swenson, *Phys. Rev.* **1966**, *149*, 231.

[248] R. P. Huebener, R. T. Kampwirth, R. L. Martin, T. W. Barbee, R. B. Zubeck, *J. Low Temp. Phys.* **1975**, *19*, 247.

[249] P. Townsend, J. Sutton, *Phys. Rev.* **1962**, *128*, 591.

[250] S. Kondo, *J. Mater. Res.* **1992**, *7*, 853.

[251] X. Zhang, A. Engel, Q. Wang, A. Schilling, A. Semenov, M. Sidorova, H. W. Hübers, I. Charaev, K. Ilin, M. Siegel, *Phys. Rev. B* **2016**, *94*, 174509.

[252] A. W. Smith, T. W. Clinton, C. C. Tsuei, C. J. Lobb, *Phys. Rev. B* **1994**, *49*, 12927.

[253] S. Kubo, *J. Appl. Phys.* **1988**, *63*, 2033.

[254] A. Y. Rusanov, M. B. S. Hesselberth, J. Aarts, *Phys. Rev. B* **2004**, *70*, 024510.

[255] J. Nagamatsu, N. Nakagawa, T. Muranaka, Y. Zenitani, J. Akimitsu, *Nature* **2001**, *410*, 63.

[256] D. G. Hinks, H. Claus, J. D. Jorgensen, *Nature* **2001**, *411*, 457.

[257] D. K. Finnemore, J. E. Ostenson, S. L. Bud’ko, G. Lapertot, P. C. Canfield, *Phys. Rev. Lett.* **2001**, *86*, 2420.

## 8. List of symbols

$\alpha$ – Absorptance.

$A$ – Material-depended coefficient.

$B$ – Material-depended exponent.

$c$ – Speed of light.

$\delta T$ – Thermal fluctuations.

$\delta I$ – Electronic fluctuations.

$\Delta$ – Superconducting electron-electron binding-energy.

$\Delta_0$ – Superconducting electron-electron binding-energy at $T = 0K$.

$d$ – Thickness.

$D$ – Diffusivity.

$e$ – Euler’s constant.

$g_1$ – Electron-electron interaction constant.

$\eta_{abs}$ – Probability of photon absorption ($\eta_{abs||}, \eta_{abs\perp}$ - absorption probability for photons with parallel and perpendicular polarization, respectively).

$\eta_{cpl}$ – Device-source coupling coefficient.

$\eta_{qe}$ – Quantum efficiency.

$h$ – Planck's constant.

$\hbar$ – Reduced Planck's constant

$I_b$ – Bias current.

$I_{b_n}$ – Normalized bias current.

$I_c$ – Critical current.

$J_c$ – Critical current density.

$k_B$ – Boltzmann constant.

λ – Wavelength.

$\lambda^*$ – Electric-impurity coupling constant.

$l$ – Length.

$L$ - Inductivity.

$L_k$ – Kinetic inductance.

$L_{k_\square}$ – Sheet kinetic inductance.

$\mu^*$ – Magnetic-impurity coupling constant.

$n$ – Energy-loss factor.

$N(0)$ – Density of states at the Fermi level (metallic).

$q_e$ – Electron charge.

ρ – Resistivity.

$R_n$ – Normal-state resistance.

$R_{\mathrm{shunt}}$ – Shunt resistance.

$R_{\square}$ – Sheet resistance.

$\sigma_{\mathrm{n}}$ – Conductivity.

$\varsigma$ – Multiplication efficiency of quasiparticles.

$\tau$ – Timing jitter.

$\tau_{\mathrm{d}}$ – Fundamental device jitter.

$\tau_{\mathrm{dd}}$ – Geometric and design contribution to the jitter.

$\tau_{\mathrm{di}}$ – Jitter contribution from intrinsic material properties.

$\tau_{\mathrm{e}}$ – Electronics contribution to the jitter.

$\tau_{\mathrm{mfp}}$ – Electron mean-free-path time

$\tau_{\mathrm{s}}$ – Uncertainty in timing of the single-photon source.

$\tau_{\mathrm{th}}$ – Electron thermalization time.

$\Theta_{\mathrm{D}}$ – Debay temperature.

$t_{\mathrm{dead}}$ – Dead time.

$t_{\mathrm{rise}}$ – Rise time.

$t_{\mathrm{reset}}$ – Reset time.

$T_{\mathrm{c}}$ – Superconducting critical temperature.

$T_{\mathrm{c}}^{0}$ – Unperturbed (bulk) superconducting critical temperature.

$T_{E}$ – Transverse electric polarization.

$w$ – Width.

$Z_{\mathrm{shunt}}$ – Shunt impedance.

## 9. List of abbreviations

ARC – Anti-reflection coating.

BCS – Bardeen, Cooper and Schrieffer's model for Cooper-pair dominated superconductivity.

BKT – Berezinskii, Kosterlitz and Thouless model for vortex-dominated superconductivity.

cps – Counts per second.

DCR – Dark-count (and false-count) rate.

DE – Device efficiency.

FWHM – Full-width half maximum.

IRF – Instrument response function.

ME – Multi element.

SDE – System detection efficiency.

SIT – Superconducting-to-insulator transition.

SNAP – Superconducting-nanowire avalanche photodetector.

SNSPD – Superconducting-nanowire single-photon detector.

SNR – Signal-to-noise ratio.

UV – Ultra-violate radiation.